\pdfoutput=1
\documentclass[11pt,twoside,a4paper,cmspaper,final,collab]{cms-tdr}
\def\svnVersion{f9538d6}\def\svnDate{2022/07/15}\def\cmsCernNoTag{CERN-EP-2021-142}\def\cmsCernDate{\today}\def\cmsMessage{Published in the Journal of High Energy Physics as \href{http://dx.doi.org/10.1007/JHEP07(2022)067}{\doi{10.1007/JHEP07(2022)067}.}}
\begin{document}\cmsNoteHeader{EXO-19-017}

\newlength\cmsTabSkip\setlength{\cmsTabSkip}{1ex}

\newcommand{\MT}{\ensuremath{M_\mathrm{T}}\xspace}
\newcommand{\MTlower}{\ensuremath{M_\mathrm{T}^\mathrm{min}}\xspace}
\newcommand{\WprimeKK}{\ensuremath{\PW_{\mathrm{KK}}}\xspace}
\newcommand{\WprimeKKn}{\ensuremath{\PW^{(n)}_{\mathrm{KK}}}\xspace}
\newcommand{\WprimeKKtwo}{\ensuremath{\PW^\mathrm{(2)}_{\mathrm{KK}}}\xspace}
\newcommand{\slep}[1][]{$\widetilde {\ell}$}
\newcommand{\MW}{M_{\PW}}
\newcommand{\MWp}{M_{\PWpr}}
\newcommand{\gW}{g_{\PW}}
\newcommand{\gWp}{g_{\PWpr}}
\newcommand{\Mstau}{M_{\sTau}}
\newcommand{\mH}{\mu_{\PH}}
\newcommand{\sigPT}{\sigma_{\pt}}
\newcommand{\xsecWp}{\sigma_{\PWpr}}

\providecommand{\cmsTable}[1]{\resizebox{\textwidth}{!}{#1}}

\cmsNoteHeader{EXO-19-017} 
\title{Search for new physics in the lepton plus missing transverse momentum final state in proton-proton collisions at \texorpdfstring{$\sqrt{s} = 13\TeV$}{sqrt(s) = 13 TeV}}

\date{\today}

\abstract{
  A search for physics beyond the standard model (SM) in final states with an electron or muon and missing transverse momentum is presented.
The analysis uses data from proton-proton collisions at a centre-of-mass energy of 13\TeV, collected with the CMS detector at the LHC in 2016--2018 and corresponding to an integrated luminosity of 138\fbinv.
No significant deviation from the SM prediction is observed.
Model-independent limits are set on the production cross section of {\PWpr} bosons decaying into lepton-plus-neutrino final states. 
Within the framework of the sequential standard model, with the combined results from the electron and muon decay channels a {\PWpr} boson with mass less than 5.7\TeV is excluded at 95\% confidence level.
Results on a SM precision test, the determination of the oblique electroweak $W$ parameter, are presented using LHC data for the first time.
These results together with those from the direct {\PWpr} resonance search are used to extend existing constraints on composite Higgs scenarios.
This is the first experimental exclusion on compositeness parameters 
using results from LHC data other than Higgs boson measurements. 
}

\hypersetup{
pdfauthor={CMS Collaboration},%
pdftitle={Search for new physics in the lepton plus missing transverse momentum final state in \texorpdfstring{$\Pp\Pp$}{pp} collisions at \texorpdfstring{$\sqrt{s} = 13\TeV$}{sqrt(s) = 13 TeV}},
pdfsubject={CMS},
pdfkeywords={CMS, Physics, Exotica, Muon, Electron, RPV-SUSY, Universal Extra Dimensions, Extra Dimensions, W boson, Composite Higgs,  W prime}
}

\maketitle

\section{Introduction}
\label{sec:intro}

Experimental evidence from the last half-century has established the standard model (SM) as the foundational theory of particle physics.
However, several fundamental aspects of Nature cannot be explained by the SM~\cite{PDG2020}. 
To address a variety of open issues, many models beyond-the-SM (BSM) have been proposed.
To resolve the hierarchy problem~\cite{PhysRevD.11.2558,PhysRevD.11.566}, some of these models propose additional heavy gauge bosons as remnants of electroweak (EW) symmetry breaking.
To achieve gauge coupling unification, other models invoke extended gauge sectors~\cite{PhysRevD.82.035011} that can give rise to extra spatial dimensions~\cite{ARKANIHAMED1998263,Appelquist:2000nn}.

This paper focuses on a search for a new heavy charged gauge boson (\PWpr) decaying into an energetic charged lepton (electron or muon) and a neutrino.
As neutrinos cannot be detected directly with the CMS detector, their presence is inferred from an imbalance in the transverse momentum, \ptvec, in the event; 
this imbalance is referred to as missing transverse momentum \ptvecmiss~\cite{JECref}, and is defined as the negative vector sum of the \pt of all reconstructed particles in an event. 
The transverse mass (\MT) of the charged lepton (\Pell) and the \ptvecmiss is used as the main discriminating variable of the search, and is given by
\begin{equation} 
\MT = \sqrt{2  \pt^{\Pell}   \ptmiss  \bigl(1 - \cos [\Delta \phi(\ptvec^{\Pell},\ptvecmiss)]\bigr)},   
\label{eqn:mt}      
\end{equation}   
where $\pt^{\Pell}$ is the magnitude of the lepton \ptvec, \ptmiss is the magnitude of \ptvecmiss, and $\Delta \phi(\ptvec^{\Pell},\ptvecmiss)$ is the azimuthal opening angle between the directions of the lepton and \ptvecmiss. 

In this search, we look for the presence of a resonant signal in the high mass tail of the \MT distribution, where the contributions from background processes are small.
The shape of the \MT distribution is studied using a binned likelihood method. 
This approach is especially powerful as the examined models predict different \MT distributions for different hypotheses, and
allow BSM signals to be detected through significant deviations from SM predictions in their kinematic distributions. 
Results are interpreted in the context of several BSM scenarios, involving the production and decay of a \PWpr boson in the sequential standard model (SSM)~\cite{reference-model} or 
of a Kaluza--Klein (KK) excitation of the \PW boson in a model with split universal extra dimensions (split-UED)~\cite{JHEP04(2010)081,PhysRevD.79.091702}.
Additionally, we consider the production and flavour violating decay of a slepton mediator in the \textit{R} parity violating supersymmetric (RPV SUSY) model~\cite{FarrarFayet, DreinerRPV}.

Searches for a \PWpr boson at the CERN LHC have been reported by the ATLAS and CMS Collaborations with data collected at 7\TeV~\cite{Aad:2012dm,Chatrchyan:2012meb}, 8\TeV~\cite{ATLAS:2014wra,Khachatryan:2014tva}, and 13\TeV~\cite{Aaboud:2017efa,Khachatryan:2016jww,Sirunyan:2018mpc}.
ATLAS reported the exclusion of an SSM \PWpr below 6.0\TeV in the combined electron and muon channels at 95\% confidence level (\CL)~\cite{Aad:2019wvl} with data collected from 2015 to 2018.

So-called ``Universal'' effects from new physics are also studied using effective field theory (EFT) interpretations.
This EFT approach quantifies potential deviations from the SM expectations through the oblique EW $W$ parameter~\cite{Peskin:1991sw}. 
Constraints on this parameter are derived in this paper, establishing a lower limit on the scale for universal new physics effects. 
This is the first time using LHC data that constraints have been set on such a parameter in the $\Pell{+}\ptmiss$ final state.
These results are the most stringent to date, significantly improving the constraint obtained from LEP data~\cite{Falkowski:2015krw,LEP-2}.  

An interpretation of the \PWpr boson search results is also performed in the context of composite Higgs boson models~\cite{Giudice_2007}.
These models  predict that the Higgs boson is a composite particle emerging from a strongly interacting sector. Other new composite states are assumed to be produced, such as a \PWpr boson. 
We use the direct resonance search and the indirect information obtained from the oblique $W$ parameter determination in the $\Pell{+}\ptmiss$ final state, 
to place new constraints on composite Higgs scenarios~\cite{Wulzer_2017}. 
Further regions are excluded using recent CMS Higgs boson cross section results~\cite{Higgs_combination}.  

This analysis uses proton-proton ($\Pp\Pp$) collision data corresponding to an integrated luminosity of 138\fbinv at $\sqrt{s} = 13\TeV$, recorded using the CMS detector at the LHC during 2016--2018. 
Unless stated otherwise, the results of the 2017--2018 data analysis presented here are combined statistically with the published results~\cite{Sirunyan:2018mpc} using the 2016 data corresponding to 36\fbinv without reprocessing them in the analysis.

The paper is organized as follows.
Section~\ref{sec:detector} gives a brief description of the CMS detector. 
A description of the signal models and simulated event samples is given in Section~\ref{sec:models}. 
The event reconstruction and particle identification are given in Section~\ref{sec:object}, and the event selection procedures and background estimation are described in Section~\ref{sec:selection}. 
Systematic uncertainties are described in Section~\ref{sec:systematic}. The results and their statistical interpretation are given in Section~\ref{sec:result}. The paper is summarized in Section~\ref{sec:summary}.

Tabulated results are provided in the HEPData record for this analysis~\cite{hepdata}.

\section{The CMS detector}
\label{sec:detector}

The central feature of the CMS apparatus is a superconducting solenoid of 6\unit{m} internal diameter, providing a magnetic field of 3.8\unit{T}. 
Within the solenoid volume are a silicon pixel and strip tracker, a lead tungstate crystal electromagnetic calorimeter (ECAL), and a brass and scintillator hadron calorimeter (HCAL).
The calorimeters are each composed of a barrel and two endcap sections. 
Forward calorimeters extend the pseudorapidity ($\eta$) coverage provided by the barrel and endcap detectors. 
Muons are detected in gas-ionization detectors embedded in the steel flux-return yoke outside the solenoid.
 
Events of interest are selected using a two-tiered trigger system. The first level~\cite{Sirunyan:2020zal,Sirunyan:2021zrd}, composed of custom hardware processors, uses information from the calorimeters and muon detectors to select events at a rate of around 100\unit{kHz} within a fixed latency of about 4\mus. 
The second level, known as the high-level trigger~\cite{Khachatryan:2016bia}, consists of a farm of processors running a version of the full event reconstruction software optimized for fast processing, and reduces the event rate to around 1\unit{kHz} before data storage. 

A more detailed description of the CMS detector, together with a definition of the coordinate system used and the relevant kinematic variables, can be found in Ref.~\cite{Chatrchyan:2008aa}.

\section{Physics models and event simulation}
\label{sec:models}
Many BSM models predict the existence of new heavy particles.
Those new particles of interest to this analysis are mediators that decay to a charged lepton and a neutrino.
Several theoretical interpretations are investigated by this analysis.
In this Section, we describe these models and their assumptions, together with the Monte Carlo event generators used for the simulation of signal events.
The simulations of the SM background events used in this analysis are also presented.

\subsection{Sequential Standard Model \texorpdfstring{\PWpr}{W'} boson} 
\label{sec:models-ssm}
\begin{figure}[hbtp]
\centering
\includegraphics[width=.9\textwidth]{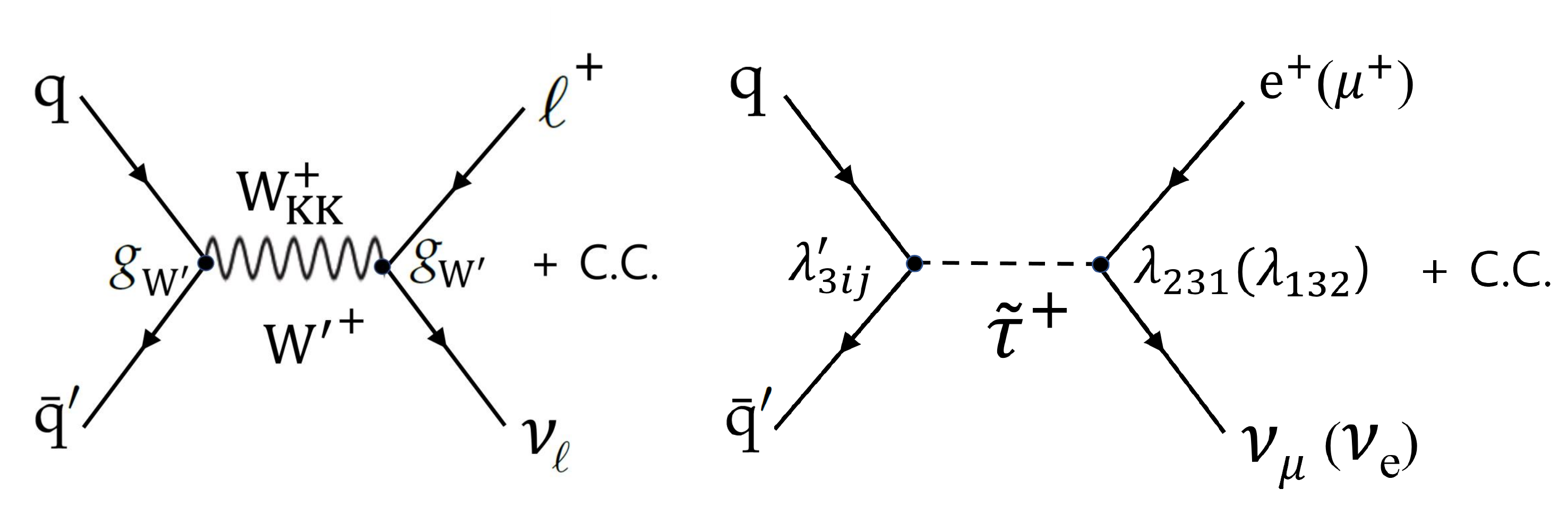}
\caption{Feynman diagram for the production and decay of a new heavy boson, an SSM \PWpr or a Kaluza--Klein excitation mode of W (\WprimeKK) (left). The coupling strength, $\gWp$, is allowed to vary. 
In RPV SUSY, a \PGt slepton (\PSGt) could also act as a mediator (right) with the corresponding coupling strength, \Lam, for the decay. This coupling strength is allowed to be different between the two final states, denoted by $\Lam_{231}$ and $\Lam_{132}$ for the electron and muon final states, respectively.
Here, C.C. stands for charge conjugation.}
\label{fig:feynman}
\end{figure}

The SSM~\cite{reference-model} has been used as a benchmark model in which the \PWpr boson is considered to be a heavy analogue of the SM \PW boson.
In the SSM, the \PWpr boson exhibits the same couplings as those of the SM \PW boson. 
In this analysis, the $\PQt\PAQb$ decay channel is allowed, while diboson decays are considered to be suppressed.
These assumptions yield a predicted branching fraction ($\mathcal{B}$) of about 8.5\% for each of the leptonic channels~\cite{reference-model}.
The expected intrinsic width of a 1\TeV \PWpr boson is about 33\GeV.
The leading order Feynman diagram showing the production of a charged \PWpr boson in a quark-antiquark ($\PQq\PAQq^{\prime}$) interaction is depicted in Fig.~\ref{fig:feynman} (left).

The signal events for the SSM $\PWpr\to\Pe/\PGm{+}\PGn$ channels are generated using \PYTHIA v8.230~\cite{Sjostrand:2014zea} at leading order (LO) accuracy in quantum chromodynamics (QCD) with the NNPDF3.1 NNLO \cite{NNPDF31} parton distribution function (PDF) set. The masses of the \PWpr boson ($\MWp$) are varied in steps of 0.2\TeV from 0.2 to 6.4\TeV.
Regarding the shape of the \MT distribution, as $\MWp$ increases, the fraction of off-shell production increases in the low \MT region as predicted by the PDFs.
A mass-dependent $K$ factor~\cite{Vogt:2002eu}, which is defined as the ratio of the next-to-next-to-LO (NNLO) to the LO cross section, is calculated using \FEWZ v3.1~\cite{fewz3} with the NNPDF3.1 NNLO PDF set, and varies from 1.1 ($\MWp = 0.2$ and 6.0\TeV) to 1.3 ($\MWp = 3.0\TeV$).
This $K$ factor for the signal is used to achieve QCD NNLO precision in production cross sections. 

\subsection{Varying coupling strength in the Sequential Standard Model}
\label{sec:models-gc}
The \PWpr boson coupling strength, $\gWp$, can be given in terms of the SM weak coupling strength $\gW$.
The coupling ratio, $\gWp/\gW,$ is set to unity in the SSM model; however, various coupling strengths could be possible.
Variations in the \PWpr boson's coupling will affect its width and consequently its \MT distribution.
Signal samples for a range of coupling ratios were generated using \MGvATNLO~\cite{Alwall:2014hca} at LO with the NNPDF2.3 LO~\cite{NNPDF23} PDF set for the electron and muon channels.
The coupling ratios range from $10^{-2}$ to 3.
These signals exhibit different widths as well as different cross sections.

The generated distributions of \PYTHIA samples are reweighted to include the decay width dependence and provide the appropriate \MT distributions.
For a coupling ratio of unity, the theoretical LO cross sections apply. 
For all other coupling ratios, the theoretical cross sections scale with the square of the coupling strength, $\gWp$.

\subsection{Split-UED model}
\label{sec:models-ued}
Other models in which a mediator may decay into a charged lepton and a neutrino are characterized by the existence of universal extra dimensions.
In particular, we consider the split-UED model~\cite{Appelquist:2000nn, PhysRevD.79.091702, JHEP04(2010)081}.
This model extends the conventional notion of space-time to include an additional spatial dimension of radius $R$.
Here, all SM particles have a corresponding Kaluza--Klein (KK) partner. The mediator, \WprimeKKn, couples to SM fermions; here, $n$ refers to the KK excitation mode of the \WprimeKKn boson and takes only positive integer values.
More compact dimensions, \ie smaller values of $R$, require larger \WprimeKKn boson masses as given by the relation $M(\WprimeKKn) = \sqrt{\smash[b]{\MW^{2} + (n/R)^{2}}}$, where $\MW$ is the \PW boson mass.
The split-UED model requires a bulk mass parameter $\mu$ for the fermion field in five dimensions.
The special case of minimal UED corresponds to the limit of vanishing bulk mass.
The \WprimeKKn couplings to SM fermions depend on $\mu$.
Additionally, the product of the cross section for \WprimeKKn boson production and its branching fraction to SM fermions goes to zero as $\mu$ goes to zero.
One of the important features of this model is the conservation of KK-parity.
The $n =$ odd (even) KK modes must have odd (even) KK-parity, and the lightest odd KK-parity particle cannot directly couple to the even modes.
Therefore, the conservation of KK-parity forbids coupling of KK-odd modes of the \WprimeKKn boson to SM fermions (zero mode).
For the $n = 2$ mode, the \WprimeKKtwo boson decay to leptons is kinematically similar to the SSM \PWpr boson decay (Fig.~\ref{fig:feynman} left). 
This allows the limits obtained from the $\PWpr\to\Pe\PGn$ and $\PWpr\to\PGm\PGn$ searches to be reinterpreted directly in terms of the $M(\WprimeKKtwo)$, provided the differences in the widths of the resonances have little impact on the results.

Signal samples are produced at the generator level using \PYTHIA at LO with the NNPDF2.3 LO PDF set. 
This is done for the ranges of 0.2--3.0\TeV for the $1\!/\!R$ parameter and of 0.05--10.00\TeV for $\mu$.
The $1\!/\!R$ range corresponds to an $M(\WprimeKKtwo)$ range of 0.4--6.0\TeV.
The mass-dependent $K$ factors from the SSM \PWpr boson interpretation are used.

\subsection{\texorpdfstring{$\textit{R}$}{R} parity violating SUSY with a slepton mediator} 
\label{sec:models-rpv}
It is possible for a SUSY slepton (\slep) by virtue of \textit{R} parity violation to undergo a lepton flavour violating decay to a charged lepton and a neutrino~\cite{DreinerRPV, FarrarFayet}.
The analysis focuses on a \PGt slepton (\sTau) decaying to $\Pe\PGnGm$ or $\PGm\PGne$ channels as illustrated in Fig.~\ref{fig:feynman} (right).
The coupling at the production vertex is always a function of the hadronic-leptonic RPV coupling $\Lamp_{3ij}$ ($i, j$ = 1, 2, 3).
This is the coupling strength related to the third generation, specifically the \sTau.
The coupling at the decay vertex is governed by $\Lam_{231}$ ($\Lam_{132}$) for the decay to $\Pe\PGnGm$ ($\PGm\PGne$).

The signal samples, generated using \MGvATNLO at LO with the NNPDF2.3 LO PDF set, are for \sTau masses, $\Mstau$, of 0.4--6.0\TeV and coupling strength parameters $\Lam_{231}$, $\Lam_{132}$, and $\Lamp_{3ij}$ of 0.05--0.50.
The cross sections for $\Lam_{231}$ (or $\Lam_{132}$)=$\Lamp_{3ij}$= 0.1 are used as references.
As examples, at these couplings the cross sections are 82, 0.92, and 0.018\unit{fb} for both the electron and muon channels when $\Mstau$ = 1, 3, and 5\TeV.

\subsection{Universal new physics through constraints on the \texorpdfstring{$W$}{W} parameter}
\label{sec:models-W-parameter}
One possible way to incorporate effects from new physics is to consider dimension-6 operators involving only \PH, \PW, and \PB fields.
Here, \PH is the Higgs boson doublet, \PW is the weak isospin triplet, and \PB is the hypercharge field, which can be expressed as the mixture of photon and \PZ boson fields. 
This BSM extension would involve new physics effects related to EW symmetry breaking, without the introduction of either new strong interactions or flavour violations up to the scale of new physics currently considered. 
In this context, there are four parameters, $S$, $T$, $W$, and $Y$~\cite{Peskin:1991sw}, quantifying the universal deviations. They correspond to operators that modify the propagators of the SM EW vector bosons both on pole ($S$ and $T$) and off-pole ($W$ and $Y$). The terms accompanying $W$ and $Y$ grow with the centre-of-mass energy of the process under study, thus LHC is particularly suited to study observables like $W$ and $Y$, in the tail of relevant distributions.
These parameters are constrained by the EW precision measurements, mainly from the CERN LEP and LHC colliders~\cite{Farina_2017}.
These parameters are further constrained by this analysis.

In the specific case of four-fermion contact interactions with difermion final states, LHC data is sensitive to the oblique $W$ and $Y$ parameters through the interpretation of the dilepton mass distribution. 
In the SM assumption, the values of the $W$ and $Y$ parameters are zero. In the case of nonvanishing $W$ and $Y$ parameter values, and in the $\Pell\PGn$ channel, deviations (either an excess or a deficit) in the \MT distribution would be manifested, especially in the high-\MT region.
This expected effect is implemented in the signal events via a reweighting technique (details in Section~\ref{sec:W-parameter_limit}) applied on the events from the simulated SM background samples.
These predictions are then compared to data to place constraints on the $W$ parameter. 
This procedure assumes that the new resonance lies beyond the current kinematic limit. 
Deviations from the SM predictions in the mass region between the SM \PW boson and the hypothetical new resonance could, however, still be observed.

\subsection{Composite Higgs boson models}
\label{sec:models-Higgs-composite}
A reinterpretation of the results from the \PWpr boson search is done in the framework of composite Higgs boson models. 
These BSM models predict the Higgs boson to be a composite particle associated with strong interaction dynamics at a higher energy scale~\cite{Giudice_2007}. 
Two parameters characterize the dynamics: $m_{*}$ and $g_{*}$, the mass scale of the new spectrum of composite resonances and the coupling associated to the new interaction, respectively.
The input for this reinterpretation comes in two complementary ways.
The first uses the limits on the \PWpr coupling, $\gWp$, presented in Section~\ref{sec:models-gc} and with values later shown in Section~\ref{sec:coupling_ratio_limit}, obtained from the direct \PWpr search. 
The second uses the constraints on the oblique $W$ parameter presented in Section~\ref{sec:models-W-parameter} and with values shown in Section~\ref{sec:W-parameter_limit}. 

In the first case, the \PWpr boson is assumed to be a state composed of more fundamental particles.
The gauge coupling to the new constituents, $g_{*}$, is screened by the presence of the resonance so that it can be seen at low energy in the form of $g^2/g_{*}$~\cite{Wulzer_2017}, \ie
\begin{equation}
\gWp = \frac{g^{2}}{g_{*}},
\label{eqn:direct_Wparam}
\end{equation}
where $g$ is the $\mathrm{SU(2)}_{\PL}$ SM gauge coupling and $\gWp$ is the \PWpr boson coupling strength to fermions, present at both the resonance production and decay vertices. 

In the second case, the value we obtain for the $W$ parameter in this analysis is used to quantify deviations from the SM using the relation~\cite{Farina_2017}
\begin{equation}
{g_{*}}^{2} = \frac{g^{2}\,\MW^{2}}{W m_{*}^{2}}.
\label{eqn:indirect_Wparam}
\end{equation}
Using a third approach, additional constraints are placed in the $m_{*}$--$g_{*}$ plane when applying current measurements of the Higgs boson cross section from a combination of production mode and decay channels~\cite{Higgs_combination}. 
In this case, the new interactions would modify the SM predictions for Higgs production and decay, generating an extra contribution to the Higgs kinetic term in the effective Lagrangian~\cite{Giudice_2007}. The corresponding modification would scale as 
\begin{equation}
\Delta{\mH} = \frac{{g_{*}}^{2}\,v^{2}}{m_{*}^{2}}, 
\label{eqn:Higgs_xsec}
\end{equation}
where $\mH$ is the measured Higgs boson cross section relative to that of its SM cross section and $v$ is the EW symmetry breaking scale.

Diboson decay channels are allowed in this model, following the \PWpr description in the Heavy Vector Triplet (HVT) model~\cite{HVT_model}, in contrast to the physics models described in Sections~\ref{sec:models-ssm} to~\ref{sec:models-W-parameter}, where they are assumed to be suppressed.

\subsection{Background simulation}
\label{sec:simulation}

Simulated event samples are used to estimate the event yield of the SM background processes. 
In the following we describe the simulation of the background samples used to analyze the data obtained in 2017--2018.
The generation of background samples for the analysis of the 2016 data is described in Ref.~\cite{Sirunyan:2018mpc}.
For the 2017--2018 samples, the parton showering and hadronization are simulated with \PYTHIA v8.230~\cite{Sjostrand:2014zea}, and the CP5 underlying event tune~\cite{Sirunyan:2019dfx} with NNPDF3.1 NNLO~\cite{NNPDF31} PDF modelling is used.

The main irreducible background process is the SM $\PW\to\Pell\PGn$ decay, with $\Pell=\Pe, \PGm$, and \PGt.
The background from $\PW\to\PGt\PGn$ decay is taken into consideration when the \PGt lepton decays into an electron or a muon.
To estimate the dominant \PW boson background precisely, a combination of three different samples is used.

For the high-mass region ($\MW > 0.1\TeV$), the off-shell W background sample is generated at LO with \PYTHIA, ensuring enough simulated events in the high mass region, up to 7\TeV.
To normalize the LO cross section to EW next-to-LO (NLO) and the QCD NNLO precision, $K$ factors are calculated as a function of $\MW$.
The EW NLO corrections are derived using the {\footnotesize MCSANC} v1.2 event generator~\cite{mcsanc_v1,mcsanv_2} with the NNPDF3.1\_nnlo\_as0118\_luxqed PDF set~\cite{Bertone_2018, Manohar_2016}. 
The QCD NNLO corrections are computed with the \FEWZ v3.1 package~\cite{fewz3}.
These corrections are combined with an additive method~\cite{Butterworth_2016}. 
The $K$ factor applied is largest at low mass, with a value of $1.20\pm0.04$ at $\MW = 0.2\TeV$, falling to $0.89\pm0.19$ at high mass ($\MW = 6\TeV$).

For the low-mass region ($\MW < 0.1\TeV$), the inclusive on-shell \PW{+}jets sample and the high-\pt \PW{+}jets sample are used. 
Both samples are generated at LO with \MGvATNLO v2.4.2~\cite{Alwall:2014hca, Frederix:2012ps} and the simulated jets are matched to the matrix element and parton shower produced by \PYTHIA, following the MLM approach~\cite{Alwall:2007fs}.
The on-shell sample is used in the low-\HT ($< 0.1\TeV$) region and the high-\pt sample is used at low masses and in the high-\HT ($> 0.1\TeV$) region, where \HT is defined as the scalar \pt sum of all jets in an event.
The on-shell \PW sample is scaled to QCD NLO precision using \MGvATNLO, following the FxFx scheme~\cite{Frederix:2012ps}, and the high-\pt \PW sample is corrected with the \PW boson \pt-dependent NLO QCD{+}NLO EW $K$ factors, using \MGvATNLO~\cite{Lindert:2017olm}.

The backgrounds from \ttbar, single top quark ($t$-channel and tW-channel), Drell--Yan ($\PZ/\PGg^{*}\to\Pell\Pell$, with $\Pell=\Pe, \PGm$, and \PGt), and $\PW\PW$ samples are generated at QCD NLO precision with \POWHEG v2~\cite{Nason:2004rx, Frixione:2007vw, Alioli:2010xd} and interfaced with \PYTHIA for the parton showers.
The \ttbar and $\PW\PW$ samples are scaled to QCD NNLO precision~\cite{PhysRevLett.110.252004,Czakon:2017wor,PhysRevLett.113.212001}.
The $s$-channel single top quark production is generated with \MGvATNLO.

The inclusive $\PW\PZ$ and $\PZ\PZ$ samples are generated with \PYTHIA and scaled to QCD NLO precision~\cite{Campbell_2011,Cascioli_2014} in comparing the samples with data.
The multijet background is estimated using simulated \PYTHIA samples generated with sufficient numbers of events at high lepton \pt values, and normalized to data as described in Section~\ref{sec:selection}.

In the electron channel, 
\PGg{+}jets samples are generated at LO using \MGvATNLO with MLM matching. 
These samples are used to estimate the effects of photons misidentified as electrons.
In the muon channel, several background samples that give nonnegligible contributions are taken into account in the low-\MT (${<}0.4\TeV$) region.
These backgrounds are $\PZ/\PGg^{*}${+}jets ($\PZ/\PGg^{*}\to\PGn\PGn$), generated at QCD NLO with \POWHEG, and $\PW\PGg$ ($\PW\PGg\to\Pell\PGn\PGg$ with $\Pell=\Pe, \PGm$, and \PGt) samples generated at LO using \MGvATNLO. 
Their contributions are negligible in the electron channel because of the much larger trigger thresholds for electrons used during 2017--2018 data-taking, as described in Section~\ref{sec:selection}.

The simulation of the CMS detector response is performed using the \GEANTfour~\cite{Agostinelli:2002hh,ALLISON2016186} framework.
All simulated samples used in the analysis are normalized to each total integrated luminosity of the three years and combined together.
The average number of additional $\Pp\Pp$ collisions in the same or nearby bunch crossings (pileup) was 32 in 2017--2018, corresponding to a total inelastic $\Pp\Pp$ cross section of 69.2\unit{mb}~\cite{Sirunyan:2018nqx}. 
All simulated samples are weighted to have the same pileup multiplicity distributions as measured in data.

\section{Event reconstruction and particle identification}
\label{sec:object}

The reconstruction algorithms are similar to those used in an earlier analysis of 13\TeV data~\cite{Sirunyan:2018mpc}.
The signature of interest is characterized by a single high-\pt lepton and significant missing transverse momentum.

The candidate vertex with the largest value of summed physics-object $\pt^{2}$ is taken to be the primary $\Pp\Pp$ interaction vertex (PV). 
The physics objects used for this determination are the jets, clustered using the jet finding algorithm~\cite{Cacciari:2008gp,Cacciari:2011ma} with the tracks assigned to candidate vertices as inputs, and the associated \ptmiss, taken as the negative vector \pt sum of those jets~\cite{JECref}.

Information from all components of the CMS detector is used by a particle-flow (PF) technique~\cite{Sirunyan:2017ulk} to reconstruct and identify each individual particle in an event. 
Photon energy is obtained from the ECAL measurement. 
The electron energy is determined from a combination of the electron momentum at the PV as determined by the tracker, the energy of the corresponding ECAL cluster, and the sum of energy from all photon bremsstrahlung processes spatially compatible with originating from the electron track. 
The energy of muons is obtained from the curvature of the corresponding track. 
The charged hadron energy is determined from a combination of the momentum measured in the tracker and the matching ECAL and HCAL energy deposits, corrected for the response function of the calorimeters to hadronic showers. 
Finally, the energy of neutral hadrons is obtained from the associated corrected ECAL and HCAL energies.

Electrons are reconstructed from clusters of energy deposits in the ECAL.
These are matched to hits in the tracker and the candidates are required to pass a set of identification and isolation criteria~\cite{Khachatryan:2016zqb, EGM-17-01, Sirunyan:2018fpa} optimized for high energy values.
Electrons with transverse energy (\ET) larger than 35\GeV and $\abs{\eta}<1.40$ ($1.57<\abs{\eta}<2.50$) in the barrel (endcap) are required and the measured energy is taken directly from the ECAL cluster without using track information.
High-\ET electrons can saturate the ECAL readout electronics, biasing the determination of shower-shape.
Therefore, subdetector-based isolation is used instead of PF isolation.
The electron is required to be isolated in a cone of radius $\DR=\sqrt{(\Delta\eta)^2{+}(\Delta\phi)^2}=0.3$ in both the calorimeters and the tracker. 
In the calorimeters, the energy sum is required to be less than 3\% of the electron's \pt. 
The HCAL has two readout depths in the endcap. 
Only the first longitudinal depth is used for the HCAL isolation because the second depth can be affected by detector noise. 
In the tracker, the \pt sum must be less than 5\GeV, including only well-measured tracks consistent with originating from the same vertex as the electron.
Energy deposits in the HCAL around the direction of the electron, corrected for noise, pileup, and leakage of electrons through the inter-module gaps, are required to be less than 5\% of the reconstructed electron energy, in order to reject jets.
For high energy electrons, the last contribution dominates.
Other requirements are applied on the number of lost hits in the inner tracker layers and on the impact parameter relative to the centre of the luminous region in the transverse plane, $d_{xy}$. 
The former must have no more than one missing hit and the latter is required to be less than 0.2 (0.5)\mm for the barrel (endcap). 
Taken together, these requirements suppress background sources such as electrons originating from photon conversion, hadronic activity misidentified as electrons, and electrons from semileptonic decays of \PQb or \PQc quarks.

Muons are reconstructed by matching tracks from the inner tracker with hits, or segments, in the muon system.
The information from the tracker and muon system is combined in such a way as to ensure good \pt resolution and low sensitivity to energy losses due to muon bremstrahlung. 
The muon \pt is determined using a specific combination (TuneP) of different algorithms optimized for $\pt>200\GeV$~\cite{HighpTMupaper}.
Muon candidates with $\abs{\eta}<2.4$ are required to satisfy the track fit and matching quality requirements to ensure a good \pt measurement and suppression of potential muon misidentification from any hadron shower remnants reaching the muon system.
For a reliable \pt measurement, we require that $\sigPT{/}\pt<0.3$, where $\sigPT$ is the uncertainty in \pt from the TuneP reconstruction.
For good track quality, the muon track must have at least one hit in the pixel detector, and at least six tracker layers with hits, to suppress muons from nonprompt meson decays.
The muon track is also required to have segments with hits in at least two consecutive muon detector planes, to reduce contamination from hadronic punch-through. 
To reduce the background due to cosmic ray muons, we require that $d_{xy}$ and the longitudinal impact parameter, $d_{z}$, be less than 0.02 and 0.5\mm, respectively. 
These parameters are defined relative to the PV. 
The muon isolation in the tracker requires the scalar \pt sum of all tracks originating from the interaction vertex within a cone of $\DR=0.3$ around the muon direction, 
excluding the muon itself, to be less than 10\% of the muon \pt.

Jets are clustered from PF candidates using the anti-\kt algorithm~\cite{Cacciari:2008gp, Cacciari:2011ma} with a $\DR$ of 0.4. Jet momentum is determined as the vector sum of all particle momenta in the jet, and is found from simulation to be, on average, within 5--10\% of the true momentum over the whole \pt spectrum and detector acceptance. 
Pileup can contribute additional tracks and calorimetric energy depositions, increasing the apparent jet momentum. 
To mitigate this effect, tracks identified to be originating from pileup vertices are discarded and an offset correction is applied to correct for remaining contributions. 
Jet energy corrections are derived from simulation studies so that the average measured energy of jets becomes identical to that of particle level jets.
In situ measurements of the momentum balance in dijet, \PGg{+}jet, \PZ{+}jet, and multijet events are used to determine any residual differences between the jet energy scale in data and in simulation, and appropriate corrections are made~\cite{Khachatryan:2016kdb}. 
Additional selection criteria are applied to each jet to remove jets mismeasured because of instrumental effects or reconstruction failures.
Jets originating from the hadronization of \PQb quarks are identified by a secondary vertex algorithm, DeepCSV~\cite{Sirunyan:2017ezt}. 
This \PQb tagging algorithm is applied to the leading jet with a tight working point that has an identification efficiency of about 50\% for \PQb quark jets and misidentification probabilities of 2.0\% for \PQc quark jets and 0.1\% for light-flavour or gluon jets.
To correct the simulated events, these efficiencies are measured as a function of jet \pt and $\eta$ in data samples enriched in \ttbar and multijet events.

The vector \ptvecmiss is computed as the negative vector \pt sum of all the PF candidates in an event, and its magnitude is denoted as \ptmiss~\cite{JECref}. 
The \ptvecmiss must satisfy reconstruction quality criteria chosen to reduce detector noise effects.
It is modified to account for the corrections to the jet energy scale~\cite{collaboration_2011,Khachatryan:2016kdb} in the event. 
In both the electron and muon channels, further corrections are applied to account for possible differences in the lepton \pt relative to that obtained from the PF algorithm.

The lepton reconstruction and identification efficiencies for both channels are measured using the ``tag-and-probe'' method~\cite{CMS:2011aa} with samples of high mass Drell--Yan pairs.
Scale factors (SFs) are applied to the simulated events to account for any differences in the selection efficiencies between observed and simulated data.
The SFs for the electron identification, isolation, and reconstruction efficiencies are 0.967 (0.973) for barrel (endcap) in 2017, and 0.969 (0.984) in 2018~\cite{EGM-17-01}.
The SFs for the muon identification efficiencies are in the range 0.974--0.994, as a function of $\eta$, for both 2017 and 2018~\cite{HighpTMupaper}.
The dependence on $\eta$ reflects the geometry of the muon detectors system and the effect of radiative processes associated with muon interactions in the material of the detector at very high \pt.
The muon isolation is well reproduced in simulated samples and the SF is consistent with unity, within systematic uncertainties, for both 2017 and 2018 data.

\section{Event selection}
\label{sec:selection}

Events are selected in the high-level trigger by requiring the presence of one isolated high-\pt lepton. 
Electron events require a single electromagnetic cluster having $\ET>200\GeV$ and $\abs{\eta}<2.5$.
Muon events require a single muon with $\pt>50\GeV$ and $\abs{\eta}<2.4$. 

In addition to satisfying the tight offline identification criteria described in Section~\ref{sec:object}, electrons (muons) are required to have $\pt>240$ (53)\GeV, so that they are on the efficiency plateau of the trigger.
The trigger efficiency for the electron channel is measured in simulation and in an independent data sample recorded with single-muon triggers for each data-taking year.
The SF is determined to be 0.995 (0.994) for barrel (endcap) in 2017, and 0.997 (0.998) for barrel (endcap) in 2018.
In addition, the SFs for the level-1 trigger timing drift~\cite{Sirunyan:2020zal} are applied for the electron channel in 2017. 
This timing drift caused a partial loss of events in the trigger, as the reconstructed clusters in the ECAL could be wrongly assigned to the previous bunch crossing, an occurrence referred to as ``trigger prefiring''.
The prefiring mostly impacts events with large electromagnetic activity at large $\abs{\eta}$ values (2.0--3.0).
The trigger SF values for the muon channel, applied as a function of $\eta$, range 0.911--1.011 in 2017--2018~\cite{Sirunyan:2021zrd, HighpTMupaper}.

To reduce the Drell--Yan, \ttbar, and multiboson backgrounds, 
events are required not to have a second lepton, either electron or muon, with $\pt>25\GeV$.
For a high efficiency, this second lepton is identified with a set of loose selection criteria.
The backgrounds from \ttbar and single top quark events in both semileptonic and dileptonic decay modes are sources of \ptmiss and high-\pt prompt leptons.
In the muon channel, the contribution from \ttbar background is further reduced by suppressing events containing more than five jets ($\pt>25\GeV$, $\abs{\eta}<2.5$ and $\DR(\text{jet},\Pell)>0.5$) and events with the leading jet consistent with originating from a \PQb quark~\cite{Sirunyan:2017ezt}.
This requirement is not needed in the electron channel since the higher electron \pt threshold applied in the selection already significantly reduces the \ttbar contribution. 
The agreement of data with simulation lies within uncertainties for the jet multiplicity distributions for up to 6 jets.
For higher jet multiplicities, an additional 10\% uncertainty in the normalization of the \ttbar sample is included. 

The multijet background has the largest cross section among all background processes, but it is efficiently reduced thanks to the lepton high-\pt cut and the isolation criteria.  
The shape of the lepton \pt distribution of the multijet background is taken from the simulated samples and normalized to data using a control region enriched in events with energetic leptons from QCD production of multijets. 
This region is obtained by applying the same selection criteria as in the analysis, but inverting the lepton isolation requirement and limiting the amount of \ptmiss in the event. 

The lepton \ptvec and \ptvecmiss are expected to be nearly back-to-back in the transverse plane and balanced in \pt. 
To incorporate these characteristics in the analysis, the ratio of the lepton \pt to \ptmiss and the angle between the directions of the lepton and \ptvecmiss provide additional kinematic criteria to select events.
The ratio $\pt/\ptmiss$ is required to be between 0.4 and 1.5, and the angular difference must be $\Delta\phi(\ptvec,\ptvecmiss)>2.5$.

\begin{figure}[hbtp]
\centering
\includegraphics[width=0.6\textwidth]{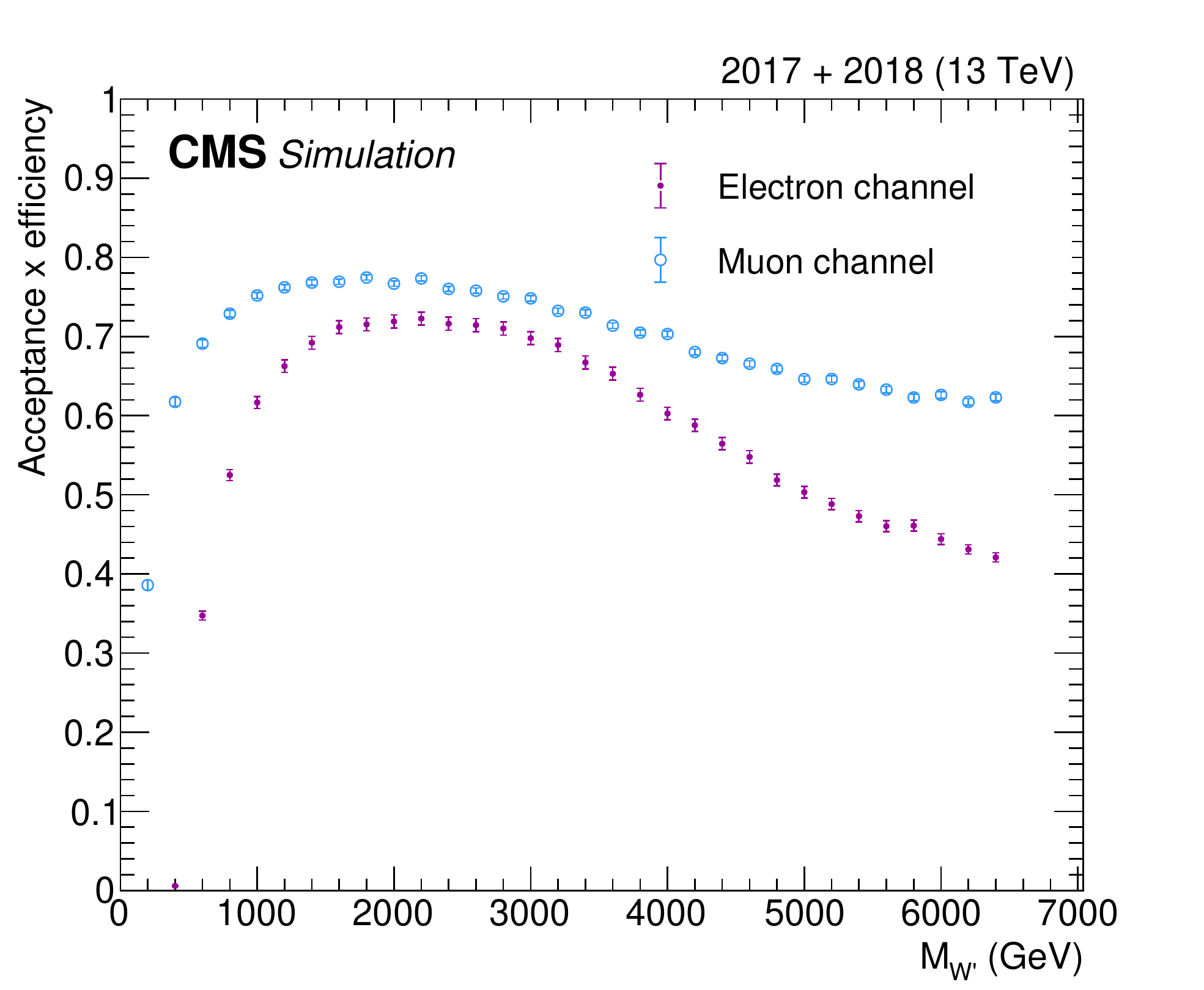}
\caption{Product of acceptance and efficiency for the SSM \PWpr signal, as a function of the \PWpr mass, after all selection criteria are applied for the electron (filled purple markers) and the muon (open blue markers) channels.}
\label{fig:signaleff_ELandMU}
\end{figure}

For simulated events passing all of the selection criteria, the signal efficiency for an SSM \PWpr boson with no requirement on the reconstructed \MT 
in the event reaches a maximum value of 0.70 (0.77) in the electron (muon) decay channel, for an $\MWp$ range of 1.5 to 2.4\TeV.
This value decreases gradually to about 0.4 (0.6) in the electron (muon) channel for high masses ($\MWp\sim6\TeV$).
The difference in the efficiencies between the channels is due to the difference in the \pt thresholds applied.
Off-shell production of the \PWpr boson resonance will manifest itself in the lower \MT region.
The effect increases as the \PWpr mass increases.
The lower \pt threshold of the muon trigger gives rise to a higher efficiency in the muon channel than the electron channel, for low and very high masses.
The signal efficiency as a function of $\MWp$ is shown in Fig.~\ref{fig:signaleff_ELandMU} for both channels.

\begin{figure}[htbp]
\centering
\includegraphics[width=.49\textwidth]{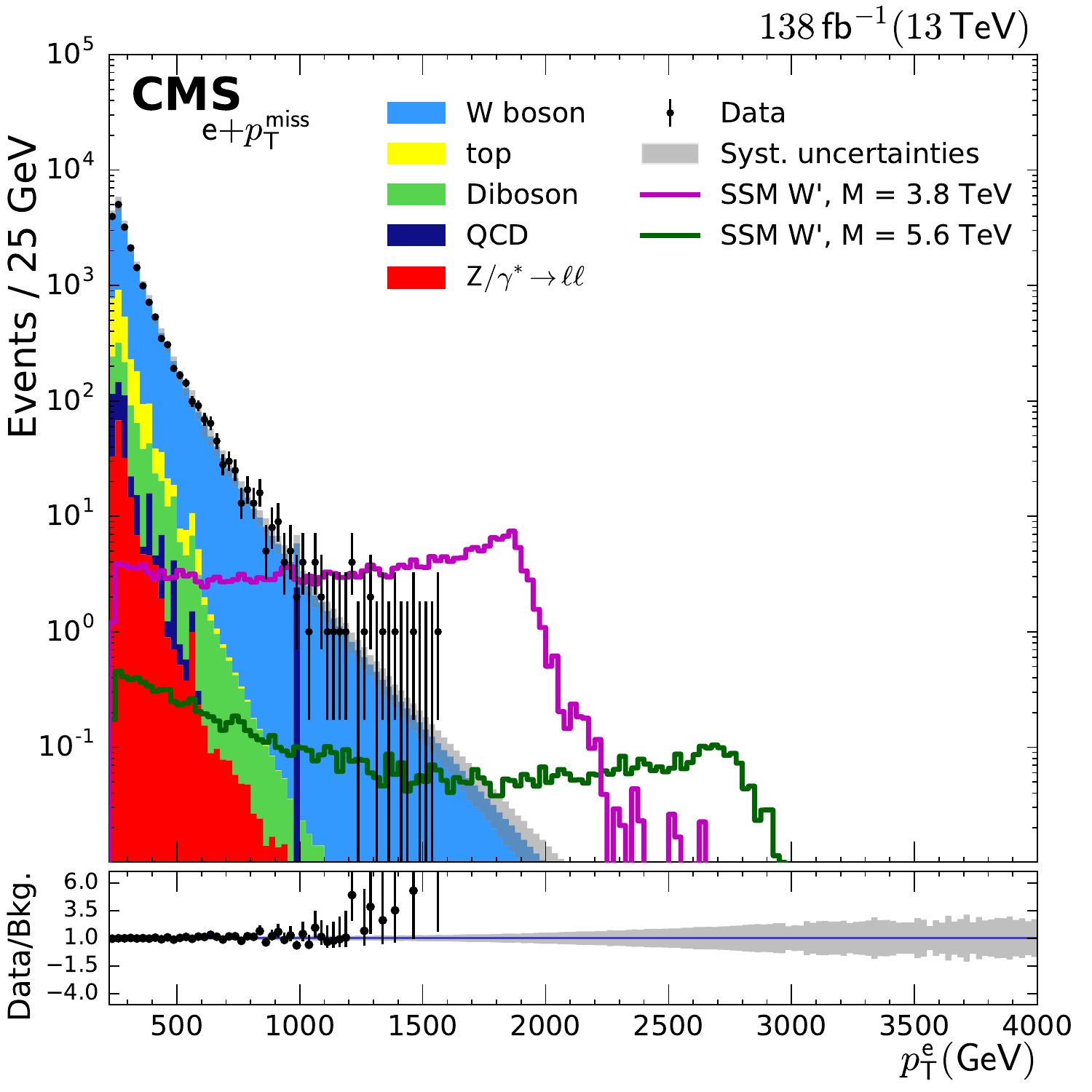}
\includegraphics[width=.49\textwidth]{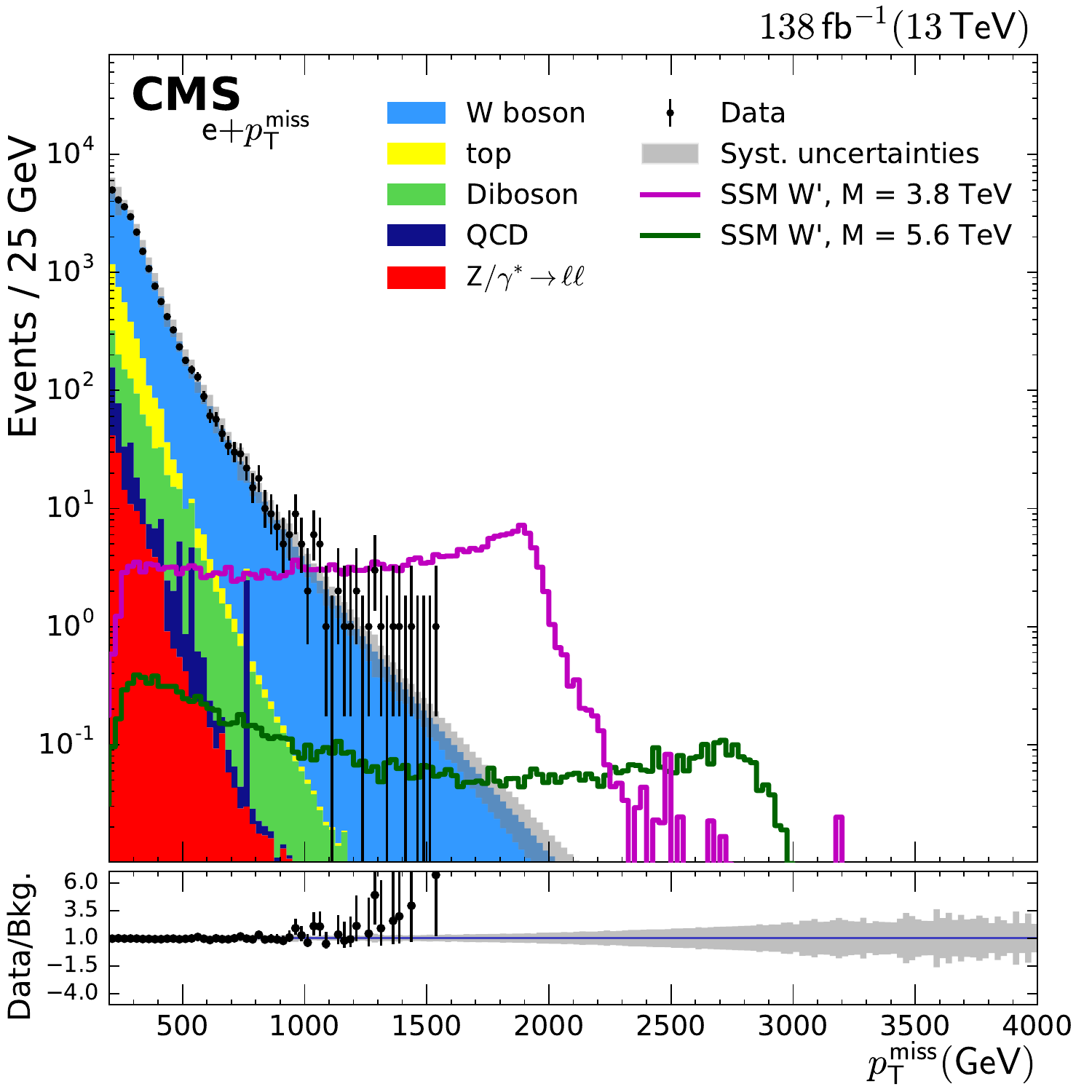} 
\includegraphics[width=.49\textwidth]{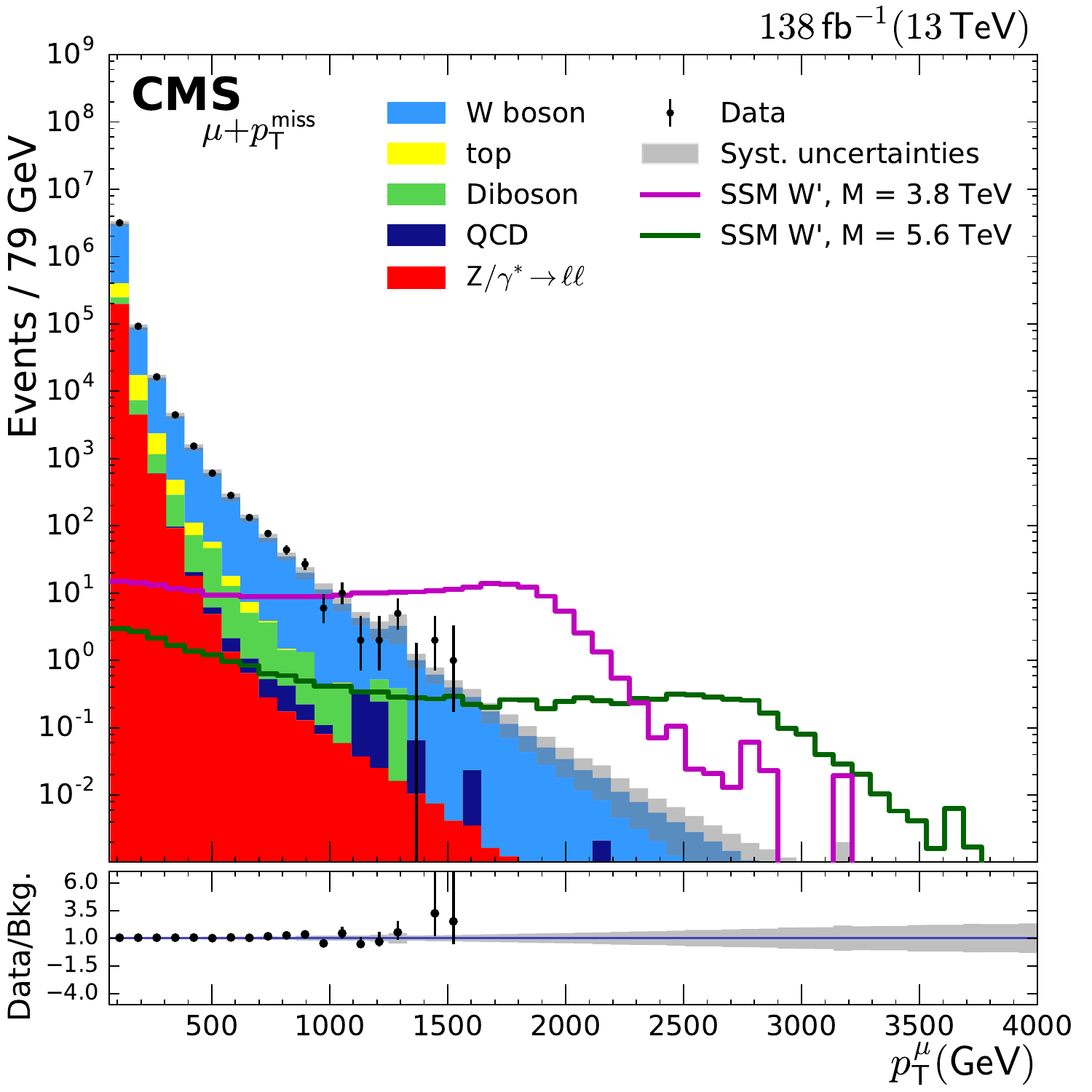}
\includegraphics[width=.49\textwidth]{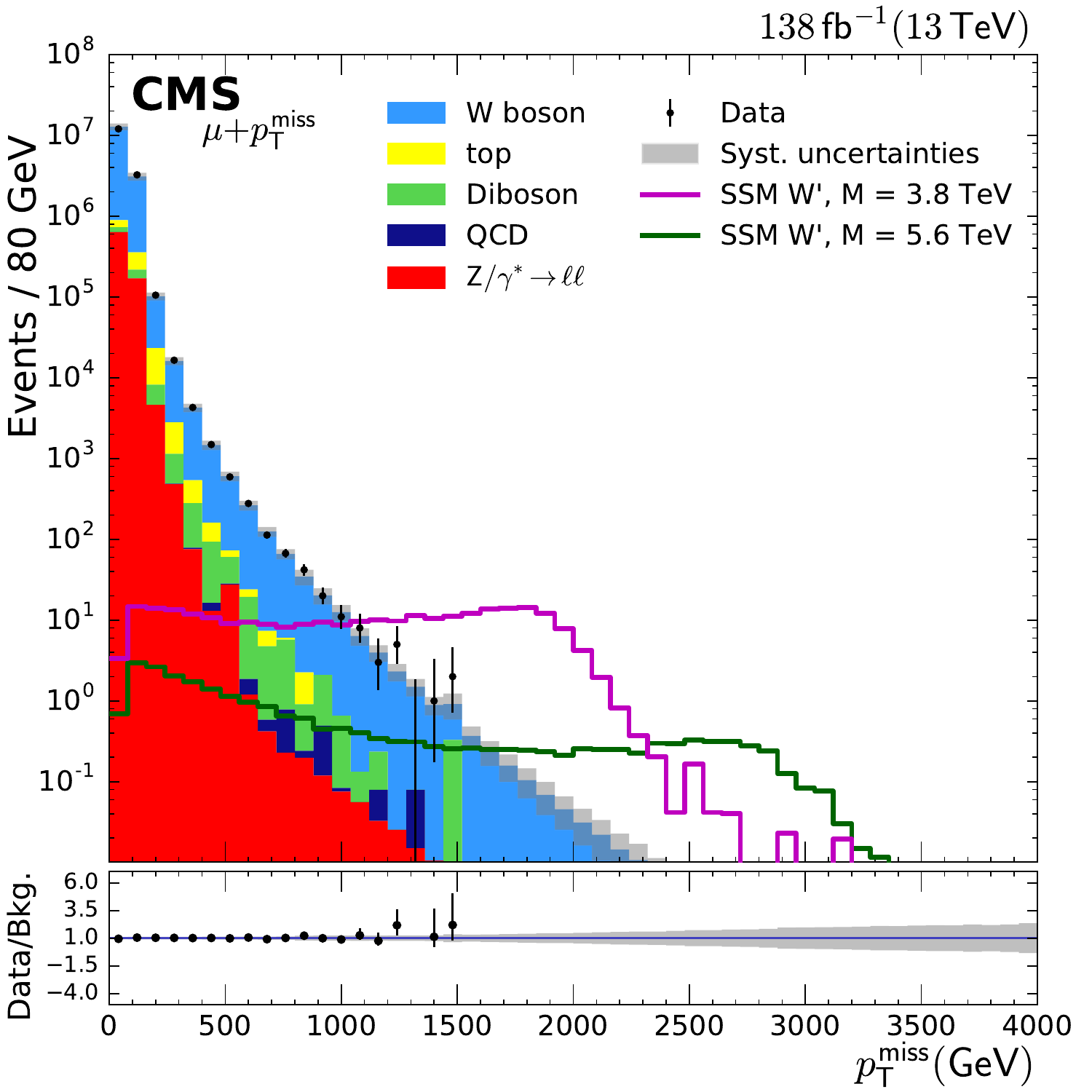}
\caption{
The distributions for lepton \pt (left) and \ptmiss (right) for the electron (upper) and muon (lower) channels after applying the full selection criteria, for the combined 2016--2018 data sets. 
Two signal distributions are presented, corresponding to SSM \PWpr boson masses of 3.8 and 5.6\TeV. 
The lower panels show the ratios of data to the SM prediction, and the shaded bands represents the systematic uncertainty.
}
\label{fig:ptmet}
\end{figure}

The distributions of the electron and muon \pt (left panels) and the \ptmiss in the electron and muon channels (right panels), after applying the full selection criteria, are shown in Fig.~\ref{fig:ptmet} for the combined 2016--2018 data sets.
The \pt and \ptmiss distributions use different bin sizes for the electron channel (25\GeV) and the muon channel (80\GeV), respectively.

\begin{figure}[htbp]
\centering
\includegraphics[width=0.49\textwidth]{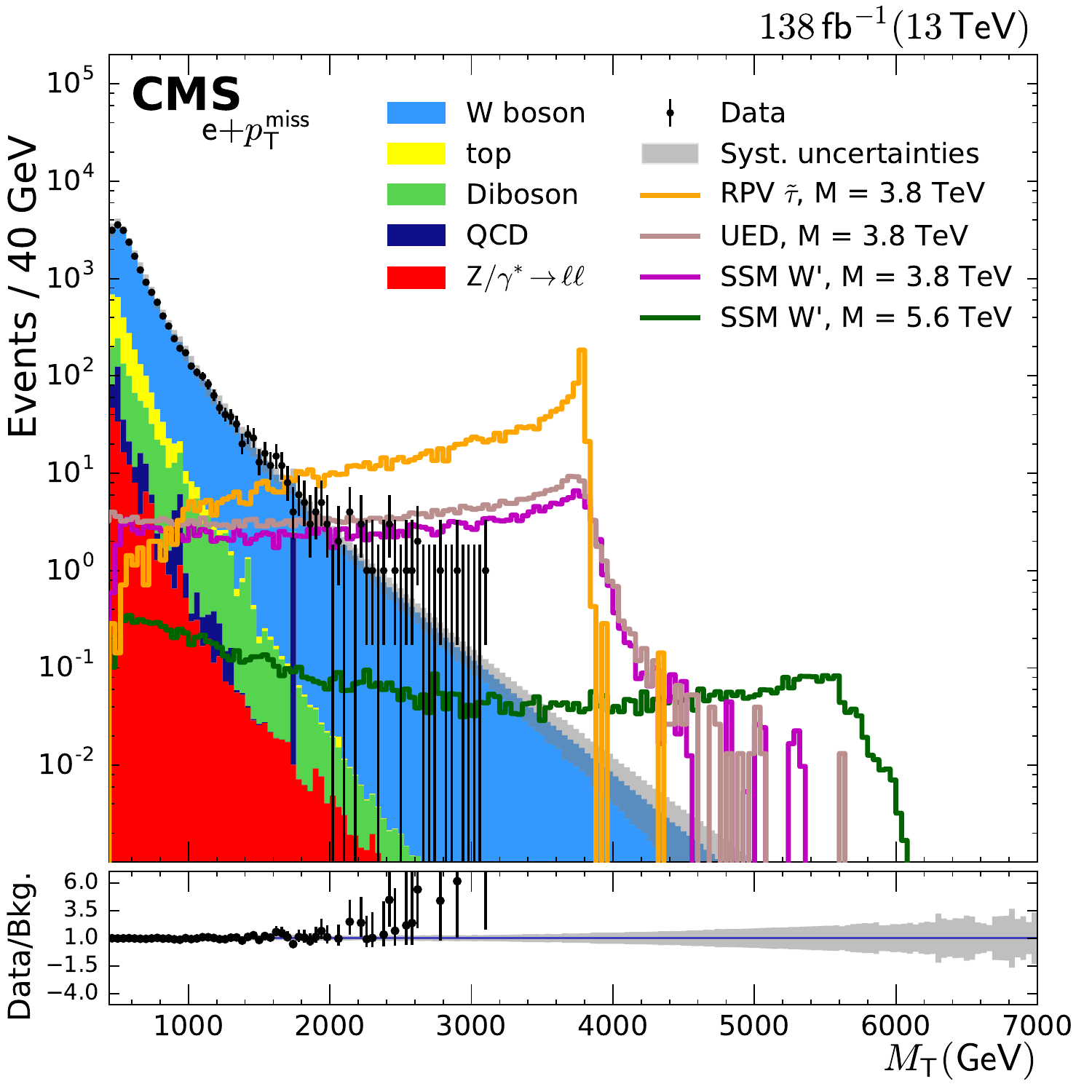}
\includegraphics[width=0.49\textwidth]{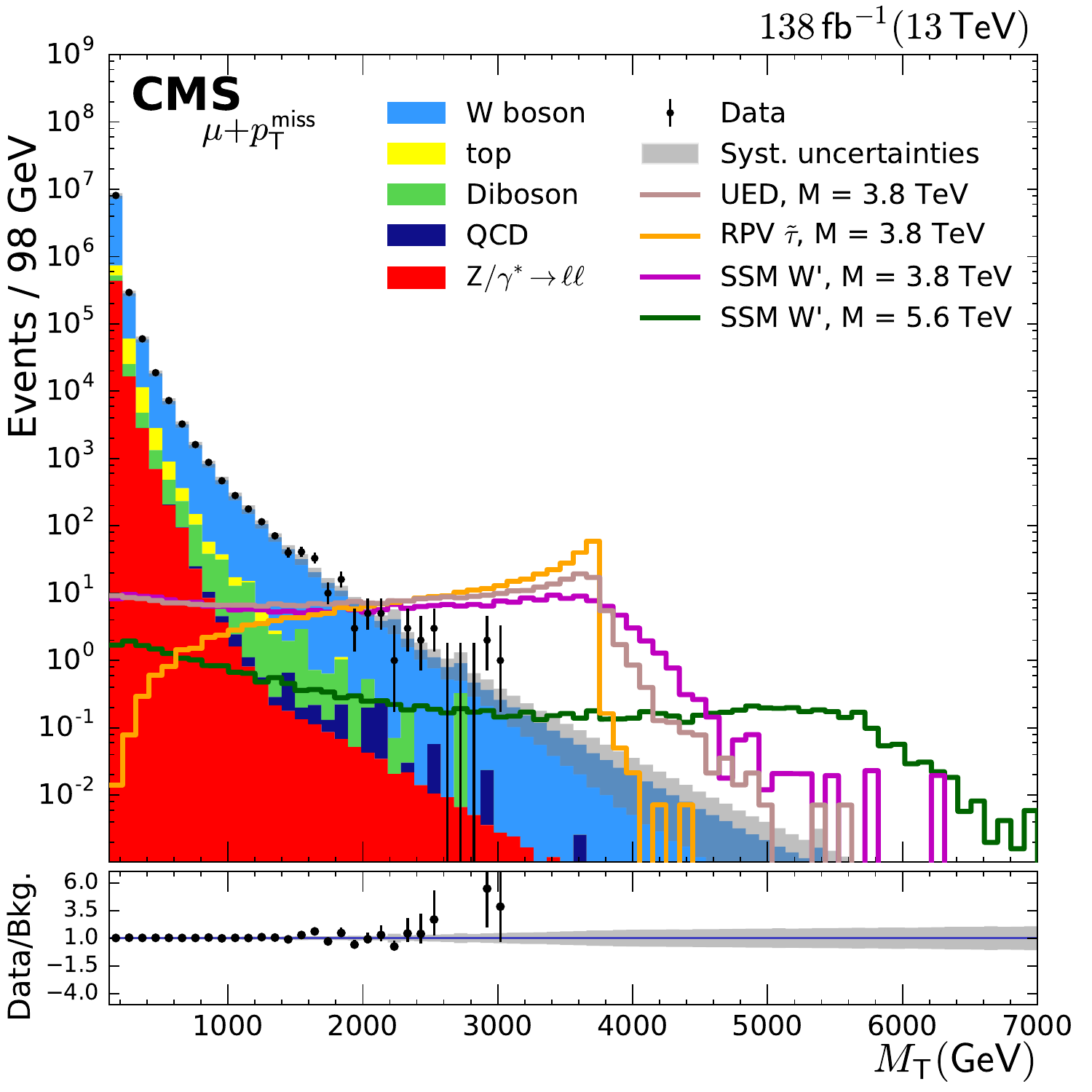}
\caption{
The distributions for \MT for the electron (left) and muon (right) channels after applying the full selection criteria, for the combined 2016--2018 data sets. 
Contributions from the SSM \PWpr, split-UED \WprimeKKtwo, and RPV SUSY \sTau signals at the masses of 3.8 and 5.6\TeV are also indicated.
The lower panels show the ratios of data to the SM prediction, and the shaded bands represent the systematic uncertainty.
}
\label{fig:mt}
\end{figure}

The resulting \MT distributions of the electron (left) and muon (right) channels are shown in Fig.~\ref{fig:mt}.
The minimum value chosen for \MT depends on the trigger threshold.
It is 250\GeV (2016) and 500\GeV (2017--2018) for the electron and 120\GeV for the muon channels, respectively.
The \MT distribution, which is used as an input to set limits, is binned according to the resolution.

The observed and expected numbers of events for a selected set of \MT thresholds are shown in Table~\ref{tab:evt_num} for the electron and muon channels, from the combination of the 2016--2018 data sets.
The highest \MT values of events observed in the electron and muon channels are 3.1 and 2.9\TeV, respectively.

\begin{table}[htpb]
\centering
\topcaption{The observed and expected number of events in the electron (upper) and muon (lower) channels, collected during three years (2016--2018), for selected values of \MT threshold. 
Predicted numbers of SSM \PWpr events are given, for $\MWp = 3.8$ and 5.6\TeV.
The statistical and systematic uncertainties are added in quadrature to provide the total uncertainty.
}
\begin{tabular}{@{}l r@{}l r@{}l r@{}l r@{}l@{}}
\hline
\MT (TeV)     & \multicolumn{2}{c} {$>$1.0} & \multicolumn{2}{c} {$>$2.0} & \multicolumn{2}{c} {$>$3.0} & \multicolumn{2}{c} {$>$4.0} \\
\hline
Electron data        & \multicolumn{2}{c} {831}  & \multicolumn{2}{c} {23}  & \multicolumn{2}{c} {1}  & \multicolumn{2}{c} {0} \\
Total SM backgrounds & {835$\pm$}&{64}           &{21.1$\pm$}&{2.5}         &{1.16$\pm$}&{0.24}       &{0.066$\pm$}&{0.029}    \\
$\MWp = 3.8\TeV$     & {211$\pm$}&{35}           &{155$\pm$}&{29}           &{93$\pm$}&{20}           &{1.95$\pm$}&{0.68}      \\
$\MWp = 5.6\TeV$     & {8.0$\pm$}&{2.1}          &{4.8$\pm$}&{1.7}          &{3.5$\pm$}&{1.5}         &{2.5$\pm$}&{1.3}        \\ [\cmsTabSkip]
Muon data            & \multicolumn{2}{c}{829}   & \multicolumn{2}{c} {21}  & \multicolumn{2}{c} {0}  & \multicolumn{2}{c} {0} \\
Total SM backgrounds &  {805$\pm$}&{83}          &{21.7$\pm$}&{2.9}         &{1.05$\pm$}&{0.34}       &{0.089$\pm$}&{0.040}    \\
$\MWp = 3.8\TeV$     &  {192$\pm$}&{28}          &{141$\pm$}&{24}           &{80$\pm$}&{19}           &{6.4$\pm$}&{1.8}        \\
$\MWp = 5.6\TeV$     & {11.0$\pm$}&{1.6}         &{6.6$\pm$}&{1.1}          &{4.6$\pm$}&{1.1}         &{3.2$\pm$}&{0.9}        \\ 
\hline
\end{tabular}
\label{tab:evt_num}
\end{table}

\clearpage

\section{Systematic uncertainties}
\label{sec:systematic}

Several sources of uncertainty affect the shape, or normalization, of the \MT distribution of the lepton plus \ptvecmiss system, impacting the background and signal description. 
Unless otherwise specified, to estimate the systematic uncertainty on the overall normalization and on the shape of the \MT distribution, the full analysis is repeated, with the values of the parameters associated with that source shifted up and down by one standard deviation.  

For the dominant \PW boson background, the theoretical uncertainties in the PDF are estimated for each \MT bin of the differential distribution.  
The uncertainties are obtained using the PDF4-\\LHC \cite{Butterworth_2016} procedure accounting for the NNPDF3.1 NNLO~\cite{NNPDF31} PDF set of replicas. 
The effect on the background event yield due to this PDF uncertainty increases gradually to 4, 40, and 120\% at $\MW=2$, 4, and 6\TeV in this analysis. 
The uncertainties in the \PW boson cross section at high mass related to the QCD{+}EW higher-order corrections are taken to be 4--19\%, depending on the $\MW$.
These uncertainties get larger with increasing $\MW$.

The global uncertainties in the integrated luminosity measurement are estimated with a precision of 2.3 and 2.5\%~\cite{cmscollaboration2021precision, CMS-PAS-LUM-17-004, CMS:2019jhq} for the 2017 and 2018 data-taking periods, respectively.
The uncertainty related to pileup modelling is evaluated for variations of 4.6\% in the total inelastic $\Pp\Pp$ cross section relative to the nominal value of 69.2\unit{mb}~\cite{Sirunyan:2018nqx}.
The resulting shift in weights is propagated through the analysis and the corresponding \MT spectra are used as inputs in the statistical treatment of the data. 
The variation of the yields induced by this procedure is less than 0.5\%.

The uncertainties related to the lepton energy or momentum measurement and identification efficiency for high-\pt leptons in data and simulation are estimated from dedicated measurements performed with \PZ bosons with a large transverse boost.
For the electron channel, the uncertainty associated with the electron energy scale and resolution corrections is taken as 0.05--0.10 (0.1--0.3)\% in the ECAL barrel (endcap)~\cite{EGM-17-01}.
For the muon channel, the uncertainty in the muon \pt scale and resolution is estimated by studying the curvature of the muon tracks in different regions of $\eta$ and $\phi$, using energetic muons.
These tracks are taken from cosmic ray data and from dimuon events from high-\pt \PZ boson decays in the collision data, together with the corresponding simulation samples~\cite{HighpTMupaper}.
Based on this study, the scale correction for high-\pt muons is assigned an uncertainty for $\pt>200\GeV$, which varies as a function of $\eta$.
This uncertainty amounts to a potential \pt shift of 5--10\% for $\pt>2\TeV$ in the endcaps ($\abs{\eta}>1.2$), and around 5\% in the barrel ($\abs{\eta}<1.2$), during 2017. 
During 2018, these values were reduced to 2\% for $\pt>2\TeV$, independent of muon $\eta$.
For muons with $\pt>100\GeV$, the uncertainty associated with \pt resolution is negligible in the barrel region and taken to be 1\% in the endcaps.
These uncertainties are propagated to the \ptmiss and \MT distributions.

The uncertainty in the trigger SF in the electron channel is determined to be 0.5 (1.1)\% for the barrel (endcap).
In addition, the uncertainty derived from the ECAL endcap level-1 trigger timing drift effect~\cite{Sirunyan:2020zal} is taken to be a maximum of 20\% (prefiring probability) and the statistical uncertainty associated with the particular bin.
The uncertainty in the identification and isolation SF for electrons in the barrel is 1\% for $\pt<90\GeV$. 
This uncertainty increases linearly from 1 to 3\% for \pt in the range of 90--1000\GeV and corresponds to 3\% for $\pt>1\TeV$. 
In the endcap the SF uncertainty is 2\% for $\pt<90\GeV$. 
This uncertainty increases linearly from 2 to 5\% for \pt in the range of 90--300\GeV, and is 5\% for $\pt>300\GeV$.

In the muon channel, the uncertainty in the trigger SF is estimated to be 0.5--8.0\%, depending on \pt and $\eta$. The corresponding prefiring probability is less than 0.4\% and its effect on the trigger efficiency is included in the systematic uncertainty.
The SF of the muon isolation is assigned an uncertainty of 0.1\% for both 2017 and 2018 data.
For the high-\pt muon identification, the uncertainty in the SF is 0.4\% for $\abs{\eta}<2.0$ and 2.1\% for $\abs{\eta}>2.0$.

The uncertainty in the determination of \ptmiss is derived from the individual uncertainties assigned to the PF objects~\cite{Khachatryan:2016kdb, JECref}.
The PF objects considered are jets, \Pe, \PGm, \PGt, \PGg, and unclustered energy. The TuneP muon \pt determination described in Section~\ref{sec:object} is taken into account.

For energetic electrons and muons the procedure has been explained above. 
The effect on the \ptmiss measurement of the uncertainty in the jet energy scale is estimated by shifting the jet energy by $\pm1$ standard deviation in simulation.
The resulting uncertainty is 2--5\%, which depends on \pt and $\eta$~\cite{Khachatryan:2016kdb}. 
In addition, an uncertainty of 10\% in the \pt is used for the unclustered energy.
The uncertainty associated with \ptmiss is derived by summing quadratically the uncertainties in the separate contributions to the PF calculation of this quantity.

The dominant uncertainties in the electron and muon channels are related to the description of the SM background used to compare to the data. 
On the experimental side, the most important uncertainty is associated with the measurement of the lepton energy or momentum, arising from the detector resolution and the energy scale calibration. 
From the theoretical side, the main uncertainties come from the choice of the PDF set and the determination of $K$ factors covering higher-order corrections for high mass off-shell \PW boson cross sections. 

All systematic uncertainties are taken to be correlated between the three data-taking years.
The exceptions to this are the object-related (leptons, jet, \ptmiss) uncertainties, which derive from statistically independent sources, and integrated luminosity uncertainties.
These exceptions are treated as partially correlated between different years, because they include some components that are fully correlated year-to-year and other components that are uncorrelated year-to-year.

\section{Results}
\label{sec:result}
The \MT distributions for each channel observed in data and predicted by the SM simulation are compared in Fig.~\ref{fig:mt} and Table~\ref{tab:evt_num}.
They agree within the statistical and systematic uncertainties.
Thus, no significant deviation from the SM expectation is observed.
These observations are interpreted in the context of the models described in Section~\ref{sec:models}.

The \MT distributions shown in Fig.~\ref{fig:mt} are used as the basis for interpreting the results in the framework of the SSM and the split-UED model. 
We perform a binned shape fit using a maximum likelihood to the \MT spectrum.
Upper limits at 95\% \CL on the product of the resonance production cross section and the branching fraction to a charged lepton and a neutrino as a function of $\MWp$ are calculated using a Bayesian method~\cite{PDG2020} with a uniform positive prior probability distribution for the signal cross section.
Each source of systematic uncertainty described in Section~\ref{sec:systematic} is considered as a nuisance parameter in the expected signal and background yields. 
The rate uncertainties (\eg luminosity) that affect only the normalization of the simulation are modelled with log-normal distributions, and the uncertainties  affecting the shapes are modelled using a template-morphing method with Gaussian distributions~\cite{conway2011incorporating}.
The statistical uncertainties from the events in the simulated signal and the SM background are considered by incorporating them into the fit as nuisance parameters using the Barlow--Beeston ``lite'' approach~\cite{BARLOW1993219, conway2011incorporating}.

Limits on the SSM \PWpr boson and on the $\WprimeKKtwo$ boson in split-UED models are calculated using a binned maximum-likelihood fit method for the electron and muon channels.
These limits are set using the \MT distributions shown in Fig.~\ref{fig:mt} above 400 (120)\GeV for electrons (muons) obtained after the complete event selection.
A model-independent (MI) limit is determined using a single bin of all events above a threshold $\MTlower$ summed together.
The MI upper limit at 95\% \CL on $\xsecWp\mathcal{B}(\PWpr\to\Pell\PGn)\mathcal{A}\epsilon$ ($\Pell = \Pe$ or $\PGm$) is set as a function of the $\MTlower$.
If the background yield is low enough, this method provides comparable sensitivity to that of the binned likelihood fit method.
The resulting MI limit can be reinterpreted using other models having the same final states.
An example for the RPV SUSY model is given in Section~\ref{sec:rpv-limit}.
Additionally, constraints on the $W$ parameter and on the composite Higgs sector are computed.

\subsection{Exclusion limit on the SSM \texorpdfstring{\PWpr}{W'} boson model}
\label{sec:SSM_limit}
\begin{figure}[hbtp]
\centering
\includegraphics[width=0.49\textwidth]{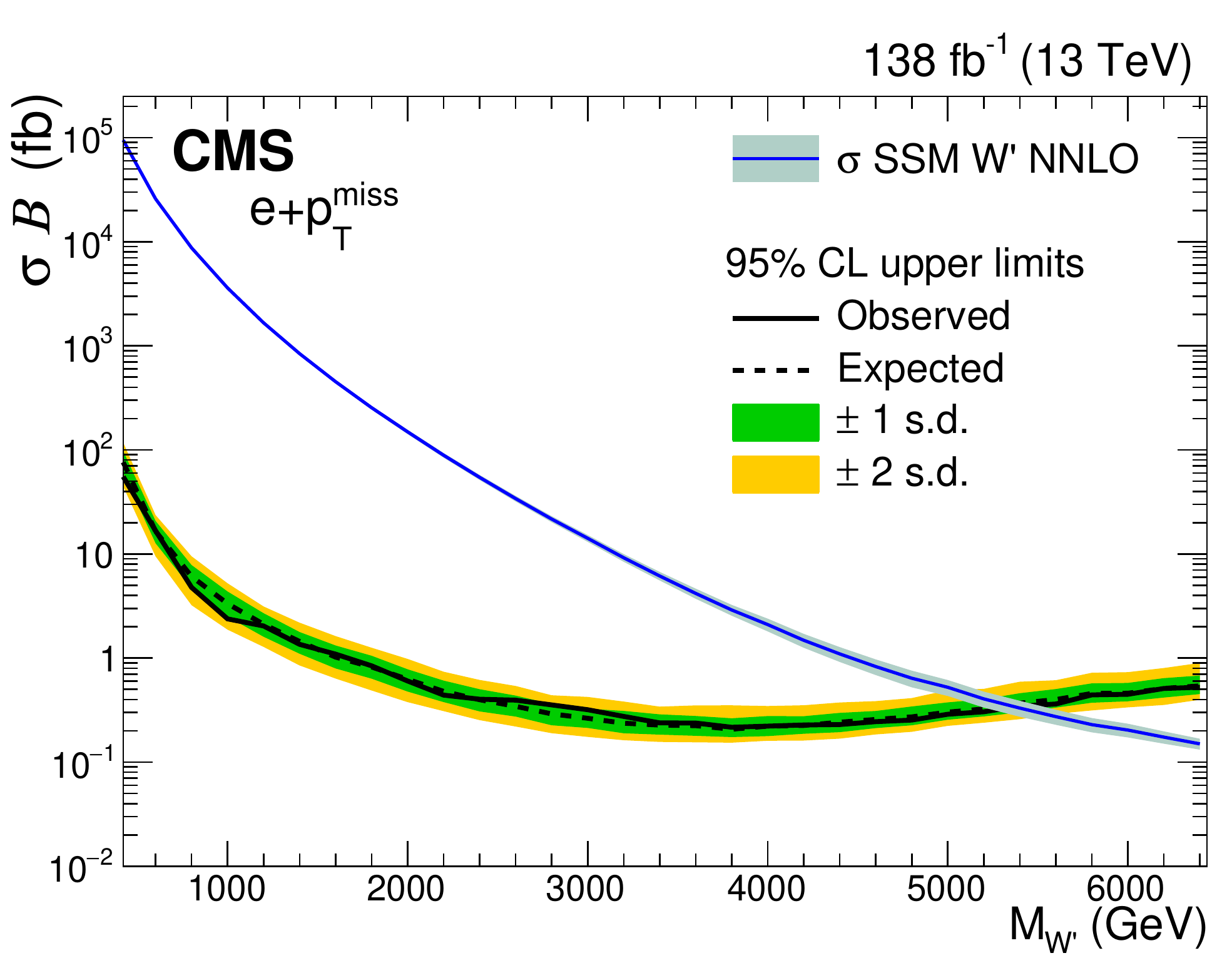}
\includegraphics[width=0.49\textwidth]{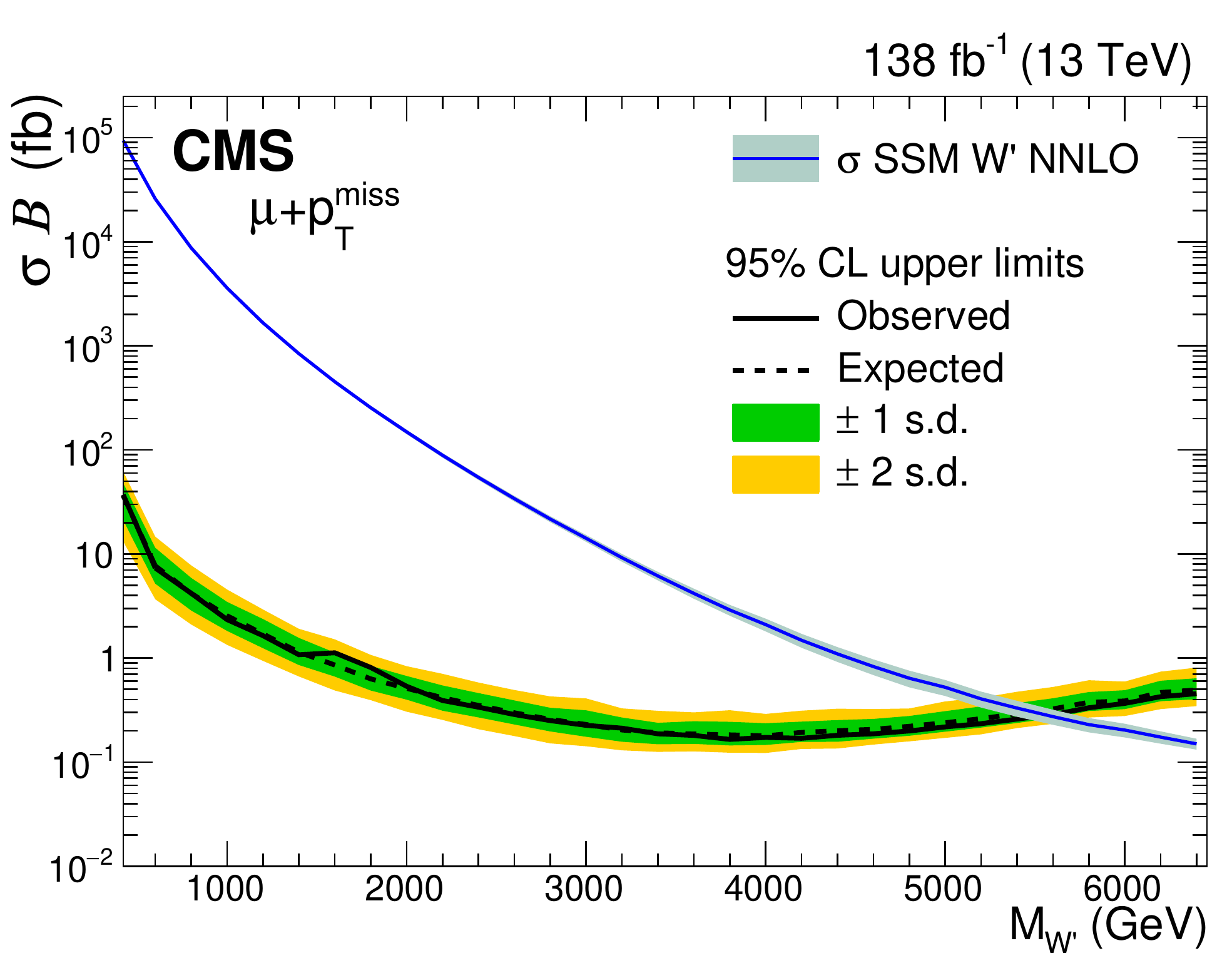}
\includegraphics[width=0.49\textwidth]{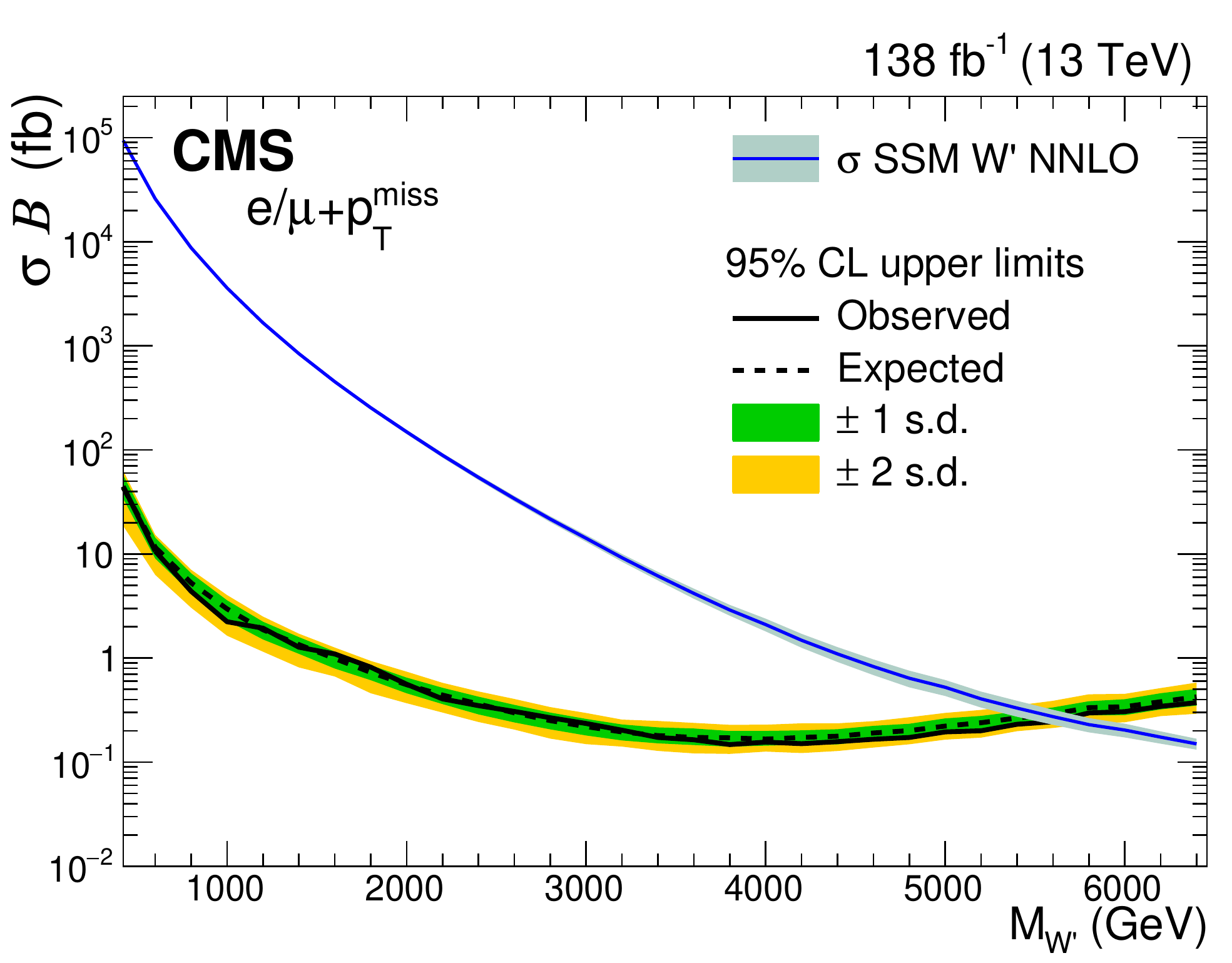}
\caption{The observed (solid line) and expected (dashed line) upper limits at 95\% \CL on $\xsecWp\mathcal{B}(\PWpr\to\Pell\PGn)$ 
for an SSM \PWpr boson model, as a function of the \PWpr boson mass, for the electron (upper left), muon (upper right) channels, 
and the combination of both channels (lower).
The shaded bands represent the one and two standard deviation uncertainty bands for the expected limits. 
The theoretical predictions for the SSM at QCD NNLO precision are shown, with the narrow grey bands indicating the uncertainty associated with the choice of PDFs and \alpS.
}
\label{fig:ssm-limit}
\end{figure}

Limits on the product $\xsecWp\mathcal{B}$ in the SSM, as a function of $\MWp$, are shown in Fig.~\ref{fig:ssm-limit}.
The indicated theoretical cross sections are scaled to QCD NNLO precision.
The PDF and strong coupling constant (\alpS) uncertainties as a function of $\MWp$ are shown as a narrow band around the theoretical cross section line.
The intersection of the central value of the theoretical prediction for $\xsecWp\mathcal{B}$ and the limit curve gives the exclusion limit on $\MWp$.

As shown in Fig.~\ref{fig:ssm-limit}, $\MWp$ below 5.4\TeV (5.3\TeV expected) and 5.6\TeV (5.5\TeV expected) are excluded in the electron (upper left) and muon (upper right) channels, respectively.
For the very low $\MWp$ region, the muon channel becomes more sensitive because of the much higher trigger threshold in the electron channel as described in Section~\ref{sec:selection}.
The lower exclusion limit for the combination of the two channels increases to 5.7\TeV (5.6\TeV expected), as shown in Fig.~\ref{fig:ssm-limit} (lower). 
The one and two standard deviation bands include both the statistical and systematic uncertainties described in Section~\ref{sec:systematic}. 
The results are summarized in Table~\ref{tab:ssmLimits}.

\begin{table}[htpb]
\centering
\topcaption{Observed and expected lower limits at 95\% \CL on the SSM \PWpr boson mass for the electron and muon channels and for the combination of the two.
}
\begin{tabular}{l c c c}
\hline
\multirow{2}{*}{Channel} & \multicolumn{2}{l}{$\MWp$ lower limit (\TeVns{})} \\
                         &  Observed & Expected \\  \hline
Electron                 & 5.4       &  5.3 \\
Muon                     & 5.6       &  5.5 \\
Combination              & 5.7       &  5.6 \\ 
\hline
\end{tabular}
\label{tab:ssmLimits}
\end{table}

\subsection{Exclusion limit on the model-independent cross section}
\label{sec:MI_limit}
\begin{figure}[hbtp]
\centering
\includegraphics[width=0.49\textwidth]{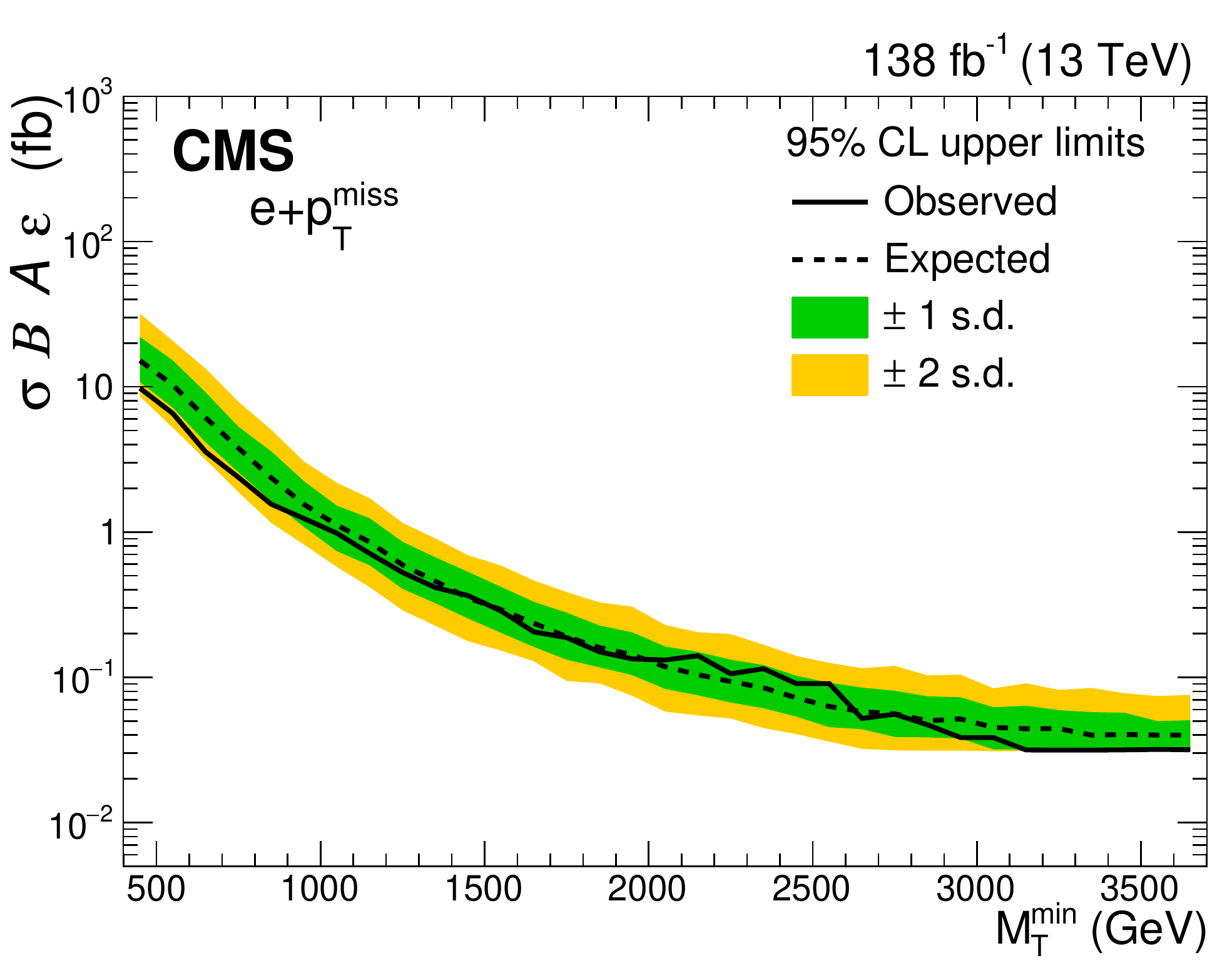} 
\includegraphics[width=0.49\textwidth]{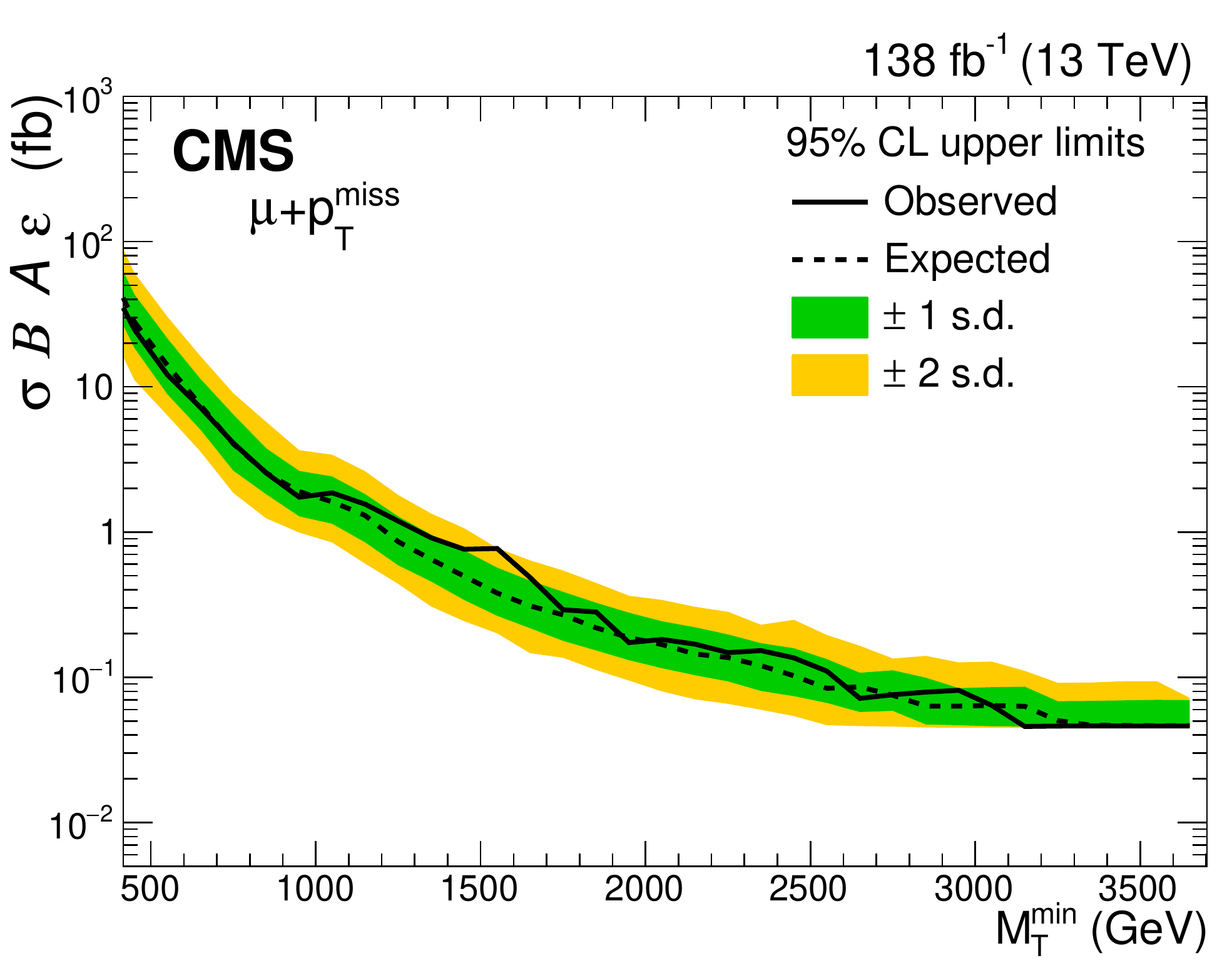}
\includegraphics[width=0.49\textwidth]{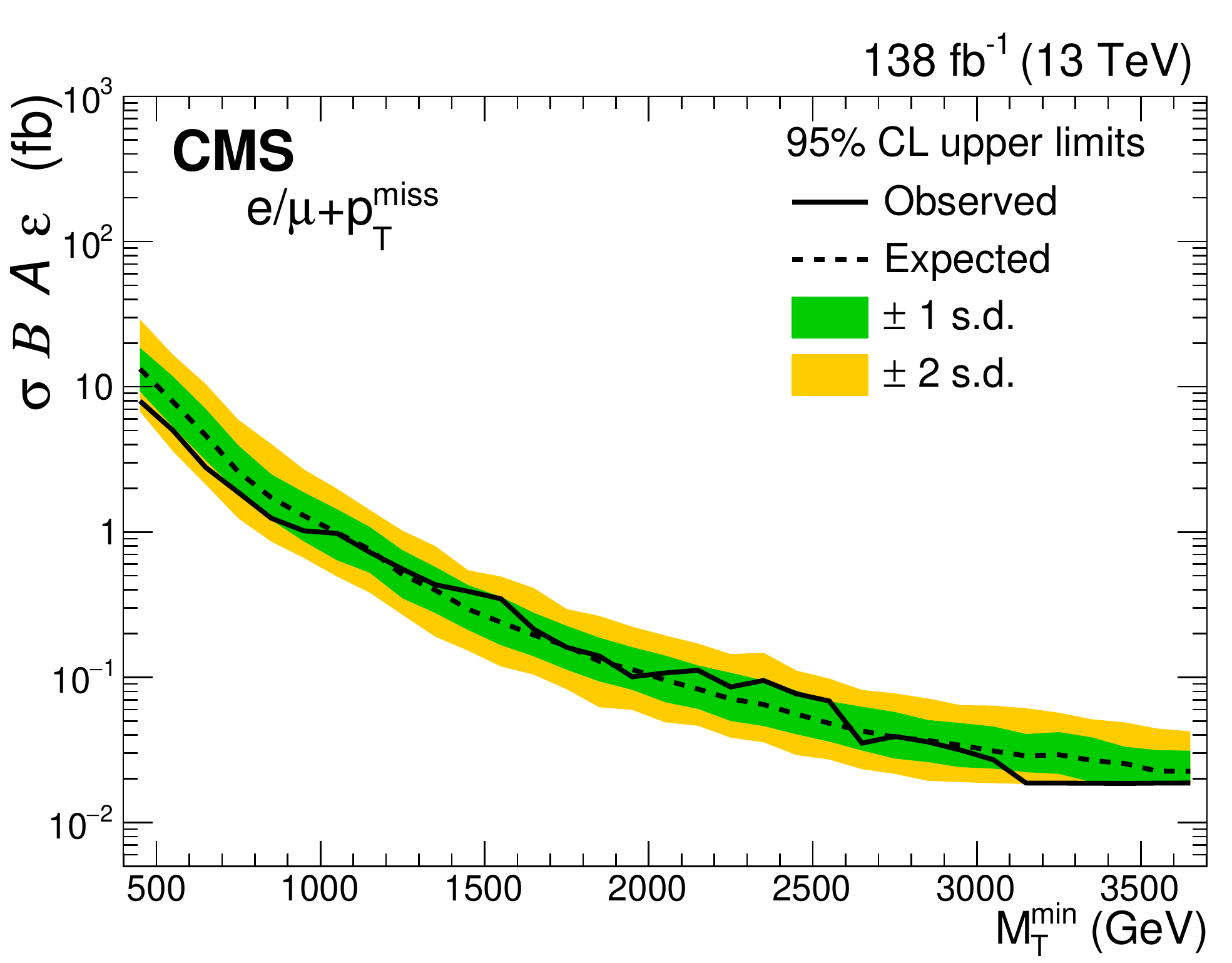}
\caption{The 95\% \CL observed (solid line) and expected (dashed line) model-independent cross section limits as functions of the $\MTlower$ threshold. These are shown for the electron (upper left) and muon (upper right) channels and their combination (lower). 
The one and two standard deviation uncertainty bands for the expected limits are shown.
}
\label{fig:mi-limit}
\end{figure}

The model-independent limits on $\sigma\mathcal{B}\mathcal{A}\epsilon$
for a new particle decaying to a lepton and a neutrino are shown in Fig.~\ref{fig:mi-limit} for the electron (left), muon (right), and the combined channels (lower).
The limits depend strongly on the $\MTlower$ threshold.
For the combination of the two channels, the MI limit ranges from about 10\unit{fb} at a $\MTlower$ of 500\GeV to about 0.02\unit{fb} at $\MTlower$ of 3.6\TeV.

The exclusion limit for any specific physics model that predicts a massive charged mediator with the same final states can be determined using the MI limit in Fig.~\ref{fig:mi-limit}.
For a specific physics model, one first finds the model-dependent efficiency for the signal, namely the product of the acceptance and the efficiency, $\mathcal{A}\epsilon$, which includes the effects of the kinematic selections, the geometrical acceptance, and the trigger threshold.
The effect of the threshold $\MTlower$ on the signal is expressed by the fraction $f_{\MT}(\MTlower)$ of generated signal events that pass the requirement $\MT > \MTlower$, where \MT is calculated at generator level.

A limit on the production cross section times the branching fraction for the specific physics model in the fiducial phase space of the measurement, $[\sigma\mathcal{B}\mathcal{A}\epsilon]_{\mathrm{excl}}$, can be obtained as a function of $\MTlower$. 
This is done by dividing the excluded cross section of the MI limit 
$[\sigma\mathcal{B}\mathcal{A}\epsilon]_{\mathrm{MI}}(\MTlower)$ given in Fig.~\ref{fig:mi-limit} by the calculated fraction $f_{\MT}(\MTlower)$: 
\begin{equation}
[\sigma\mathcal{B}\mathcal{A}\epsilon]_{\mathrm{excl}}(\MTlower) = \frac{[\sigma\mathcal{B}\mathcal{A}\epsilon]_{\mathrm{MI}}(\MTlower)}{f_{\MT}(\MTlower)}.
\label{eqn:milimit}
\end{equation}
To interpret a new signal model using the MI limit, first, a $(\sigma\mathcal{B})_{\mathrm{excl}}$ is determined as a minimum value of the $[\sigma\mathcal{B}\mathcal{A}\epsilon]_{\mathrm{excl}}(\MTlower)$ by scanning $\MTlower$.
Then, a limit on the model parameters can be set by comparing $(\sigma\mathcal{B})_{\mathrm{excl}}$ with the theoretical cross section $(\sigma\mathcal{B})_{\mathrm{theo}}$.
Models where $(\sigma\mathcal{B})_{\mathrm{theo}} > (\sigma\mathcal{B})_{\mathrm{excl}}$ can be excluded.

\subsection{Exclusion limits on the \texorpdfstring{\PWpr}{W'} boson coupling strength}
\label{sec:coupling_ratio_limit}
The cross section exclusion limit depends on the width and the mass range of the potential signal.
A limit can be derived on the ratio of coupling strengths, $\gWp/\gW$, from the relationship between the coupling strength of a \PWpr and the resonance decay width.

\begin{figure}[hbtp]
\centering
\includegraphics[width=0.49\textwidth]{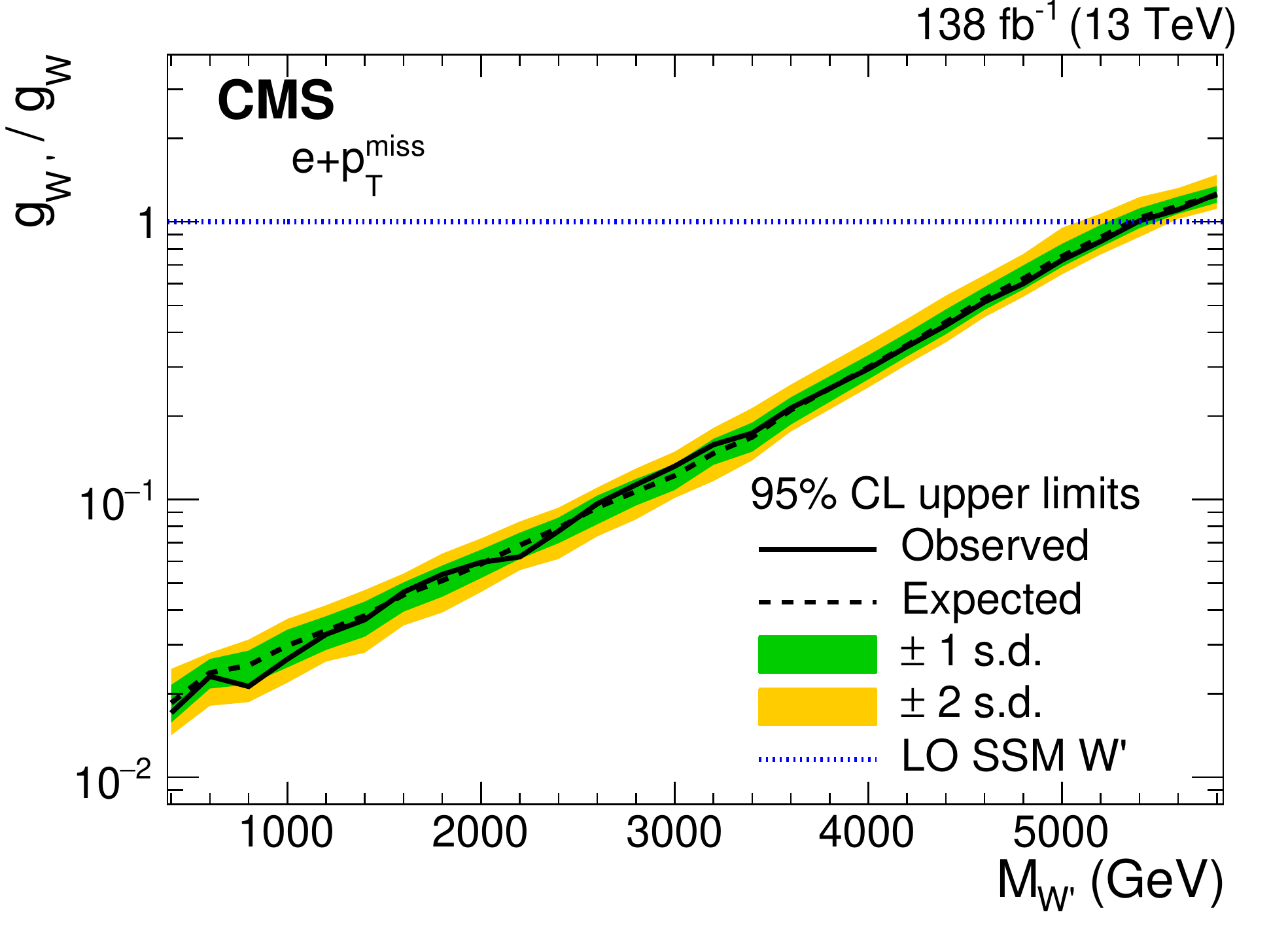} 
\includegraphics[width=0.49\textwidth]{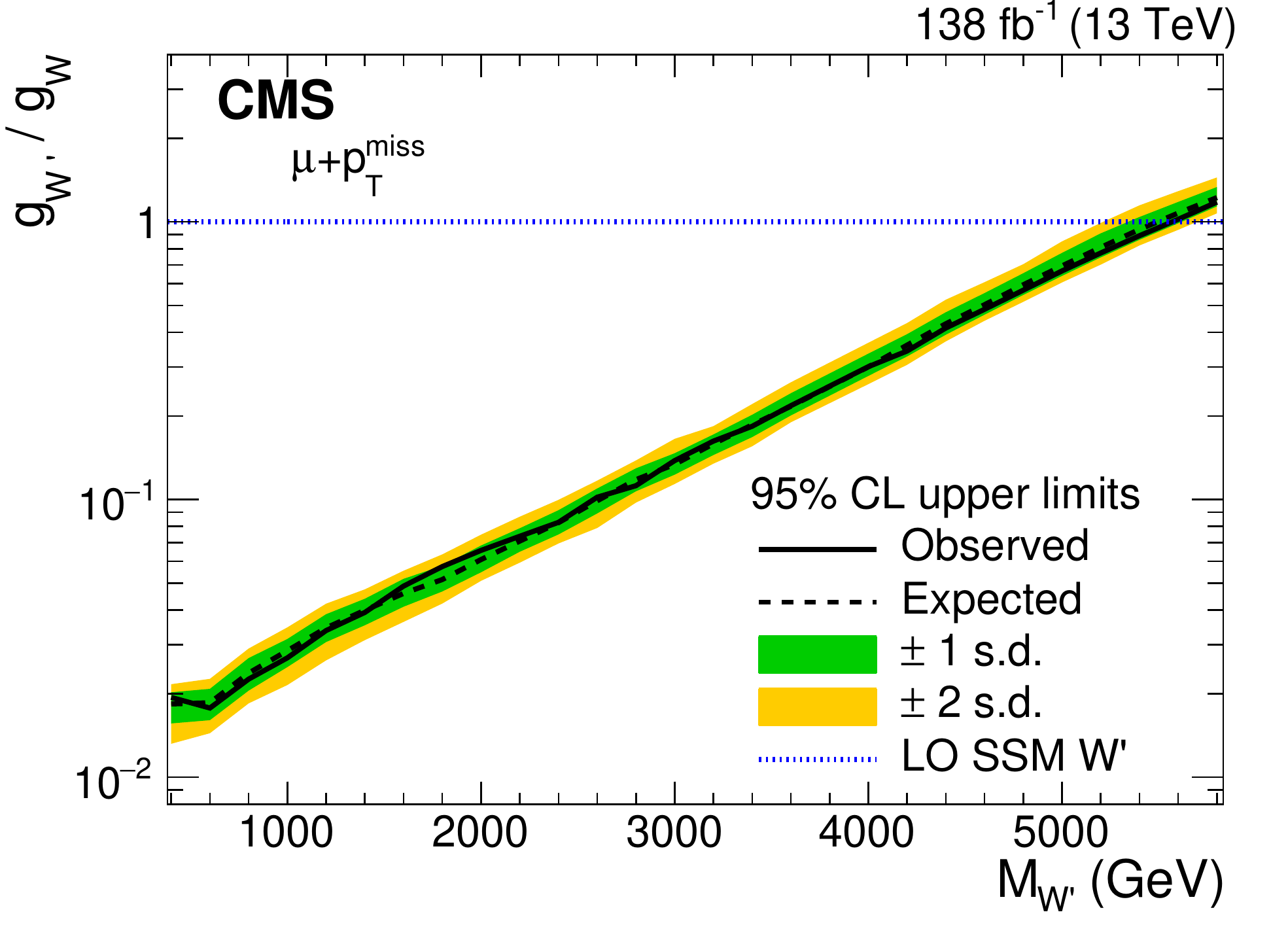}
\caption{The observed (solid line) and expected (dashed line) upper limits at 95\% \CL on the coupling strength ratio, $\gWp/\gW$ as functions of the mass of the \PWpr boson. 
These are shown for the electron (left) and muon (right) channels.
The one and two standard deviation uncertainty bands for the expected limits are shown. 
The area above the limit curve is excluded.
The dotted line represents the case of SSM couplings, $\gWp$, being equal to $\gW$, the SM coupling. 
}
\label{fig:gc-limit}
\end{figure}

Two steps are required to derive a coupling ratio limit for every $\MWp$.
The first one consists in setting, for each $\MWp$ point, an upper limit on $\xsecWp \mathcal{B}$ as a function of a coupling ratio. 
The second step is to find an intersection point of the $\xsecWp \mathcal{B}$ limit line with the theoretical cross section for every coupling ratio.
The theoretical cross sections used are calculated to LO in QCD precision for the $\MWp$ in this model. 
Coupling ratios where the limit is less than the theoretical prediction are excluded.
This procedure is repeated for every $\MWp$, and the corresponding intersection points provide the input for the result. 
The resulting exclusion limit on the coupling strength ratio, as a function of the $\MWp$, is shown in Fig.~\ref{fig:gc-limit} for the electron (left) and muon (right) channels.
No combination of channels is performed in this case since there is no assumption that the couplings have to be equal for both decay channels.
The area above the limit line is excluded. 
For the $\MWp\approx{1\TeV}$, coupling ratios above $2.7\times 10^{-2}$ ($3.0\times 10^{-2}$ expected) and $2.7\times 10^{-2}$ ($2.9\times 10^{-2}$ expected) are excluded in the electron and muon channels, respectively.

\subsection{Exclusion limits on the \texorpdfstring{\PWpr}{W'} boson in the split-UED model}
\label{sec:sUEDlimit}
\begin{figure}[hbtp]
\centering
\includegraphics[width=0.49\textwidth]{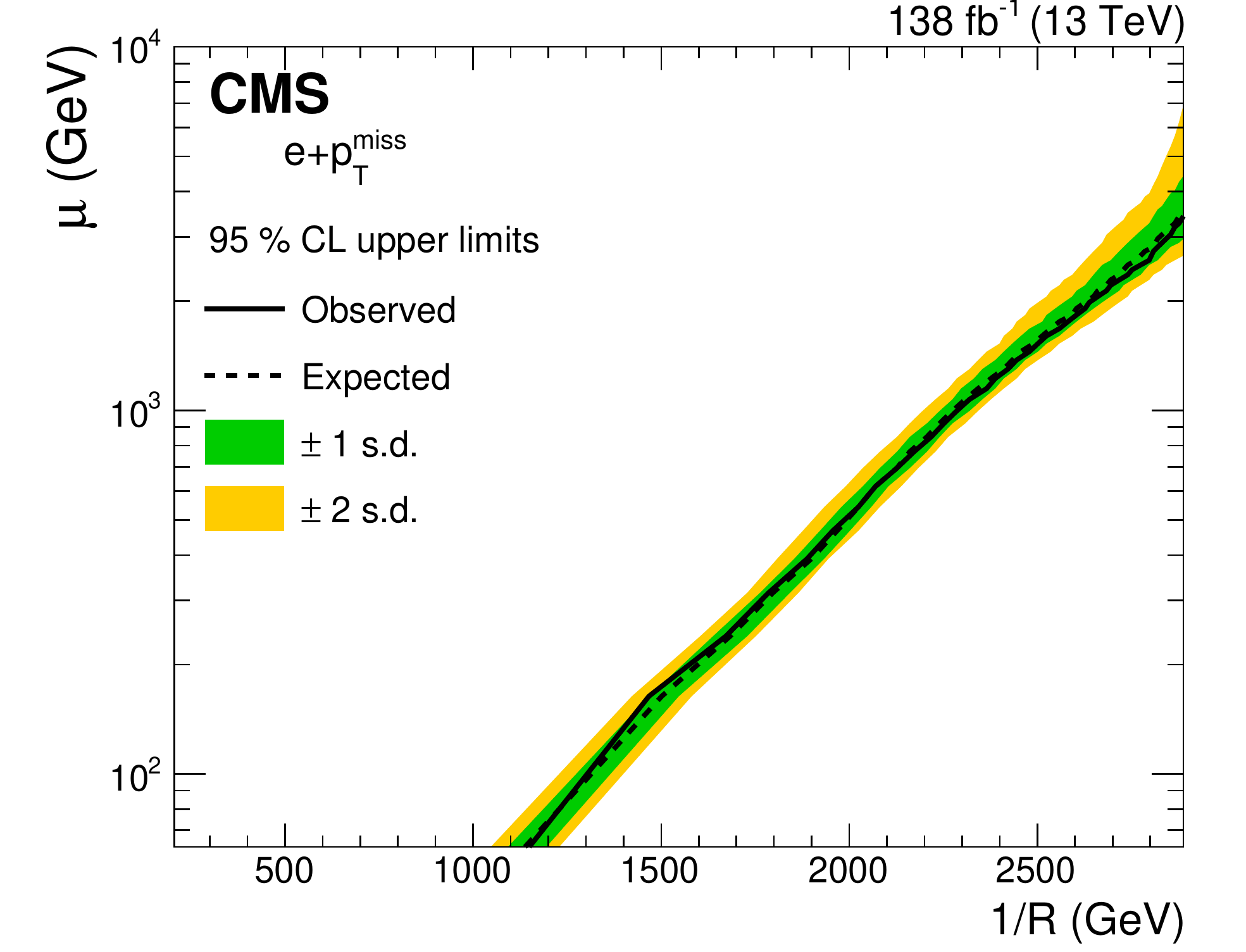} 
\includegraphics[width=0.49\textwidth]{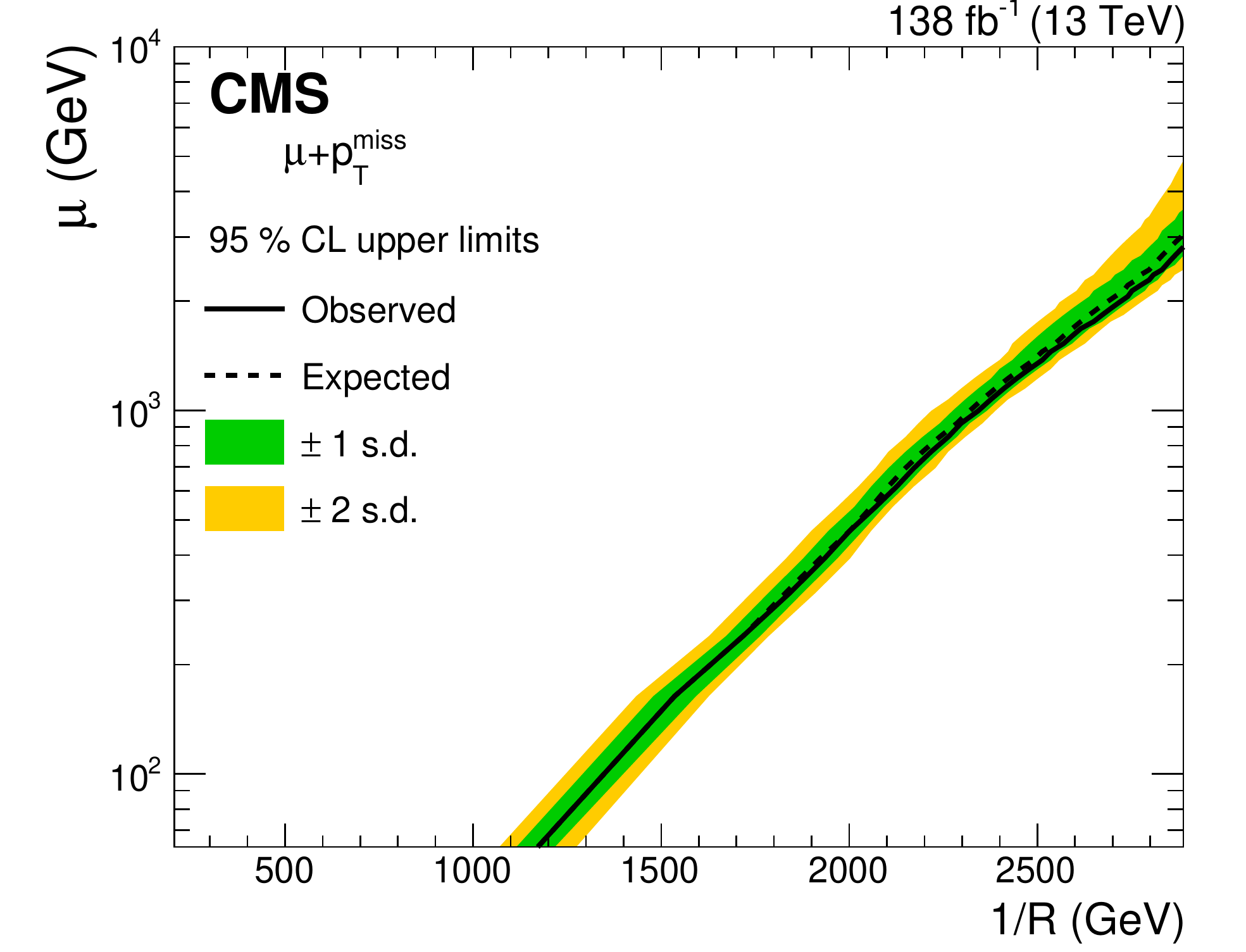}
\includegraphics[width=0.49\textwidth]{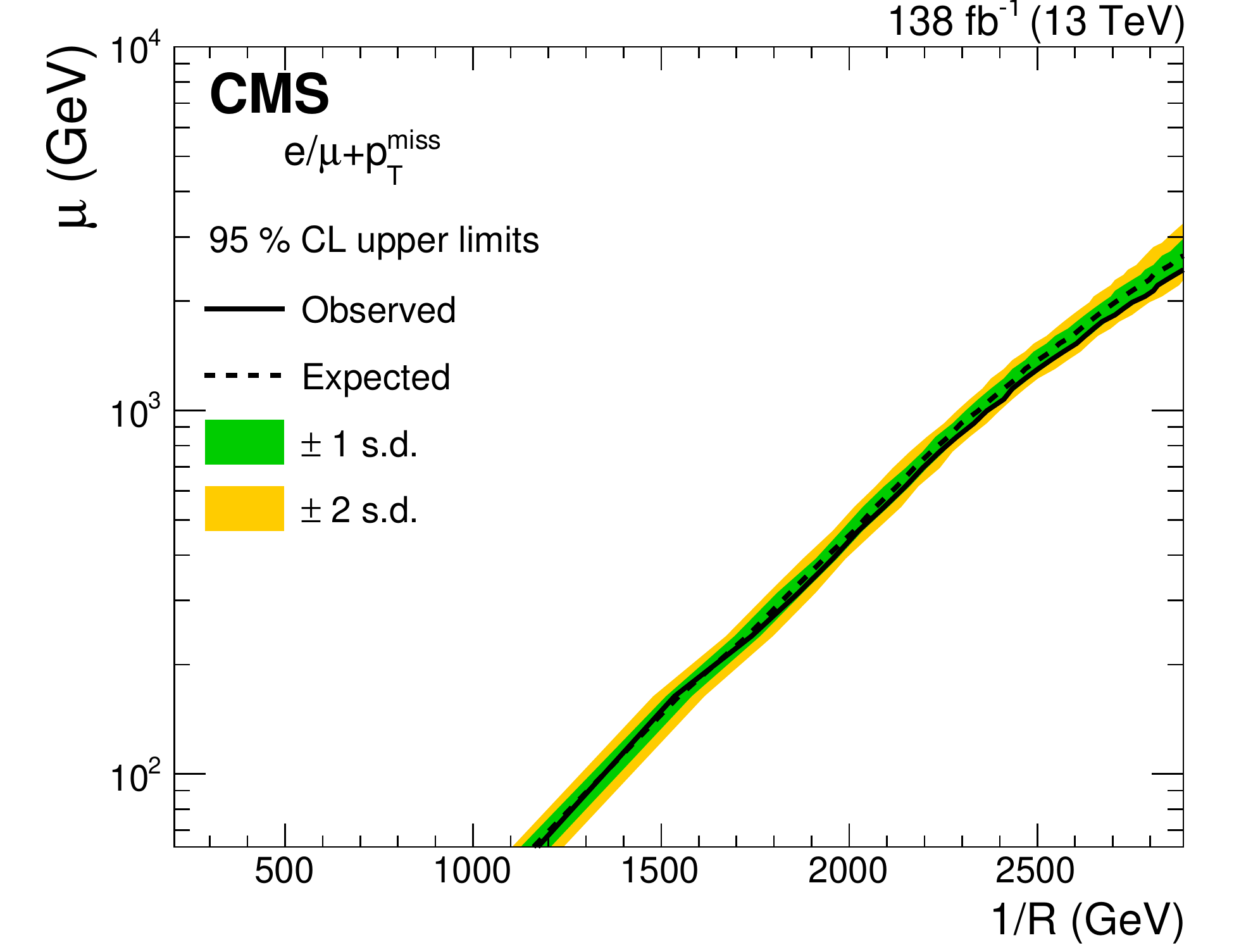}
\caption{Exclusion limits in the 2D plane ($1\!/\!R,\mu$) for the split-UED interpretation for the $n = 2$ case.
These are shown for the electron (upper left), muon (upper right), and the combination of both (lower) channels for the 2016--2018 data sets.
The expected limit is depicted as a black dashed line. 
The one and two standard deviation uncertainty bands for the expected limits are shown. 
The experimentally excluded region is the entire area to the left of the solid black line. 
}
\label{fig:ued-limit}
\end{figure}

To place limits on the split-UED model, the SSM \PWpr boson signal is reinterpreted using the extra dimension radius, $R$, and the bulk mass parameter of the five-dimensional fermion field, $\mu$.
For the lowest KK-even mode, $n = 2$, the decay of the \WprimeKKtwo to leptons is kinematically similar to the SSM \PWpr boson decay, and the differences in the widths of the resonances have little impact on the results. The impact is at most 2\% for $\mu<100\GeV$, and is negligible for higher values of $\mu$.
Taking into account the similar signal shapes and efficiencies allows the lower mass limits on the SSM \PWpr boson to be reinterpreted as lower limits on $1\!/\!R$, for a given value of $\mu$.
Separate limits on the electron and muon channels, and on their combination are shown in Fig.~\ref{fig:ued-limit}.
For the given value $\mu = 2\TeV$, the observed (expected) lower limits on $1\!/\!R$ are 2.7 (2.6)\TeV for electrons, 2.7 (2.7)\TeV for muons, and 2.8 (2.7)\TeV for the combination of the two channels.

\subsection{Exclusion limits on \texorpdfstring{\textit{R}}{R} parity violating SUSY}
\label{sec:rpv-limit}
\begin{figure}[hbtp]
\centering
\includegraphics[width=0.49\textwidth]{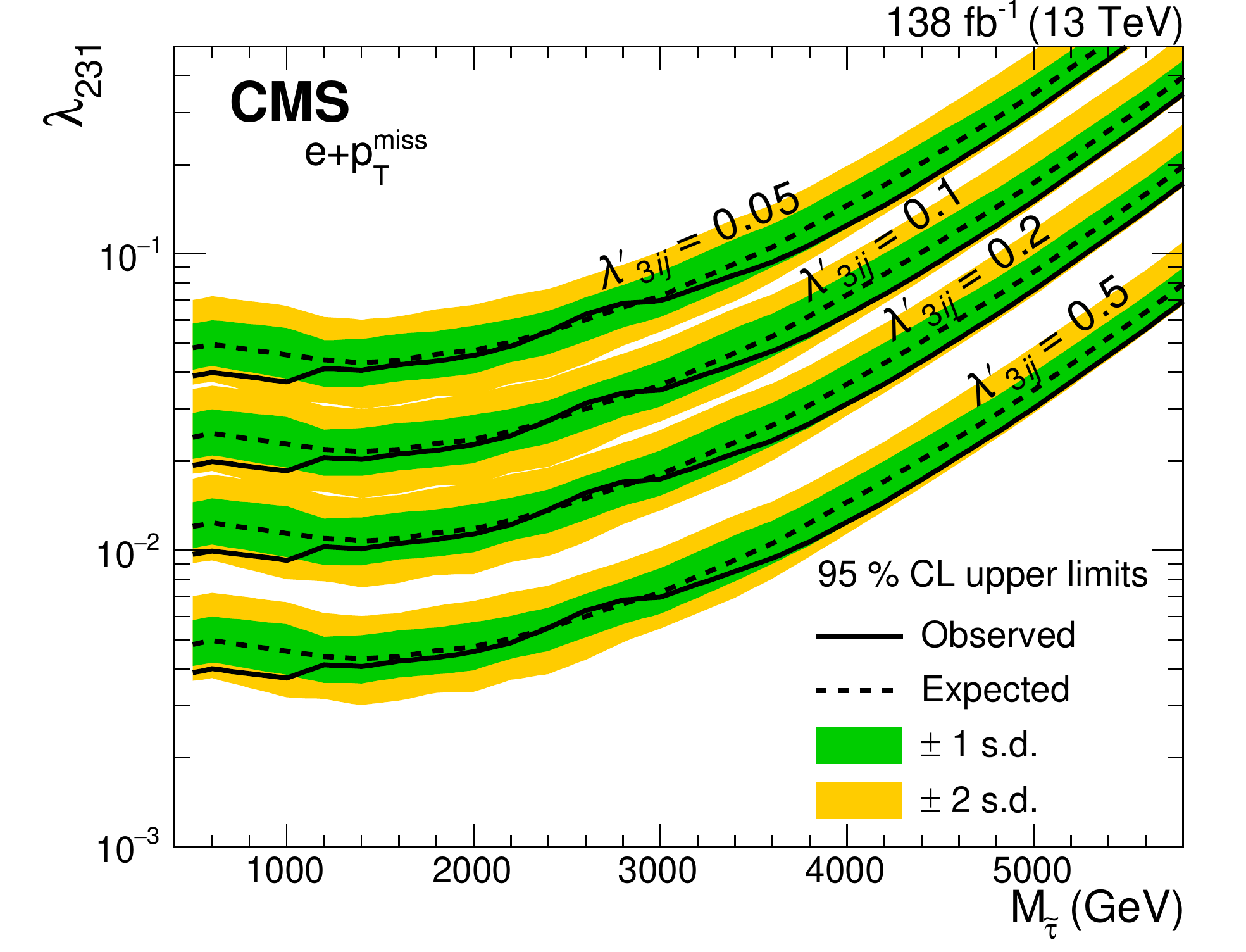} 
\includegraphics[width=0.49\textwidth]{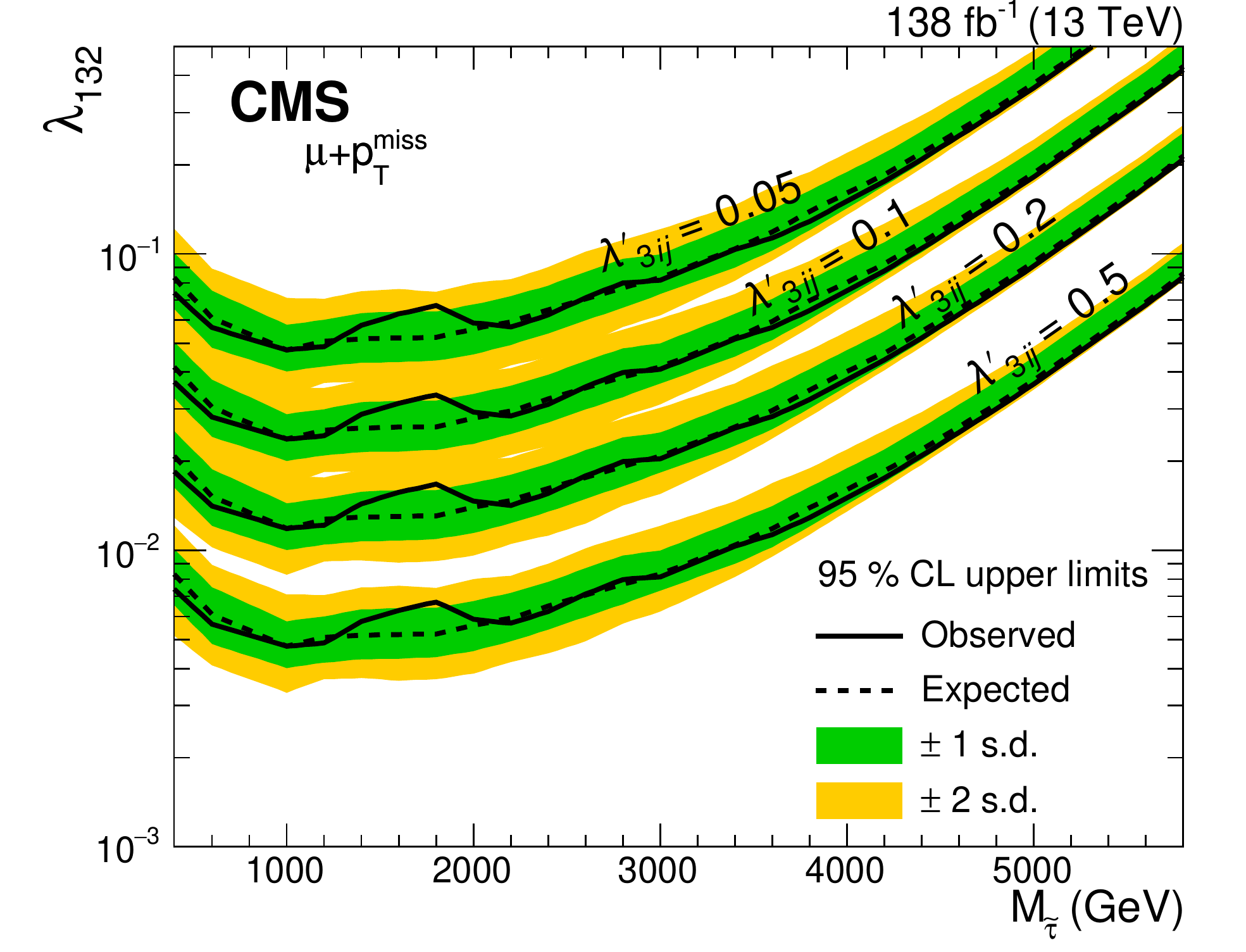} 
\caption{The observed (solid line) and expected (dashed line) upper limits at 95\% \CL on the couplings $\Lam_{231}$ (left) and $\Lam_{132}$ (right), for several values of $\Lamp_{3ij}$ ($i, j$ = 1, 2, 3), as a function of the mass of the mediator \sTau in the RPV SUSY model.
The one and two standard deviation uncertainty bands for the expected limits are shown.
The areas above the limit curves are excluded.
}
\label{fig:rpv-limit}
\end{figure}

The RPV SUSY interpretation assumes a \sTau mediator with distinct RPV SUSY couplings at the production and decay vertices.  
By using the MI limit in Fig.~\ref{fig:mi-limit}, the generated \MT distribution and the corresponding $f_{\!\MT}$ fractions, the $(\sigma\mathcal{B})_{\mathrm{excl}}$ value was calculated as described in Section~\ref{sec:MI_limit}. 
The cross sections calculated by \MGvATNLO~ at LO for the different coupling values were used to convert $(\sigma\mathcal{B})_{\mathrm{excl}}$ into limits on RPV SUSY couplings.
While $\Lamp_{3ij}$ is common to both decay channels, the coupling at the decay vertex is either $\Lam_{231}$ (for the $\Pe\PGnGm$ channel) or $\Lam_{132}$ (for the $\PGm\PGne$ channel). 
Upper exclusion limits on $\Lam_{231}$ and $\Lam_{132}$, as a function of the \sTau mediator mass, $\Mstau$, are shown in Fig.~\ref{fig:rpv-limit} for a number of $\Lamp_{3ij}$ coupling values. 
For the $\Mstau = 1\TeV$ and $\Lamp_{3ij} = 0.5$, values of $\Lam_{231}$ above $3.7\times 10^{-3}$ ($4.6\times 10^{-3}$ expected) and $\Lam_{132}$ above $4.7\times 10^{-3}$ ($4.7\times 10^{-3}$ expected) are excluded.

\subsection{Universal new physics through constraints on the \texorpdfstring{$W$}{W} parameter}
\label{sec:W-parameter_limit}
The comparison between the charged lepton plus a neutrino \MT spectra in data and the expected SM background contributions allows the oblique $W$ and $Y$ parameters~\cite{Farina_2017} to be constrained. 
For nonzero $W$ and $Y$ values, the modification of the \MT distributions of the $\Pell\PGn$ system at generator level, relative to that of the SM, can be obtained via event reweighting.
The weight is defined as
\begin{equation}
\left|\frac{P_{\PW}}{P_{\PW}^{(0)}}\right|^{2} = \left( 1+{\frac{(2t^{2}-1)W}{1-t^{2}}} + {\frac{t^{2}Y}{1-t^{2}}} - {\frac{W(q^{2}-m_{\PW}^{2})}{m_{\PW}^{2}}}\right)^{2},
\label{eqn:Wparam}
\end{equation}
where $P_{\PW}$ is the expression for the propagator of the $\qqbar\to\PW\to\Pell\PGn$ process that depends on the $W$ and $Y$ parameters, $P_{\PW}^{(0)}$ is the SM \PW boson propagator ($W=Y=0$), $t$ is the tangent of the SM weak mixing angle ($t^{2}\approx0.3$), and $q^{2}$ is the square of the invariant mass of the $\Pell\PGn$ system at the hard scattering level.

For high \MT values, in practice, the weight depends solely on the oblique $W$ parameter and is independent of the oblique $Y$ parameter. 
This region is where most of the signal sensitivity occurs. 
Thus, the $\Pell\PGn$ analysis allows the determination of the $W$ parameter value. 

As mentioned in the Section~\ref{sec:intro}, the 2016 data set has not been reanalyzed.
The combined electron and muon channel data taken during 2017--2018, corresponding to an integrated luminosity of 101\fbinv, are used to constrain the $W$ parameter. 

To find the oblique $W$ parameter value that best reproduces data, a binned negative log likelihood fit has been implemented using {\small MINUIT}~\cite{MINUIT}. 
The distribution used as input in the fit is the binned \MT distribution for data and for the SM prediction as previously presented, where the $\PW\to\Pell\PGn$ contribution is reweighted as explained.   
The oblique $W$ parameter appears in the fit through the weight of each event in the SM \PW boson background sample.
Systematic uncertainties are included as nuisance parameters in the fit with  log-normal distributions. 
The best-fit value of the parameter and an approximate 68\% \CL confidence interval are extracted following the procedure described in Section~\ref{sec:models-W-parameter} of Ref.~\cite{Khachatryan:2014jba}.

We assume that only the SM $\PW\to\Pell\PGn$ process is modified by new physics at high energy scales, disregarding any potential effect on the Drell--Yan $\PZ\to\Pell\Pell$ background process, which contributes only around 4\% of the selected $\PW\to\Pell\PGn$ events.
In addition, and according to Ref.~\cite{Farina_2017}, the sensitivity of \PZ boson processes to the $W$ parameter is approximately one half that of \PW boson processes for similar integrated luminosity and $\sqrt{s}$ values.
This effect is statistically translated to having an excess (or deficit) of 1\% of the \PW boson sample, which is included as a systematic uncertainty in the fit.

\begin{figure}[hbtp]
\centering
\includegraphics[width=0.6\textwidth]{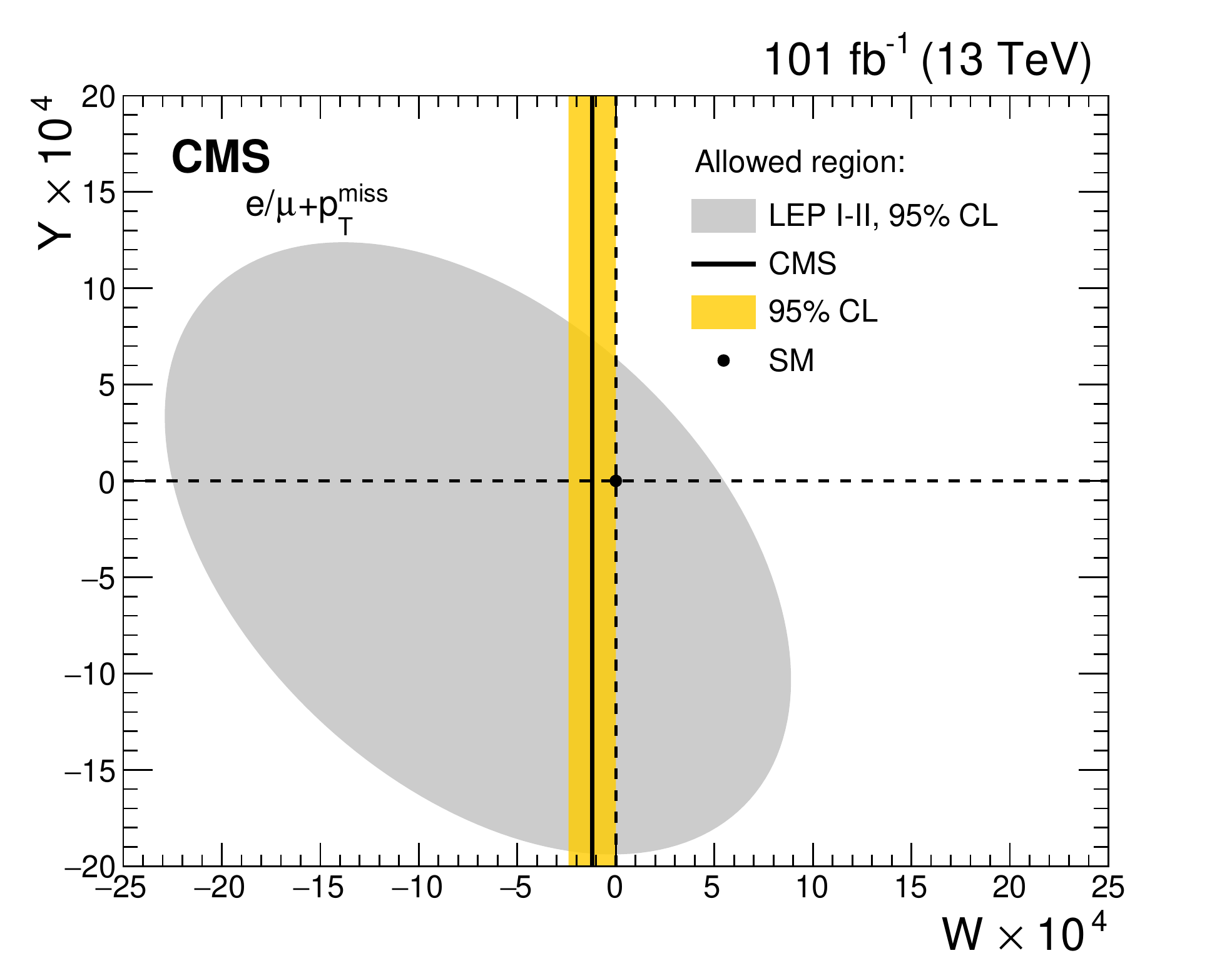}
\caption{Region in the oblique $(W, Y)$ parameter phase space allowed by the current analysis at 95\% \CL, obtained by combining electron and muon channel distributions. 
Comparison of the result from the current analysis with the area derived from LEP experiments (grey-shaded area) is presented.
 }
\label{fig:Wparamconstrained}
\end{figure}

By combining the electron and muon channel distributions from the 2017--2018 data sets, a fit value of $W = -1.2^{+0.5}_{-0.6}\times 10^{-4}$ is found.
The uncertainties reflect both statistical and systematic components, and they are in agreement with sensitivity studies performed using pseudodata instead of the experimental data themselves. The outcome of the fit and the region allowed (band at 95\% \CL) by these results in the oblique $(W,Y)$ parameter space together with the allowed region derived~\cite{Farina_2017} from LEP results~\cite{Falkowski:2015krw} at 95\% \CL are shown in Fig.~\ref{fig:Wparamconstrained}. 
The SM value of $W = 0$ lies within the 2 sigma region constrained at 95\% \CL by the current analysis.
This result improves by an order of magnitude on previous limits for the value of the oblique $W$ parameter.

\subsection{Constraints on the composite Higgs boson model}
\label{sec:Higgs-composite-res}
Results obtained in this search are reinterpreted in the context of composite Higgs boson models. 
Constraints are placed in the $m_{*}$--$g_{*}$ plane, where $m_{*}$ is the mass scale of new composite resonances generated in these models and $g_{*}$ the coupling of the new interaction, respectively.

\begin{figure}[hbtp]
\centering
\includegraphics[width=0.6\textwidth]{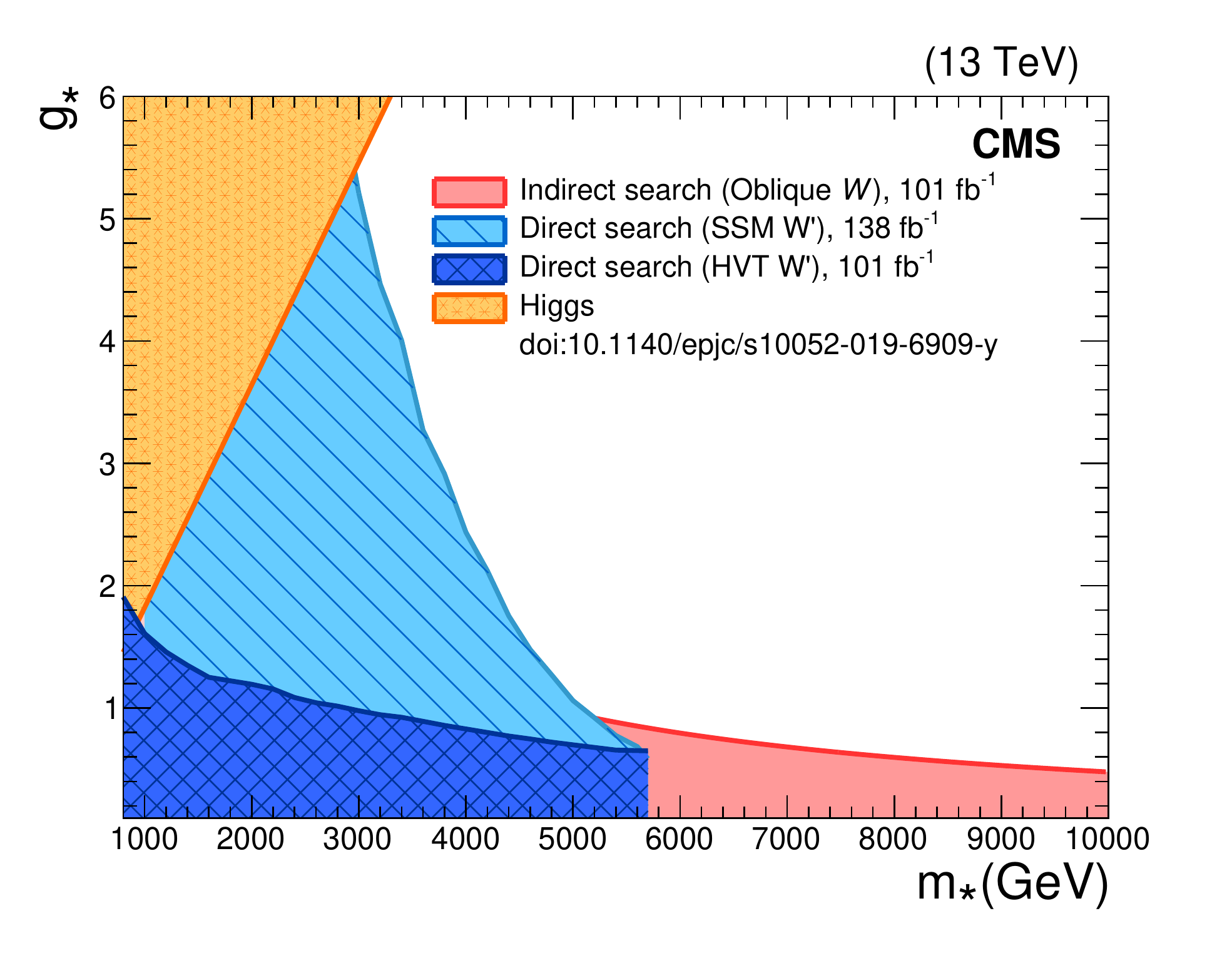}
\caption{Regions in the $m_{*}$--$g_{*}$ plane excluded at 95\% \CL by the results derived in the framework of the 
different models considered in the current analysis. 
Constraints from the limit on the oblique $W$ parameter are shown in red. 
Constraints from the limit on the \PWpr boson coupling strength are shown in light blue for the SSM \PWpr hypothesis (only coupling to fermions), and in dark blue for the HVT \PWpr hypothesis (coupling to bosons as well).
In addition, the exclusion coming from current CMS constraints on the Higgs boson cross section is shown as a orange shaded area.
}
\label{fig:Hcomp_exclusion_all}
\end{figure}

The coupling strength ratio of the \PWpr boson to the SM \PW boson, $\gWp/\gW$, is related to the $g_{*}$ coupling through Eq.~(\ref{eqn:direct_Wparam}). The limits derived in Section~\ref{sec:coupling_ratio_limit} yield the constraints shown in Fig.~\ref{fig:Hcomp_exclusion_all} (light blue shaded area) where the equivalence between $\MWp$ and $m_{*}$ is assumed.
For clarity, only the observed limit is shown, the expected one being very similar as it is deduced from the expected limit on the coupling strength ratio in Fig.~\ref{fig:gc-limit}.  
Data sets from 2016--2018 are used, combining the electron and muon channels.
The shaded region is excluded, given the assumption of an SSM \PWpr, with only couplings to fermions used in the extraction of the coupling ratio limit.  

In order to accommodate couplings to bosons and fermions (Heavy Vector Triplet model, HVT~\cite{HVT_model}), a reweighting procedure has been introduced. 
It implements the ratio of Breit--Wigner distributions of corresponding fermionic plus bosonic widths relative to that of the fermionic width, and takes into account the different cross section values. 
The region excluded in the $m_{*}$--$g_{*}$ plane, achieved by combining the results from the electron and muon channels for the 2017--2018 data sets, is shown by the dark blue area in Fig.~\ref{fig:Hcomp_exclusion_all}.
This region is less restrictive than the light blue one because of the smaller branching fraction being tested with the current final state and the reduced amount of data. 
However, the interpretation is more general in this HVT framework.

Being quantities extracted from a direct search, these constraints are kinematically limited by $\sqrt{s}$.
The EFT approach followed to set constraints on the oblique $W$ parameter at 95\% \CL extends
beyond the limit established by the direct search. 
The constraint from the oblique $W$ parameter, derived from the combination of the electron and muon channels, is also used to set limits in the $m_{*}$--$g_{*}$ plane, following Eq.~(\ref{eqn:indirect_Wparam}).
Numerically, this implies $g_{*} < (4770\GeV)/m_{*}$. 
Figure~\ref{fig:Hcomp_exclusion_all} shows the region (red shaded) excluded by these results at 95\% \CL. 
This approach extends the constraints up to high resonance masses, since the EFT approach remains valid in this region.

In addition, current CMS constraints on the deviation ($\Delta\mH$) of the Higgs boson production cross section from the SM value also set limits in the $m_{*}$--$g_{*}$ plane of the composite Higgs boson sector. 
Considering the value measured by CMS in a combination of several Higgs boson decay modes~\cite{Higgs_combination} of $\Delta\mH < 0.20$ at 95\% \CL, and following Eq.~(\ref{eqn:Higgs_xsec}), the exclusion region $g_{*} > (0.00182\GeV^{-1}) m_{*}$ is established. 
The excluded region coming from the Higgs boson cross section constraint, together with the regions already presented, is shown in Fig.~\ref{fig:Hcomp_exclusion_all} (orange triangular area in the upper-left part of the plot).
Taking all exclusion regions into account, an energy scale for Higgs boson compositeness 
can be excluded below $m_{*} = 3\TeV$ under the SSM \PWpr boson assumption, or alternatively, below 1\TeV for an HVT interpretation of the \PWpr resonance. 

All limits obtained in this paper are summarized in Table~\ref{tab:limit_summary}.

\begin{table}[htb!]
 \centering
 \topcaption{Summary of all 95\% \CL exclusion limit results with various theoretical model interpretations in the electron and muon channels, and the combination of both channels. 
The results in EFT and the composite Higgs boson model are obtained using 2017--2018 data sets only.
The final allowed ranges for the model parameters are in the last column, except for the oblique $W$ parameter, where a measurement is provided, and the Composite Higgs scenario, as explained in the text.
 }
 \renewcommand{\arraystretch}{1,3}
 \cmsTable{
   \begin{tabular}{llcl}
   \hline
   {Model}                       & {Parameter}                   & {Channel}& {Observed (expected) limit}     \\ \hline
   \multirow{3}{*}{SSM \PWpr}           & $\MWp$	    	               & \Pe		 & $\MWp >$ 5.4 (5.3)\TeV                 \\
   			                & ($\gWp/\gW=1$)                       & \PGm	         & $\MWp >$ 5.6 (5.5)\TeV                 \\ 
   			                & 			               & \Pe + \PGm	 & $\MWp >$ 5.7 (5.6)\TeV                 \\ [\cmsTabSkip]
   SSM \PWpr		                & $\gWp/\gW$  	                       & \Pe 	         & $\gWp/\gW <$ 2.7 (3.0)$\times10^{-2}$    \\ 
   with various $g_{\PWpr}$             & (if $\MWp\approx1\TeV)$              & \PGm            & $\gWp/\gW <$ 2.7 (2.9)$\times10^{-2}$    \\ [\cmsTabSkip]
   \multirow{2}{*}{Split-UED \WprimeKKtwo} &  $1\!/\!R$ 	               & \Pe  	         & $1\!/\!R >$ 2.7 (2.6)\TeV              \\
   			                & (if $\mu=2\TeV$)                     & \PGm	         & $1\!/\!R >$ 2.7 (2.7)\TeV              \\ 
   			                & 			               & \Pe + \PGm      & $1\!/\!R >$ 2.8 (2.7)\TeV              \\ [\cmsTabSkip] 
   \multirow{2}{*}{RPV SUSY \sTau}      & $\Lam_{\text{decay=231,132}}$        & \Pe             & $\Lam_{231} <$ 3.7 (4.6)$\times10^{-3}$ \\
                                        & (if $\Lamp_{3ij}=0.5$, $\Mstau\approx1\TeV$) & \PGm    & $\Lam_{132} <$ 4.7 (4.7)$\times10^{-3}$ \\ [\cmsTabSkip]
   EFT                                  & Oblique $W$ parameter                & \Pe + \PGm      & $W = -1.2^{+0.5}_{-0.6}\times10^{-4}$  \\ [\cmsTabSkip]
   \multirow{2}{*}{Composite Higgs}     & Regions in $m_{*}$--$g_{*}$ plane in SSM & \Pe + \PGm   & $m_{*} >$ 3\TeV            \\ 
                                        & Regions in $m_{*}$--$g_{*}$ plane in HVT & \Pe + \PGm   & $m_{*} >$ 1\TeV            \\ 
   \hline
   \end{tabular}
 }
\label{tab:limit_summary}
\end{table}

\section{Summary}
\label{sec:summary}

A search for a deviation relative to standard model (SM) expectations in events with a final state consisting of a lepton (electron or muon) and missing transverse momentum in proton-proton collisions at a centre-of-mass energy of 13\TeV has been performed.
The analysis strategy is similar to that of the previous study using data corresponding to an integrated luminosity of 36\fbinv~\cite{Sirunyan:2018mpc}.
This search uses data collected by the CMS detector in 2016--2018 corresponding to 138\fbinv of total integrated luminosity.
In addition to employing four times larger data set, this search considers two new interpretations.

No evidence for new physics is observed when examining the transverse mass distributions.
These observations are interpreted as limits on the parameters of several models.
The exclusion limits at 95\% confidence level (\CL) on the mass of a sequential standard model (SSM) \PWpr boson are found to be 5.4 and 5.6\TeV for the electron and muon channels, respectively.
The 95\% \CL exclusion limit from the combination of both channels is 5.7\TeV.
Variations in the coupling strength of the SSM \PWpr boson are also examined.
Models for which the ratio of the coupling strength of the \PWpr boson to the SM \PW boson is at the level of $2\times 10^{-2}$ are excluded for \PWpr masses up to 0.5\TeV.
For higher masses the constraint on the coupling weakens, approaching $\gWp/\gW = 1$ at the value of $\MWp$ corresponding to the exclusion limit obtained in the SSM analysis.
The result is also interpreted in the split universal extra dimensions model.
The inverse radius of the extra dimension, $1\!/\!R$, is constrained to be larger than 2.8\TeV if the bulk mass parameter of the five-dimensional fermion field, $\mu = 2\TeV$.

Model-independent limits are also provided.
These can be used to constrain parameters of several models through reinterpretations. 
As an example, the limits have been interpreted in the context of an \textit{R}-parity violating supersymmetric model.
This interpretation provides limits on the coupling strengths at the decay vertex as a function of the mediator \sTau mass, for various coupling values, $\Lamp_{3ij}$, at the production vertex.

In addition to these results, new interpretations of an $\Pell\PGn$ final state analysis using LHC data are considered for the first time.
One is the measurement of the oblique $W$ parameter, using the combined electron and muon channels.
The oblique $W$ parameter fitted value is $W = -1.2^{+0.5}_{-0.6}\times 10^{-4}$, in agreement within uncertainties with the SM value. 
This is the most precise measurement to date of the oblique $W$ parameter, reducing its allowed range by more than an order of magnitude relative to previous bounds derived from LEP results.

Another new interpretation is made within the framework of composite Higgs boson models.
The 95\% \CL exclusion limit for the combined channels is set in the $m_{*}$--$g_{*}$ plane, where $m_{*}$ indicates the mass scale of compositeness and $g_{*}$ the coupling strength of the new composite sector.
In this case, interpretations of several of the presented results have been used in a complementary way, as different inputs for setting the limit.
Each of the inputs is sensitive to a different region in the $m_{*}$--$g_{*}$ plane. 
Constraints coming from the limit on the \PWpr coupling strength dominate the low-$m_{*}$ region and are kinematically limited by the collider energy. 
The high-$m_{*}$ region is probed by the indirect oblique $W$ parameter interpretation.
Finally, we set a lower limit of 3 (1)\TeV for the mass scale of such a composite Higgs boson, when an SSM (Heavy Vector Triplet) \PWpr boson is assumed, extending existing constraints from previous measurements.

\begin{acknowledgments}
  We congratulate our colleagues in the CERN accelerator departments for the excellent performance of the LHC and thank the technical and administrative staffs at CERN and at other CMS institutes for their contributions to the success of the CMS effort. In addition, we gratefully acknowledge the computing centres and personnel of the Worldwide LHC Computing Grid and other centres for delivering so effectively the computing infrastructure essential to our analyses. Finally, we acknowledge the enduring support for the construction and operation of the LHC, the CMS detector, and the supporting computing infrastructure provided by the following funding agencies: BMBWF and FWF (Austria); FNRS and FWO (Belgium); CNPq, CAPES, FAPERJ, FAPERGS, and FAPESP (Brazil); MES and BNSF (Bulgaria); CERN; CAS, MoST, and NSFC (China); MINCIENCIAS (Colombia); MSES and CSF (Croatia); RIF (Cyprus); SENESCYT (Ecuador); MoER, ERC PUT and ERDF (Estonia); Academy of Finland, MEC, and HIP (Finland); CEA and CNRS/IN2P3 (France); BMBF, DFG, and HGF (Germany); GSRI (Greece); NKFIA (Hungary); DAE and DST (India); IPM (Iran); SFI (Ireland); INFN (Italy); MSIP and NRF (Republic of Korea); MES (Latvia); LAS (Lithuania); MOE and UM (Malaysia); BUAP, CINVESTAV, CONACYT, LNS, SEP, and UASLP-FAI (Mexico); MOS (Montenegro); MBIE (New Zealand); PAEC (Pakistan); MSHE and NSC (Poland); FCT (Portugal); JINR (Dubna); MON, RosAtom, RAS, RFBR, and NRC KI (Russia); MESTD (Serbia); MCIN/AEI and PCTI (Spain); MOSTR (Sri Lanka); Swiss Funding Agencies (Switzerland); MST (Taipei); ThEPCenter, IPST, STAR, and NSTDA (Thailand); TUBITAK and TAEK (Turkey); NASU (Ukraine); STFC (United Kingdom); DOE and NSF (USA).

\hyphenation{Rachada-pisek} Individuals have received support from the Marie-Curie programme and the European Research Council and Horizon 2020 Grant, contract Nos.\ 675440, 724704, 752730, 758316, 765710, 824093, 884104, and COST Action CA16108 (European Union); the Leventis Foundation; the Alfred P.\ Sloan Foundation; the Alexander von Humboldt Foundation; the Belgian Federal Science Policy Office; the Fonds pour la Formation \`a la Recherche dans l'Industrie et dans l'Agriculture (FRIA-Belgium); the Agentschap voor Innovatie door Wetenschap en Technologie (IWT-Belgium); the F.R.S.-FNRS and FWO (Belgium) under the ``Excellence of Science -- EOS" -- be.h project n.\ 30820817; the Beijing Municipal Science \& Technology Commission, No. Z191100007219010; the Ministry of Education, Youth and Sports (MEYS) of the Czech Republic; the Deutsche Forschungsgemeinschaft (DFG), under Germany's Excellence Strategy -- EXC 2121 ``Quantum Universe" -- 390833306, and under project number 400140256 - GRK2497; the Lend\"ulet (``Momentum") Programme and the J\'anos Bolyai Research Scholarship of the Hungarian Academy of Sciences, the New National Excellence Program \'UNKP, the NKFIA research grants 123842, 123959, 124845, 124850, 125105, 128713, 128786, and 129058 (Hungary); the Council of Science and Industrial Research, India; the Latvian Council of Science; the Ministry of Science and Higher Education and the National Science Center, contracts Opus 2014/15/B/ST2/03998 and 2015/19/B/ST2/02861 (Poland); the Funda\c{c}\~ao para a Ci\^encia e a Tecnologia, grant CEECIND/01334/2018 (Portugal); the National Priorities Research Program by Qatar National Research Fund; the Ministry of Science and Higher Education, projects no. 0723-2020-0041 and no. FSWW-2020-0008 (Russia); MCIN/AEI/10.13039/501100011033, ERDF ``a way of making Europe", and the Programa Estatal de Fomento de la Investigaci{\'o}n Cient{\'i}fica y T{\'e}cnica de Excelencia Mar\'{\i}a de Maeztu, grant MDM-2017-0765 and Programa Severo Ochoa del Principado de Asturias (Spain); the Stavros Niarchos Foundation (Greece); the Rachadapisek Sompot Fund for Postdoctoral Fellowship, Chulalongkorn University and the Chulalongkorn Academic into Its 2nd Century Project Advancement Project (Thailand); the Kavli Foundation; the Nvidia Corporation; the SuperMicro Corporation; the Welch Foundation, contract C-1845; and the Weston Havens Foundation (USA).
\end{acknowledgments}

\bibliography{auto_generated}

\providecommand{\href}[2]{#2}\begingroup\raggedright\begin{thebibliography}{10}%
\makeatletter
\providecommand{\hrefCMSnoop }[0]{\@secondoftwo}%
\makeatother
\providecommand{\doi}{\texttt{doi:}\begingroup \urlstyle{tt}\Url}

\bibitem{PDG2020}
\hrefCMSnoop {}{{Particle Data Group}, P.~A. Zyla {et~al.}, ``Review of
  particle physics'',} \textit{ Prog. Theor. Exp. Phys.} \textbf{ 2020} (2020)
  083C01,
  \href{http://dx.doi.org/10.1093/ptep/ptaa104}{\doi{10.1093/ptep/ptaa104}}.

\bibitem{PhysRevD.11.2558}
\hrefCMSnoop {}{R.~N. Mohapatra and J.~C. Pati, ````{Natural}'' left-right
  symmetry'',} \textit{ Phys. Rev. D} \textbf{ 11} (1975) 2558,
  \href{http://dx.doi.org/10.1103/PhysRevD.11.2558}{\doi{10.1103/PhysRevD.11.2558}}.

\bibitem{PhysRevD.11.566}
\hrefCMSnoop {}{R.~N. Mohapatra and J.~C. Pati, ``Left-right gauge symmetry and
  an ``isoconjugate'' model of $\mathrm{CP}$ violation'',} \textit{ Phys. Rev.
  D} \textbf{ 11} (1975) 566,
  \href{http://dx.doi.org/10.1103/PhysRevD.11.566}{\doi{10.1103/PhysRevD.11.566}}.

\bibitem{PhysRevD.82.035011}
\hrefCMSnoop {}{K.~Hsieh, K.~Schmitz, J.-H. Yu, and C.-P. Yuan, ``Global
  analysis of general {SU(2)}{$\times$}{SU(2)}{$\times$}{U(1)} models with
  precision data'',} \textit{ Phys. Rev. D} \textbf{ 82} (2010) 035011,
  \href{http://dx.doi.org/10.1103/PhysRevD.82.035011}{\doi{10.1103/PhysRevD.82.035011}},
  \href{http://www.arXiv.org/abs/1003.3482}{\texttt{arXiv:1003.3482}}.

\bibitem{ARKANIHAMED1998263}
\hrefCMSnoop {}{N.~Arkani-Hamed, S.~Dimopoulos, and G.~Dvali, ``The hierarchy
  problem and new dimensions at a millimeter'',} \textit{ Phys. Lett. B}
  \textbf{ 429} (1998) 263,
  \href{http://dx.doi.org/10.1016/S0370-2693(98)00466-3}{\doi{10.1016/S0370-2693(98)00466-3}},
\href{http://www.arXiv.org/abs/hep-ph/9803315}{\texttt{arXiv:hep-ph/9803315}}.
%%CITATION = HEP=PH/9803315%%.

\bibitem{Appelquist:2000nn}
\hrefCMSnoop {}{T.~Appelquist, H.-C. Cheng, and B.~A. Dobrescu, ``Bounds on
  universal extra dimensions'',} \textit{ Phys. Rev. D} \textbf{ 64} (2001)
  035002,
  \href{http://dx.doi.org/10.1103/PhysRevD.64.035002}{\doi{10.1103/PhysRevD.64.035002}},
\href{http://www.arXiv.org/abs/hep-ph/0012100}{\texttt{arXiv:hep-ph/0012100}}.
%%CITATION = HEP-PH/0012100;%%.

\bibitem{JECref}
\hrefCMSnoop {}{{CMS Collaboration}, ``Performance of missing transverse
  momentum reconstruction in proton-proton collisions at $\sqrt{s}=13$ {TeV}
  using the {CMS} detector'',} \textit{ JINST} \textbf{ 14} (2019) P07004,
  \href{http://dx.doi.org/10.1088/1748-0221/14/07/p07004}{\doi{10.1088/1748-0221/14/07/p07004}},
\href{http://www.arXiv.org/abs/1903.06078}{\texttt{arXiv:1903.06078}}.
%%CITATION = CMS-PAS-JME-16-004;%%.

\bibitem{reference-model}
\hrefCMSnoop {}{G.~Altarelli, B.~Mele, and M.~Ruiz-Altaba, ``Searching for new
  heavy vector bosons in {$\Pp\Pap$} colliders'',} \textit{ Z. Phys. C}
  \textbf{ 45} (1989) 109,
  \href{http://dx.doi.org/10.1007/BF01556677}{\doi{10.1007/BF01556677}}.
  [Erratum: \DOI{10.1007/BF01552335}].

\bibitem{JHEP04(2010)081}
\hrefCMSnoop {}{K.~Kong, S.~C. Park, and T.~G. Rizzo, ``Collider phenomenology
  with split-{UED}'',} \textit{ JHEP} \textbf{ 04} (2010) 081,
  \href{http://dx.doi.org/10.1007/JHEP04(2010)081}{\doi{10.1007/JHEP04(2010)081}},
\href{http://www.arXiv.org/abs/1002.0602}{\texttt{arXiv:1002.0602}}.
%%CITATION = ARXIV:1002.0602;%%.

\bibitem{PhysRevD.79.091702}
C.-R. Chen\hrefCMSnoop {}{ {et~al.}, ``Dark matter and collider phenomenology
  of split-{UED}'',} \textit{ JHEP} \textbf{ 09} (2009) 078,
  \href{http://dx.doi.org/10.1088/1126-6708/2009/09/078}{\doi{10.1088/1126-6708/2009/09/078}},
\href{http://www.arXiv.org/abs/0903.1971}{\texttt{arXiv:0903.1971}}.
%%CITATION = ARXIV:0903.1971;%%.

\bibitem{FarrarFayet}
\hrefCMSnoop {}{G.~R. Farrar and P.~Fayet, ``Phenomenology of the production,
  decay, and detection of new hadronic states associated with supersymmetry'',}
  \textit{ Phys. Lett. B} \textbf{ 76} (1978) 575,
\href{http://dx.doi.org/10.1016/0370-2693(78)90858-4}{\doi{10.1016/0370-2693(78)90858-4}}.
%%CITATION = PHLTA,76B,575;%%.

\bibitem{DreinerRPV}
\hrefCMSnoop {}{H.~K. Dreiner and T.~Stefaniak, ``Bounds on {R}-parity
  violation from resonant slepton production at the {LHC}'',} \textit{ Phys.
  Rev. D} \textbf{ 86} (2012) 055010,
  \href{http://dx.doi.org/10.1103/PhysRevD.86.055010}{\doi{10.1103/PhysRevD.86.055010}},
  \href{http://www.arXiv.org/abs/1201.5014}{\texttt{arXiv:1201.5014}}.

\bibitem{Aad:2012dm}
\hrefCMSnoop {}{{{ATLAS}} Collaboration, ``{ATLAS} search for a heavy gauge
  boson decaying to a charged lepton and a neutrino in {$\Pp\Pp$} collisions at
  $\sqrt{s}=7$ {TeV}'',} \textit{ Eur. Phys. J. C} \textbf{ 72} (2012) 2241,
  \href{http://dx.doi.org/10.1140/epjc/s10052-012-2241-5}{\doi{10.1140/epjc/s10052-012-2241-5}},
\href{http://www.arXiv.org/abs/1209.4446}{\texttt{arXiv:1209.4446}}.
%%CITATION = ARXIV:1209.4446;%%.

\bibitem{Chatrchyan:2012meb}
\hrefCMSnoop {}{{CMS Collaboration}, ``Search for leptonic decays of {\PWpr}
  bosons in {$\Pp\Pp$} collisions at $\sqrt{s}=7$ {TeV}'',} \textit{ JHEP}
  \textbf{ 08} (2012) 023,
  \href{http://dx.doi.org/10.1007/JHEP08(2012)023}{\doi{10.1007/JHEP08(2012)023}},
\href{http://www.arXiv.org/abs/1204.4764}{\texttt{arXiv:1204.4764}}.
%%CITATION = ARXIV:1204.4764;%%.

\bibitem{ATLAS:2014wra}
\hrefCMSnoop {}{{ATLAS Collaboration}, ``Search for new particles in events
  with one lepton and missing transverse momentum in {$\Pp\Pp$} collisions at
  $\sqrt{s}=8$ {TeV} with the {ATLAS} detector'',} \textit{ JHEP} \textbf{ 09}
  (2014) 037,
  \href{http://dx.doi.org/10.1007/JHEP09(2014)037}{\doi{10.1007/JHEP09(2014)037}},
\href{http://www.arXiv.org/abs/1407.7494}{\texttt{arXiv:1407.7494}}.
%%CITATION = ARXIV:1407.7494;%%.

\bibitem{Khachatryan:2014tva}
\hrefCMSnoop {}{{CMS Collaboration}, ``Search for physics beyond the standard
  model in final states with a lepton and missing transverse energy in
  proton-proton collisions at $\sqrt{s}=8$ {TeV}'',} \textit{ Phys. Rev. D}
  \textbf{ 91} (2015) 092005,
  \href{http://dx.doi.org/10.1103/PhysRevD.91.092005}{\doi{10.1103/PhysRevD.91.092005}},
\href{http://www.arXiv.org/abs/1408.2745}{\texttt{arXiv:1408.2745}}.
%%CITATION = ARXIV:1408.2745;%%.

\bibitem{Aaboud:2017efa}
\hrefCMSnoop {}{{ATLAS Collaboration}, ``Search for a new heavy gauge boson
  resonance decaying into a lepton and missing transverse momentum in 36
  {$\fbinv$} of {$\Pp\Pp$} collisions at $\sqrt{s}=13$ {TeV} with the {ATLAS}
  experiment'',} \textit{ Eur. Phys. J. C} \textbf{ 78} (2018) 401,
  \href{http://dx.doi.org/10.1140/epjc/s10052-018-5877-y}{\doi{10.1140/epjc/s10052-018-5877-y}},
\href{http://www.arXiv.org/abs/1706.04786}{\texttt{arXiv:1706.04786}}.
%%CITATION = ARXIV:1706.04786;%%.

\bibitem{Khachatryan:2016jww}
\hrefCMSnoop {}{{CMS Collaboration}, ``Search for heavy gauge {$\PWpr$} boson
  in events with an energetic lepton and large missing transverse momentum at
  $\sqrt{s}=13$ {TeV}'',} \textit{ Phys. Lett. B} \textbf{ 770} (2017) 278,
  \href{http://dx.doi.org/10.1016/j.physletb.2017.04.043}{\doi{10.1016/j.physletb.2017.04.043}},
\href{http://www.arXiv.org/abs/1612.09274}{\texttt{arXiv:1612.09274}}.
%%CITATION = ARXIV:1612.09274;%%.

\bibitem{Sirunyan:2018mpc}
\hrefCMSnoop {}{{CMS Collaboration}, ``Search for high-mass resonances in final
  states with a lepton and missing transverse momentum at $\sqrt{s}=13$
  {TeV}'',} \textit{ JHEP} \textbf{ 06} (2018) 128,
  \href{http://dx.doi.org/10.1007/JHEP06(2018)128}{\doi{10.1007/JHEP06(2018)128}},
\href{http://www.arXiv.org/abs/1803.11133}{\texttt{arXiv:1803.11133}}.
%%CITATION = ARXIV:1803.11133;%%.

\bibitem{Aad:2019wvl}
\hrefCMSnoop {}{{ATLAS Collaboration}, ``Search for a heavy charged boson in
  events with a charged lepton and missing transverse momentum from {$\Pp\Pp$}
  collisions at $\sqrt{s}=13$ {TeV} with the {ATLAS} detector'',} \textit{
  Phys. Rev. D} \textbf{ 100} (2019) 052013,
  \href{http://dx.doi.org/10.1103/PhysRevD.100.052013}{\doi{10.1103/PhysRevD.100.052013}},
  \href{http://www.arXiv.org/abs/1906.05609}{\texttt{arXiv:1906.05609}}.

\bibitem{Peskin:1991sw}
\hrefCMSnoop {}{M.~E. Peskin and T.~Takeuchi, ``Estimation of oblique
  electroweak corrections'',} \textit{ Phys. Rev. D} \textbf{ 46} (1992) 381,
\href{http://dx.doi.org/10.1103/PhysRevD.46.381}{\doi{10.1103/PhysRevD.46.381}}.
%%CITATION = PHRVA,D46,381;%%.

\bibitem{Falkowski:2015krw}
\hrefCMSnoop {}{A.~Falkowski and K.~Mimouni, ``Model independent constraints on
  four-lepton operators'',} \textit{ JHEP} \textbf{ 02} (2016) 086,
  \href{http://dx.doi.org/10.1007/JHEP02(2016)086}{\doi{10.1007/JHEP02(2016)086}},
  \href{http://www.arXiv.org/abs/1511.07434}{\texttt{arXiv:1511.07434}}.

\bibitem{LEP-2}
\hrefCMSnoop {}{{ALEPH, DELPHI, L3, and OPAL Collaborations, and the LEP
  Electroweak Working Group}, ``{Electroweak} measurements in electron-positron
  collisions at {$\PW$}-boson-pair energies at {LEP}'',} \textit{ Phys. Rept.}
  \textbf{ 532} (2013) 119,
  \href{http://dx.doi.org/10.1016/j.physrep.2013.07.004}{\doi{10.1016/j.physrep.2013.07.004}},
\href{http://www.arXiv.org/abs/1302.3415}{\texttt{arXiv:1302.3415}}.
%%CITATION = ARXIV:1302.3415;%%.

\bibitem{Giudice_2007}
\hrefCMSnoop {}{G.~F. Giudice, C.~Grojean, A.~Pomarol, and R.~Rattazzi, ``The
  strongly-interacting light {Higgs}'',} \textit{ JHEP} \textbf{ 06} (2007)
  045,
  \href{http://dx.doi.org/10.1088/1126-6708/2007/06/045}{\doi{10.1088/1126-6708/2007/06/045}},
  \href{http://www.arXiv.org/abs/hep-ph/0703164}{\texttt{arXiv:hep-ph/0703164}}.

\bibitem{Wulzer_2017}
\hrefCMSnoop {}{A.~Thamm, R.~Torre, and A.~Wulzer, ``Future tests of {H}iggs
  compositeness: direct vs indirect'',} \textit{ JHEP} \textbf{ 07} (2015) 100,
  \href{http://dx.doi.org/10.1007/jhep07(2015)100}{\doi{10.1007/jhep07(2015)100}},
  \href{http://www.arXiv.org/abs/1502.01701}{\texttt{arXiv:1502.01701}}.

\bibitem{Higgs_combination}
\hrefCMSnoop {}{{CMS Collaboration}, ``Combined measurements of higgs boson
  couplings in proton-proton collisions at $\sqrt{s}=13$ {TeV}'',} \textit{
  Eur. Phys. J. C} \textbf{ 79} (2019) 421,
  \href{http://dx.doi.org/10.1140/epjc/s10052-019-6909-y}{\doi{10.1140/epjc/s10052-019-6909-y}},
\href{http://www.arXiv.org/abs/1809.10733}{\texttt{arXiv:1809.10733}}.
%%CITATION = ARXIV:1809.10733;%%.

\bibitem{hepdata}
\hrefCMSnoop {}{}{HEPD}ata record for this analysis, 2021.
\newblock
  \href{http://dx.doi.org/10.17182/hepdata.106058}{\doi{10.17182/hepdata.106058}}.

\bibitem{Sirunyan:2020zal}
\hrefCMSnoop {}{{CMS Collaboration}, ``Performance of the {CMS} {Level-1}
  trigger in proton-proton collisions at $\sqrt{s}=13$\, {TeV}'',} \textit{
  JINST} \textbf{ 15} (2020) P10017,
  \href{http://dx.doi.org/10.1088/1748-0221/15/10/P10017}{\doi{10.1088/1748-0221/15/10/P10017}},
  \href{http://www.arXiv.org/abs/2006.10165}{\texttt{arXiv:2006.10165}}.

\bibitem{Sirunyan:2021zrd}
\hrefCMSnoop {}{{CMS Collaboration}, ``Performance of the {CMS} muon trigger
  system in proton-proton collisions at $\sqrt{s}=13$ {TeV}'',} \textit{ JINST}
  \textbf{ 16} (2021) P07001,
  \href{http://dx.doi.org/10.1088/1748-0221/16/07/p07001}{\doi{10.1088/1748-0221/16/07/p07001}},
  \href{http://www.arXiv.org/abs/2102.04790}{\texttt{arXiv:2102.04790}}.

\bibitem{Khachatryan:2016bia}
\hrefCMSnoop {}{{CMS Collaboration}, ``The {CMS} trigger system'',} \textit{
  JINST} \textbf{ 12} (2017) P01020,
  \href{http://dx.doi.org/10.1088/1748-0221/12/01/P01020}{\doi{10.1088/1748-0221/12/01/P01020}},
\href{http://www.arXiv.org/abs/1609.02366}{\texttt{arXiv:1609.02366}}.
%%CITATION = ARXIV:1609.02366;%%.

\bibitem{Chatrchyan:2008aa}
\hrefCMSnoop {}{{CMS Collaboration}, ``The {CMS} experiment at the {CERN}
  {LHC}'',} \textit{ JINST} \textbf{ 3} (2008) S08004,
\href{http://dx.doi.org/10.1088/1748-0221/3/08/S08004}{\doi{10.1088/1748-0221/3/08/S08004}}.
%%CITATION = JINST,3,S08004;%%.

\bibitem{Sjostrand:2014zea}
T.~Sj{\"o}strand\hrefCMSnoop {}{ {et~al.}, ``An introduction to {PYTHIA
  8.2}'',} \textit{ Comput. Phys. Commun.} \textbf{ 191} (2015) 159,
  \href{http://dx.doi.org/10.1016/j.cpc.2015.01.024}{\doi{10.1016/j.cpc.2015.01.024}},
\href{http://www.arXiv.org/abs/1410.3012}{\texttt{arXiv:1410.3012}}.
%%CITATION = ARXIV:1410.3012;%%.

\bibitem{NNPDF31}
\hrefCMSnoop {}{{NNPDF} Collaboration, ``Parton distributions from
  high-precision collider data'',} \textit{ Eur. Phys. J. C} \textbf{ 77}
  (2017) 663,
  \href{http://dx.doi.org/10.1140/epjc/s10052-017-5199-5}{\doi{10.1140/epjc/s10052-017-5199-5}},
\href{http://www.arXiv.org/abs/1706.00428}{\texttt{arXiv:1706.00428}}.
%%CITATION = ARXIV:1706.00428;%%.

\bibitem{Vogt:2002eu}
\hrefCMSnoop {}{R.~Vogt, ``The usage of the {K} factor in heavy ion physics'',}
  \textit{ Acta Phys. Hung. A} \textbf{ 17} (2003) 75,
  \href{http://dx.doi.org/10.1556/APH.17.2003.1.9}{\doi{10.1556/APH.17.2003.1.9}},
  \href{http://www.arXiv.org/abs/hep-ph/0207359}{\texttt{arXiv:hep-ph/0207359}}.

\bibitem{fewz3}
\hrefCMSnoop {}{Y.~Li and F.~Petriello, ``Combining {QCD} and electroweak
  corrections to dilepton production in {FEWZ}'',} \textit{ Phys. Rev. D}
  \textbf{ 86} (2012) 094034,
  \href{http://dx.doi.org/10.1103/PhysRevD.86.094034}{\doi{10.1103/PhysRevD.86.094034}},
\href{http://www.arXiv.org/abs/1208.5967}{\texttt{arXiv:1208.5967}}.
%%CITATION = ARXIV:1208.5967;%%.

\bibitem{Alwall:2014hca}
J.~Alwall\hrefCMSnoop {}{ {et~al.}, ``The automated computation of tree-level
  and next-to-leading order differential cross sections, and their matching to
  parton shower simulations'',} \textit{ JHEP} \textbf{ 07} (2014) 079,
  \href{http://dx.doi.org/10.1007/JHEP07(2014)079}{\doi{10.1007/JHEP07(2014)079}},
  \href{http://www.arXiv.org/abs/1405.0301}{\texttt{arXiv:1405.0301}}.

\bibitem{NNPDF23}
\hrefCMSnoop {}{R.~D. Ball {et~al.}, ``Parton distributions with {LHC} data'',}
  \textit{ Nucl. Phys. B} \textbf{ 867} (2013) 244,
  \href{http://dx.doi.org/10.1016/j.nuclphysb.2012.10.003}{\doi{10.1016/j.nuclphysb.2012.10.003}},
\href{http://www.arXiv.org/abs/1207.1303}{\texttt{arXiv:1207.1303}}.
%%CITATION = ARXIV:1207.1303;%%.

\bibitem{Farina_2017}
M.~Farina\hrefCMSnoop {}{ {et~al.}, ``Energy helps accuracy: {Electroweak}
  precision tests at hadron colliders'',} \textit{ Phys. Lett. B} \textbf{ 772}
  (2017) 210,
  \href{http://dx.doi.org/10.1016/j.physletb.2017.06.043}{\doi{10.1016/j.physletb.2017.06.043}},
  \href{http://www.arXiv.org/abs/1609.08157}{\texttt{arXiv:1609.08157}}.

\bibitem{HVT_model}
\hrefCMSnoop {}{D.~Pappadopulo, A.~Thamm, R.~Torre, and A.~Wulzer, ``Heavy
  vector triplets: Bridging theory and data'',} \textit{ JHEP} \textbf{ 09}
  (2014) 060,
  \href{http://dx.doi.org/10.1007/jhep09(2014)060}{\doi{10.1007/jhep09(2014)060}},
  \href{http://www.arXiv.org/abs/1402.4431}{\texttt{arXiv:1402.4431}}.

\bibitem{Sirunyan:2019dfx}
\hrefCMSnoop {}{{CMS Collaboration}, ``Extraction and validation of a new set
  of {CMS} {\PYTHIA8} tunes from underlying-event measurements'',} \textit{
  Eur. Phys. J. C} \textbf{ 80} (2020) 4,
  \href{http://dx.doi.org/10.1140/epjc/s10052-019-7499-4}{\doi{10.1140/epjc/s10052-019-7499-4}},
  \href{http://www.arXiv.org/abs/1903.12179}{\texttt{arXiv:1903.12179}}.

\bibitem{mcsanc_v1}
A.~Arbuzov\hrefCMSnoop {}{ {et~al.}, ``Update of the {MCSANC} {Monte Carlo}
  integrator, v.1.20'',} \textit{ JETP Lett.} \textbf{ 103} (2016) 131,
  \href{http://dx.doi.org/10.1134/S0021364016020041}{\doi{10.1134/S0021364016020041}},
\href{http://www.arXiv.org/abs/1509.03052}{\texttt{arXiv:1509.03052}}.
%%CITATION = ARXIV:1509.03052;%%.

\bibitem{mcsanv_2}
\hrefCMSnoop {}{S.~Alioli {et~al.}, ``Precision studies of observables in {$pp
  \to$ {W} $\to l\nu$} and {$pp \to \gamma$,{Z} $\to l^{+}l^{-}$} processes at
  the {LHC}'',} \textit{ Eur. Phys. J. C} \textbf{ 77} (2017) 280,
  \href{http://dx.doi.org/10.1140/epjc/s10052-017-4832-7}{\doi{10.1140/epjc/s10052-017-4832-7}},
\href{http://www.arXiv.org/abs/1606.02330}{\texttt{arXiv:1606.02330}}.
%%CITATION = ARXIV:1606.02330;%%.

\bibitem{Bertone_2018}
\hrefCMSnoop {}{V.~Bertone, S.~Carrazza, N.~Hartland, and J.~Rojo,
  ``Illuminating the photon content of the proton within a global {PDF}
  analysis'',} \textit{ SciPost Phys.} \textbf{ 5} (2018) 008,
  \href{http://dx.doi.org/10.21468/scipostphys.5.1.008}{\doi{10.21468/scipostphys.5.1.008}},
\href{http://www.arXiv.org/abs/1712.07053}{\texttt{arXiv:1712.07053}}.
%%CITATION = ARXIV:1712.07053;%%.

\bibitem{Manohar_2016}
\hrefCMSnoop {}{A.~Manohar, P.~Nason, G.~P. Salam, and G.~Zanderighi, ``How
  bright is the proton? {A} precise determination of the photon parton
  distribution function'',} \textit{ Phys. Rev. Lett.} \textbf{ 117} (2016) 24,
  \href{http://dx.doi.org/10.1103/physrevlett.117.242002}{\doi{10.1103/physrevlett.117.242002}},
\href{http://www.arXiv.org/abs/1607.04266}{\texttt{arXiv:1607.04266}}.
%%CITATION = ARXIV:1607.04266;%%.

\bibitem{Butterworth_2016}
\hrefCMSnoop {}{J.~Butterworth {et~al.}, ``{PDF4LHC} recommendations for {LHC
  Run} {II}'',} \textit{ J. Phys. G} \textbf{ 43} (2016) 023001,
  \href{http://dx.doi.org/10.1088/0954-3899/43/2/023001}{\doi{10.1088/0954-3899/43/2/023001}},
  \href{http://www.arXiv.org/abs/1510.03865}{\texttt{arXiv:1510.03865}}.

\bibitem{Frederix:2012ps}
\hrefCMSnoop {}{R.~Frederix and S.~Frixione, ``Merging meets matching in
  {MC@NLO}'',} \textit{ JHEP} \textbf{ 12} (2012) 061,
  \href{http://dx.doi.org/10.1007/JHEP12(2012)061}{\doi{10.1007/JHEP12(2012)061}},
  \href{http://www.arXiv.org/abs/1209.6215}{\texttt{arXiv:1209.6215}}.

\bibitem{Alwall:2007fs}
J.~Alwall\hrefCMSnoop {}{ {et~al.}, ``Comparative study of various algorithms
  for the merging of parton showers and matrix elements in hadronic
  collisions'',} \textit{ Eur. Phys. J. C} \textbf{ 53} (2008) 473,
  \href{http://dx.doi.org/10.1140/epjc/s10052-007-0490-5}{\doi{10.1140/epjc/s10052-007-0490-5}},
\href{http://www.arXiv.org/abs/0706.2569}{\texttt{arXiv:0706.2569}}.
%%CITATION = ARXIV:0706.2569;%%.

\bibitem{Lindert:2017olm}
\hrefCMSnoop {}{J.~M. Lindert {et~al.}, ``Precise predictions for {$V+$} jets
  dark matter backgrounds'',} \textit{ Eur. Phys. J. C} \textbf{ 77} (2017)
  829,
  \href{http://dx.doi.org/10.1140/epjc/s10052-017-5389-1}{\doi{10.1140/epjc/s10052-017-5389-1}},
  \href{http://www.arXiv.org/abs/1705.04664}{\texttt{arXiv:1705.04664}}.

\bibitem{Nason:2004rx}
\hrefCMSnoop {}{P.~Nason, ``A new method for combining {NLO QCD} with shower
  {Monte Carlo} algorithms'',} \textit{ JHEP} \textbf{ 11} (2004) 040,
  \href{http://dx.doi.org/10.1088/1126-6708/2004/11/040}{\doi{10.1088/1126-6708/2004/11/040}},
  \href{http://www.arXiv.org/abs/hep-ph/0409146}{\texttt{arXiv:hep-ph/0409146}}.

\bibitem{Frixione:2007vw}
\hrefCMSnoop {}{S.~Frixione, P.~Nason, and C.~Oleari, ``{Matching NLO QCD}
  computations with parton shower simulations: the {POWHEG method}'',} \textit{
  JHEP} \textbf{ 11} (2007) 070,
  \href{http://dx.doi.org/10.1088/1126-6708/2007/11/070}{\doi{10.1088/1126-6708/2007/11/070}},
  \href{http://www.arXiv.org/abs/0709.2092}{\texttt{arXiv:0709.2092}}.

\bibitem{Alioli:2010xd}
\hrefCMSnoop {}{S.~Alioli, P.~Nason, C.~Oleari, and E.~Re, ``{A general
  framework for implementing NLO calculations in shower Monte Carlo programs:
  the POWHEG BOX}'',} \textit{ JHEP} \textbf{ 06} (2010) 043,
  \href{http://dx.doi.org/10.1007/JHEP06(2010)043}{\doi{10.1007/JHEP06(2010)043}},
  \href{http://www.arXiv.org/abs/1002.2581}{\texttt{arXiv:1002.2581}}.

\bibitem{PhysRevLett.110.252004}
\hrefCMSnoop {}{M.~Czakon, P.~Fiedler, and A.~Mitov, ``Total top-quark
  pair-production cross section at hadron colliders through
  $\mathcal{O}({\ensuremath{\alpha}}_{S}^{4})$'',} \textit{ Phys. Rev. Lett.}
  \textbf{ 110} (2013) 252004,
  \href{http://dx.doi.org/10.1103/PhysRevLett.110.252004}{\doi{10.1103/PhysRevLett.110.252004}},
  \href{http://www.arXiv.org/abs/1303.6254}{\texttt{arXiv:1303.6254}}.

\bibitem{Czakon:2017wor}
M.~Czakon\hrefCMSnoop {}{ {et~al.}, ``Top-pair production at the {LHC} through
  {NNLO QCD} and {NLO EW}'',} \textit{ JHEP} \textbf{ 10} (2017) 186,
  \href{http://dx.doi.org/10.1007/JHEP10(2017)186}{\doi{10.1007/JHEP10(2017)186}},
\href{http://www.arXiv.org/abs/1705.04105}{\texttt{arXiv:1705.04105}}.
%%CITATION = ARXIV:1705.04105;%%.

\bibitem{PhysRevLett.113.212001}
T.~Gehrmann\hrefCMSnoop {}{ {et~al.}, ``{${\PW^{+}\PW^{\ensuremath{-}}}$}
  production at hadron colliders in {Next-to-next-to-leading order QCD}'',}
  \textit{ Phys. Rev. Lett.} \textbf{ 113} (2014) 212001,
  \href{http://dx.doi.org/10.1103/PhysRevLett.113.212001}{\doi{10.1103/PhysRevLett.113.212001}},
  \href{http://www.arXiv.org/abs/1408.5243}{\texttt{arXiv:1408.5243}}.

\bibitem{Campbell_2011}
\hrefCMSnoop {}{J.~M. Campbell, R.~K. Ellis, and C.~Williams, ``Vector boson
  pair production at the {LHC}'',} \textit{ JHEP} \textbf{ 07} (2011) 018,
  \href{http://dx.doi.org/10.1007/JHEP07(2011)018}{\doi{10.1007/JHEP07(2011)018}},
  \href{http://www.arXiv.org/abs/1105.0020}{\texttt{arXiv:1105.0020}}.

\bibitem{Cascioli_2014}
F.~Cascioli\hrefCMSnoop {}{ {et~al.}, ``{ZZ} production at hadron colliders in
  {NNLO QCD}'',} \textit{ Phys. Lett. B} \textbf{ 735} (2014) 311,
  \href{http://dx.doi.org/10.1016/j.physletb.2014.06.056}{\doi{10.1016/j.physletb.2014.06.056}}.

\bibitem{Agostinelli:2002hh}
\hrefCMSnoop {}{{GEANT4} Collaboration, ``{\GEANTfour}---a simulation
  toolkit'',} \textit{ Nucl. Instrum. Meth. A} \textbf{ 506} (2003) 250,
\href{http://dx.doi.org/10.1016/S0168-9002(03)01368-8}{\doi{10.1016/S0168-9002(03)01368-8}}.
%%CITATION = NUIMA,A506,250;%%.

\bibitem{ALLISON2016186}
\hrefCMSnoop {}{J.~Allison {et~al.}, ``Recent developments in {\GEANTfour}'',}
  \textit{ Nucl. Instrum. Meth. A} \textbf{ 835} (2016) 186,
  \href{http://dx.doi.org/10.1016/j.nima.2016.06.125}{\doi{10.1016/j.nima.2016.06.125}}.

\bibitem{Sirunyan:2018nqx}
\hrefCMSnoop {}{{CMS Collaboration}, ``Measurement of the inelastic
  proton-proton cross section at $\sqrt{s}=13$ {TeV}'',} \textit{ JHEP}
  \textbf{ 07} (2018) 161,
  \href{http://dx.doi.org/10.1007/JHEP07(2018)161}{\doi{10.1007/JHEP07(2018)161}},
\href{http://www.arXiv.org/abs/1802.02613}{\texttt{arXiv:1802.02613}}.
%%CITATION = ARXIV:1802.02613;%%.

\bibitem{Cacciari:2008gp}
\hrefCMSnoop {}{M.~Cacciari, G.~P. Salam, and G.~Soyez, ``{The anti-$\kt$ jet
  clustering algorithm}'',} \textit{ JHEP} \textbf{ 04} (2008) 063,
  \href{http://dx.doi.org/10.1088/1126-6708/2008/04/063}{\doi{10.1088/1126-6708/2008/04/063}},
\href{http://www.arXiv.org/abs/0802.1189}{\texttt{arXiv:0802.1189}}.
%%CITATION = ARXIV:0802.1189;%%.

\bibitem{Cacciari:2011ma}
\hrefCMSnoop {}{M.~Cacciari, G.~P. Salam, and G.~Soyez, ``{FastJet} user
  manual'',} \textit{ Eur. Phys. J. C} \textbf{ 72} (2012) 1896,
  \href{http://dx.doi.org/10.1140/epjc/s10052-012-1896-2}{\doi{10.1140/epjc/s10052-012-1896-2}},
\href{http://www.arXiv.org/abs/1111.6097}{\texttt{arXiv:1111.6097}}.
%%CITATION = ARXIV:1111.6097;%%.

\bibitem{Sirunyan:2017ulk}
\hrefCMSnoop {}{{CMS Collaboration}, ``Particle-flow reconstruction and global
  event description with the cms detector'',} \textit{ JINST} \textbf{ 12}
  (2017) P10003,
  \href{http://dx.doi.org/10.1088/1748-0221/12/10/P10003}{\doi{10.1088/1748-0221/12/10/P10003}},
\href{http://www.arXiv.org/abs/1706.04965}{\texttt{arXiv:1706.04965}}.
%%CITATION = ARXIV:1706.04965;%%.

\bibitem{Khachatryan:2016zqb}
\hrefCMSnoop {}{{CMS Collaboration}, ``Search for narrow resonances in dilepton
  mass spectra in proton-proton collisions at $\sqrt{s}=13$ {TeV} and
  combination with 8 {TeV} data'',} \textit{ Phys. Lett. B} \textbf{ 768}
  (2017) 57,
  \href{http://dx.doi.org/10.1016/j.physletb.2017.02.010}{\doi{10.1016/j.physletb.2017.02.010}},
\href{http://www.arXiv.org/abs/1609.05391}{\texttt{arXiv:1609.05391}}.
%%CITATION = ARXIV:1609.05391;%%.

\bibitem{EGM-17-01}
\hrefCMSnoop {}{{CMS Collaboration}, ``Electron and photon reconstruction and
  identification with the {CMS} experiment at the {CERN LHC}'',} \textit{
  JINST} \textbf{ 16} (2021) P05014,
  \href{http://dx.doi.org/10.1088/1748-0221/16/05/p05014}{\doi{10.1088/1748-0221/16/05/p05014}},
\href{http://www.arXiv.org/abs/2012.06888}{\texttt{arXiv:2012.06888}}.
%%CITATION = ARXIV:2012.06888;%%.

\bibitem{Sirunyan:2018fpa}
\hrefCMSnoop {}{{CMS Collaboration}, ``Performance of the {CMS} muon detector
  and muon reconstruction with proton-proton collisions at $\sqrt{s}=13$
  {TeV}'',} \textit{ JINST} \textbf{ 13} (2018) P06015,
  \href{http://dx.doi.org/10.1088/1748-0221/13/06/P06015}{\doi{10.1088/1748-0221/13/06/P06015}},
\href{http://www.arXiv.org/abs/1804.04528}{\texttt{arXiv:1804.04528}}.
%%CITATION = ARXIV:1804.04528;%%.

\bibitem{HighpTMupaper}
\hrefCMSnoop {}{{CMS Collaboration}, ``Performance of the reconstruction and
  identification of high-momentum muons in proton-proton collisions at
  $\sqrt{s}=13$ {TeV}'',} \textit{ JINST} \textbf{ 15} (2020) P02027,
  \href{http://dx.doi.org/10.1088/1748-0221/15/02/P02027}{\doi{10.1088/1748-0221/15/02/P02027}},
  \href{http://www.arXiv.org/abs/1912.03516}{\texttt{arXiv:1912.03516}}.

\bibitem{Khachatryan:2016kdb}
\hrefCMSnoop {}{{CMS Collaboration}, ``Jet energy scale and resolution in the
  {CMS} experiment in {$\Pp\Pp$} collisions at 8 {TeV}'',} \textit{ JINST}
  \textbf{ 12} (2017) P02014,
  \href{http://dx.doi.org/10.1088/1748-0221/12/02/P02014}{\doi{10.1088/1748-0221/12/02/P02014}},
\href{http://www.arXiv.org/abs/1607.03663}{\texttt{arXiv:1607.03663}}.
%%CITATION = ARXIV:1607.03663;%%.

\bibitem{Sirunyan:2017ezt}
\hrefCMSnoop {}{{CMS Collaboration}, ``Identification of heavy-flavour jets
  with the {CMS} detector in {$\Pp\Pp$} collisions at 13 {TeV}'',} \textit{
  JINST} \textbf{ 13} (2018) P05011,
  \href{http://dx.doi.org/10.1088/1748-0221/13/05/P05011}{\doi{10.1088/1748-0221/13/05/P05011}},
\href{http://www.arXiv.org/abs/1712.07158}{\texttt{arXiv:1712.07158}}.
%%CITATION = ARXIV:1712.07158;%%.

\bibitem{collaboration_2011}
\hrefCMSnoop {}{{CMS Collaboration}, ``Determination of jet energy calibration
  and transverse momentum resolution in {CMS}'',} \textit{ JINST} \textbf{ 6}
  (2011) P11002,
  \href{http://dx.doi.org/10.1088/1748-0221/6/11/P11002}{\doi{10.1088/1748-0221/6/11/P11002}},
  \href{http://www.arXiv.org/abs/1107.4277}{\texttt{arXiv:1107.4277}}.

\bibitem{CMS:2011aa}
\hrefCMSnoop {}{{CMS Collaboration}, ``Measurement of the inclusive {$\PW$} and
  {$\PZ$} production cross sections in {$\Pp\Pp$} collisions at $\sqrt{s}=7$
  {TeV}'',} \textit{ JHEP} \textbf{ 10} (2011) 132,
  \href{http://dx.doi.org/10.1007/JHEP10(2011)132}{\doi{10.1007/JHEP10(2011)132}},
  \href{http://www.arXiv.org/abs/1107.4789}{\texttt{arXiv:1107.4789}}.

\bibitem{cmscollaboration2021precision}
\hrefCMSnoop {}{{CMS Collaboration}, ``Precision luminosity measurement in
  proton-proton collisions at $\sqrt{s}=13$ {TeV} in 2015 and 2016 at {CMS}'',}
  \textit{ Eur. Phys. J. C} \textbf{ 81} (2021) 800,
  \href{http://dx.doi.org/10.1140/epjc/s10052-021-09538-2}{\doi{10.1140/epjc/s10052-021-09538-2}},
\href{http://www.arXiv.org/abs/2104.01927}{\texttt{arXiv:2104.01927}}.
%%CITATION = ARXIV:2104.01927;%%.

\bibitem{CMS-PAS-LUM-17-004}
\href {https://cds.cern.ch/record/2621960}{{{CMS}} Collaboration, ``{CMS}
  luminosity measurement for the 2017 data-taking period at $\sqrt{s}=13$
  {TeV}'',} CMS Physics Analysis Summary CMS-PAS-LUM-17-004, 2018.

\bibitem{CMS:2019jhq}
\href {https://cds.cern.ch/record/2676164}{{CMS Collaboration}, ``{CMS}
  luminosity measurement for the 2018 data-taking period at $\sqrt{s}=13$
  {TeV}'',} CMS Physics Analysis Summary CMS-PAS-LUM-18-002, 2019.

\bibitem{conway2011incorporating}
\hrefCMSnoop {}{J.~S. Conway, ``{Incorporating} nuisance parameters in
  likelihoods for multisource spectra'',} 2011.
  \href{http://www.arXiv.org/abs/1103.0354}{\texttt{arXiv:1103.0354}}.

\bibitem{BARLOW1993219}
\hrefCMSnoop {}{R.~Barlow and C.~Beeston, ``Fitting using finite {Monte Carlo}
  samples'',} \textit{ Comput. Phys. Commun.} \textbf{ 77} (1993) 219,
  \href{http://dx.doi.org/10.1016/0010-4655(93)90005-W}{\doi{10.1016/0010-4655(93)90005-W}}.

\bibitem{MINUIT}
\href {http://cdssls.cern.ch/record/2296388/}{F.~James, ``{MINUIT}: function
  minimization and error analysis reference manual'',} Technical Report
  CERN-D-506, 1998.

\bibitem{Khachatryan:2014jba}
\hrefCMSnoop {}{{CMS Collaboration}, ``Precise determination of the mass of the
  {Higgs} boson and tests of compatibility of its couplings with the standard
  model predictions using proton collisions at 7 and 8 {TeV}'',} \textit{ Eur.
  Phys. J. C} \textbf{ 75} (2015) 212,
  \href{http://dx.doi.org/10.1140/epjc/s10052-015-3351-7}{\doi{10.1140/epjc/s10052-015-3351-7}},
\href{http://www.arXiv.org/abs/1412.8662}{\texttt{arXiv:1412.8662}}.
%%CITATION = ARXIV:1412.8662;%%.

\end{thebibliography}\endgroup
\cleardoublepage \appendix\section{The CMS Collaboration \label{app:collab}}\begin{sloppypar}\hyphenpenalty=5000\widowpenalty=500\clubpenalty=5000\cmsinstitute{Yerevan~Physics~Institute, Yerevan, Armenia}
A.~Tumasyan
\cmsinstitute{Institut~f\"{u}r~Hochenergiephysik, Vienna, Austria}
W.~Adam\cmsorcid{0000-0001-9099-4341}, J.W.~Andrejkovic, T.~Bergauer\cmsorcid{0000-0002-5786-0293}, S.~Chatterjee\cmsorcid{0000-0003-2660-0349}, M.~Dragicevic\cmsorcid{0000-0003-1967-6783}, A.~Escalante~Del~Valle\cmsorcid{0000-0002-9702-6359}, R.~Fr\"{u}hwirth\cmsAuthorMark{1}, M.~Jeitler\cmsAuthorMark{1}\cmsorcid{0000-0002-5141-9560}, N.~Krammer, L.~Lechner\cmsorcid{0000-0002-3065-1141}, D.~Liko, I.~Mikulec, P.~Paulitsch, F.M.~Pitters, J.~Schieck\cmsAuthorMark{1}\cmsorcid{0000-0002-1058-8093}, R.~Sch\"{o}fbeck\cmsorcid{0000-0002-2332-8784}, M.~Spanring\cmsorcid{0000-0001-6328-7887}, S.~Templ\cmsorcid{0000-0003-3137-5692}, W.~Waltenberger\cmsorcid{0000-0002-6215-7228}, C.-E.~Wulz\cmsAuthorMark{1}\cmsorcid{0000-0001-9226-5812}
% ...
\vskip\cmsinstskip
% ...
1:~Also at TU Wien, Wien, Austria\\
% ...
\end{sloppypar}
%%% END EDITABLE REGION %%%
% skeleton_end
\end{document}